\documentclass[12pt]{article}
\input{epsf}

\makeatletter
\def\Ddots{\mathinner{\mkern1mu\raise\p@
\vbox{\kern7\p@\hbox{.}}\mkern2mu
\raise4\p@\hbox{.}\mkern2mu\raise7\p@\hbox{.}\mkern1mu}}
\makeatother
\font\bm=cmmib10 at 10pt
\font\bms=cmmib10 at 7pt \textfont9=\bm \scriptfont9=\bms
\usepackage{amsfonts}
\usepackage{multirow}
\mathchardef\bupsilon= "791D
\mathchardef\bdelta= "790E
\mathchardef\bgamma= "790D
\mathchardef\balpha= "790B
\mathchardef\bbeta= "790C
\mathchardef\bTheta= "7902
\mathchardef\bzeta= "7910
\mathchardef\bOmega= "790A
\mathchardef\bGamma= "7900
\mathchardef\bDelta= "7901
\mathchardef\bPhi= "7908
\mathchardef\bphi= "791E
\mathchardef\bomega= "7921
\mathchardef\bxi= "7918
\mathchardef\bet= "7911
\mathchardef\brho= "791A
\mathchardef\btau= "791C
\mathchardef\bmu= "7916
\mathchardef\bvarpi= "7924

\def \rstar {\raisebox{-1pt}{*}}

\def \lvec{(\kern-.26em(}
\def\pmb#1{\setbox0=\hbox{#1}
\kern-.025em\copy0\kern-\wd0
\kern.05em\copy0\kern-\wd0
\kern-.025em\raise.0433em\box0 }
\mathchardef\btheta= "7912
\usepackage{amsmath}
\usepackage{amssymb}
\usepackage{authblk}
\usepackage{url}
\providecommand{\keywords}[1]{\textbf{\textit{Keywords:  }} #1}
\usepackage{graphicx}% Include figure files
\usepackage{dcolumn}% Align table columns on decimal point

\begin{document}
\title{Radiation Transfer in Cloud Layers}
\author[1]{ W. A. van Wijngaarden}
\author[2] {W. Happer}
\affil[1]{Department of Physics and Astronomy, York University, Canada}
\affil[2]{Department of Physics, Princeton University, USA}
\renewcommand\Affilfont{\itshape\small}
\date{\today}
\maketitle
\begin{abstract}
\noindent For $2n$-stream radiation transfer theory, a stack of $m = 2,3,4,\ldots$ clouds can be represented as an equivalent cloud. Individual clouds, indexed by $c=1,2,3,\ldots, m$, are characterized by $2n\times 2n$ scattering matrices $\mathcal{S}^{\{c\}}$, that describe how the cloud interacts with $2n$ streams of axially symmetric incoming radiation, propagating in upward and downward Gauss-Legendre  sample directions. 
Some of the radiation is transmitted, some is absorbed and converted to heat, and some is scattered into $2n$ outgoing streams along the directions of the incoming streams.
The clouds are also characterized by $2n\times 1$ thermal source vectors $|\dot J^{\{c\}}\}$ that  describe the thermal emission of radiation along the stream directions by cloud particulates and gas molecules. The $2n\times 2n$ scattering matrix for the equivalent cloud, $\mathcal{S}^{\{\rm ev\}}=S^{\{m\}}\rstar S^{\{m-1\}}\rstar\cdots S^{\{2\}}\rstar S^{\{1\}}$,  is the Redheffer star product of the scattering matrices $\mathcal{S}^{\{c\}}$ of the individual clouds. The $2n\times 1$ thermal source vector for the equivalent cloud,   $|\dot J^{\{\rm ev\}}\}=G^{[m,m\}}|\dot J^{\{m\}}\}+G^{[m,m-1\}}|\dot J^{\{m-1\}}\}+\cdots+G^{[m,2\}}|\dot J^{\{2\}}\}+G^{[m,1\}}|\dot J^{\{1\}}\}$, is a linear combination of the thermal source vectors $|\dot J^{\{c\}}\}$ of the individual clouds. The $2n\times 2n$  discrete Green's matrices $G^{[m,c\}}$ can be constructed from the scattering matrices $\mathcal{S}^{\{c\}}$ of individual clouds. The equivalent scattering matrix  $\mathcal{S}^{\{\rm ev\}}$ and the equivalent thermal source vector $|\dot J^{\{\rm ev\}}\}$ are analogous to the  equivalent resistance and the equivalent electromotive force (emf) of Th\'{e}venin's theorem for a network of electrical circuits.  Illustrative numerical examples are given for single clouds,  3-cloud stacks and  10-cloud stacks.  These methods are useful for modeling radiation transfer in Earth's atmosphere, which can  be represented by layers of  invisible clouds,  consisting of clear air and greenhouse gases, or visible clouds  which also include condensed water, smog, etc.
\end{abstract}
\keywords{radiative transfer, multiple scattering, reflection, absorption, emission, scatttering phase functions, equation of transfer, Gauss-Legendre quadrature, two-port networks}
\newpage
\tableofcontents
\newpage
\section{Introduction\label{hist}}
In a previous paper, {\it 2n-Stream Radiative Transfer}\,\cite{WH1}, we outlined how to use operator and matrix methods to accurately and efficiently analyze radiative transfer with angular distributions of scattering, including highly forward scattering of sunlight by cloud particulates and more nearly isotropic Rayleigh scattering by gas molecules in Earth's atmosphere.  In a subsequent paper, {\it $2n$-Stream Conservative Scattering}\,\cite{WH2}, we showed how to extend these methods to the limit of conservative scattering, where photons can only be transmitted or scattered, but not absorbed or emitted.  A third paper of this series, {\it 2n-Stream Thermal Emission from Clouds }\,\cite{WH3}, was devoted  to Kirchhoff's laws of transmission, absorption, scattering and emission of thermal radiation by clouds. The $2n$-stream method used in these papers is a generalization of the 2-stream method of Schuster\,\cite{Schuster} in his classic paper of 1905, {\it Radiation Through a Foggy Atmosphere}.

Our three previous papers on radiation transfer dealt mostly with homogeneous clouds, where the single-scattering albedo $\tilde \omega$ and the scattering phase function $p(\mu,\mu')$ were spatially uniform. In the present paper we will extend this discussion to stacks of individual clouds, $c$, each with its own  scattering matrix $\mathcal{S}^{\{c\}}$, and thermal source vector $|\dot J^{\{c\}}\}$. When scattering incoming radiation, a stack acts like a single  cloud with an equivalent scattering matrix 
$\mathcal{S}^{\{\rm ev\}}=S^{\{m\}} \rstar S^{\{m-1\}} \rstar \cdots S^{\{2\}} \rstar S^{\{1\}}$, the Redheffer star product of the scattering matrices $\mathcal{S}^{\{c\}}$ of the individual clouds.  When emitting thermal radiation, a stack acts like a single cloud with  an equivalent thermal source vector $|\dot J^{\{\rm ev\}}\}=G^{[m,m\}}|\dot J^{\{m\}}\}+G^{[m,m-1\}}|\dot J^{\{m-1\}}\}+\cdots+G^{[m,2\}}|\dot J^{\{2\}}\}+G^{[m,1\}}|\dot J^{\{1\}}\}$,  a linear combination of the thermal source vectors $|\dot J^{\{c\}}\}$ of the individual clouds. The $2n\times 2n$  discrete Green's matrices $G^{[m,c\}}$ can be constructed from the individual scattering matrices $\mathcal{S}^{\{c\}}$.
Stacks of clouds have close analogies to networks of two-port electrical circuits connected in series\,\cite{two-port}. The equivalent scattering matrix  $\mathcal{S}^{\{\rm ev\}}$ and the equivalent thermal source vector $|\dot J^{\{\rm ev\}}\}$ are analogous to the  equivalent resistance and the equivalent electromotive force (emf) of  a network of electrical circuits that follow from Th\'{e}venin's theorem\,\cite{Thevenin}. 

An instructive history of radiative transfer theory has been given by Mobley\cite{Mobley}. To our knowledge, the earliest paper where the $2n$-stream method was used to analyze multiple scattering was published in 1943 by G. C. Wick in connection with his analysis of neutron diffusion, {\it \"Uber ebene Diffusionsprobleme}\, \cite{Wick}.  Wick's work has led to an extensive literature on variants of the $2n$-stream computational methods, often described as the {\it discrete ordinate method} (DOM).  A useful review of DOM  work has been given by Ganapol\cite{Ganapol} in 2015 as   {\it The Response Matrix Discrete Ordinates Solution to the 1D Radiative Transfer Equation}. An early application of the $2n$ stream method to describe radiation transport in clouds was published by Flannery, Roberge and Rybicki\,\cite{Flannery} in 1980 as {\it The Penetration of Diffuse Ultraviolet Radiation into Stellar Clouds}.  We will frequently refer to the authoritative review of radiative transfer  in Chandrasekhar's classic book, {\it Radiative Transfer}\cite{Chandrasekhar}. The book by Thomas and Stamnes\cite{Stamnes}, {\it Radiation Transfer in the Atmosphere and Ocean} has extensive discussions of discrete ordinate methods.
\section{Radiation Intensity and Flux \label{in}}
For axial symmetry about the zenith direction, we will characterize time-independent, steady state  radiation of spatial frequency $\nu$ at an altitude $z$  above Earth's surface with  the monochromatic {\it intensity} $I(\nu,\mu,z)$, also called the {\it radiance}. One can think of the intensity as streams of photons making various angles, $\theta =\cos^{-1}\mu$, with the vertical. $I(\nu,\mu,z)\,d\mu\,d\nu$ is the radiative flux carried by photons with direction cosines between $\mu$ and $\mu+d\mu$ and with spatial frequencies between $\nu$ and $\nu+d\nu$. A representative unit of $I(\nu,\mu,z)$ is W m$^{-2}$ cm sr$^{-1}$,  where W= watts, is the unit of radiation power,  m$^2$ = square meters, is the unit of irradiated area,  cm$^{-1}$ = waves per cm, is the unit of spatial frequency of the radiation, and sr = steradian, is the unit of solid angle.   In {\bf 2.1}(3), Chandrasekhar\cite{Chandrasekhar} uses the symbol $I_{\nu}$ to denote our intensity  $I=I(\nu,\mu,z)$.
For most of the remainder of this paper we will discuss only monochromatic radiation and we will usually omit the frequency variable $\nu$ and write $I(\nu,\mu,z)=I(\mu,z)$.

The intensity $I(\mu,\tau)$ obeys the steady-state {\it equation of transfer},
\begin{equation}
\left(\mu\frac{\partial}{\partial\tau}+1\right) I(\mu, \tau) =(1-\tilde \omega) B(\tau)+\frac{\tilde\omega}{2}
\int_{-1}^1 d\mu' \,p(\mu,\mu')I(\mu',\tau).\label{in10}
\end{equation}
We neglect any variation of the intensity in horizontal spatial directions. Chandrasekhar\cite{Chandrasekhar} writes (\ref{in10}) as \S{\bf 6}(47). He uses the symbol ${\Im}_{\nu}$ to denote both source terms on the right of (\ref{in10}). He writes the emissive part of the source, which is proportional to the Planck intensity $B(\tau)$, as \S{\bf 5}(42). He writes the scattering part of the source, that is proportional to the scattering phase $p(\mu,\mu')$, as  \S{\bf 5}(41). 
The optical depth $\tau=\tau(z)$ at an altitude $z$ above the bottom of the cloud is 
\begin{equation}
\tau = \int_0^{z}dz' \,\alpha(z'),
\label{int0}
\end{equation}
The net attenuation rate, due to absorption and scattering, at the altitude $z$ is  $\alpha =\alpha(z)$. We see that $d \tau =\alpha\, d z$.

The source terms on the right of (\ref{in10}) are characterized by the  {\it single-scattering albedo} $\tilde\omega$, by the {\it Planck intensity} $B(\tau)$, and by the {\it scattering phase function} $p(\mu,\mu')$.  Both $\tilde\omega$ and $p(\mu,\mu')$ may depend on altitude $z$, or equivalently, on the vertical optical depth $\tau$.  The single-scattering albedo $\tilde\omega$ is the probability that a photon, after a collision with a molecule or cloud particulate, is elastically scattered into other directions.
A fraction $1-\tilde\omega$ of the photons is absorbed and converted to atmospheric heat. We will neglect the very small effects of Raman scattering in Earth's atmosphere, where scattered photons emerge at substantially different frequencies due to internal energy changes of the scattering molecule.
Chandrasekhar\cite{Chandrasekhar} uses the symbol $\varpi_0$, the first term in his multipole expansion \S{\bf 3}(33) of the scattering phase function, to denote the single-scattering  albedo $\tilde \omega$ of (\ref{in10}).

The phase function $p(\mu,\mu')$ of (\ref{in10}) is a symmetric and nonnegative function of the direction cosine, $\mu=\cos\theta$, of the scattered radiation and the direction cosine, $\mu'=\cos\theta'$, of incident radiation
\begin{equation}
 p(\mu,\mu')=p(\mu',\mu)\ge 0.
\label{in14}
\end{equation}
The probability $dP$ that  a collision with a cloud particulate scatters a photon, propagating with direction cosine $\mu'$,  into a photon with a direction cosine between $\mu$ and $\mu+d\mu$ is $dP=\tilde\omega\, p(\mu,\mu') d\mu/2$.
In keeping with its significance as a probability density,  the phase function satisfies the identity
\begin{equation}
\frac{1}{2}\int_{-1}^1 d\mu \,p(\mu,\mu')=1.
\label{in16}
\end{equation}

The Planck intensity $B=B(\tau)$ of (\ref{in10})  depends on the local temperature, $T$ of the scattering medium, and on the spatial frequency $\nu$ of the radiation, as described by 
\begin{equation}
B=\frac{2h_{\rm P}c^2\nu^3}{e^{\nu c\, h_{\rm P}/(k_{\rm B}T)}-1}.
\label{et10}
\end{equation}
We are using cgs units, where $h_{\rm P}$ is Planck's constant and $k_{\rm B}$ is Boltzmann's constant. We assume that the absolute air temperature, $T=T(z)$ may depend on altitude $z$. Then the Planck intensity, $B=B(\tau)$, may depend  on the altitude $z$ or equivalently, on the optical depth $\tau$.  The radiation wavelength is $\lambda=1/\nu$.

\begin{figure}[t]
%\postscriptscale{streams1}{1.1}
\includegraphics[height=100mm,width=1\columnwidth]{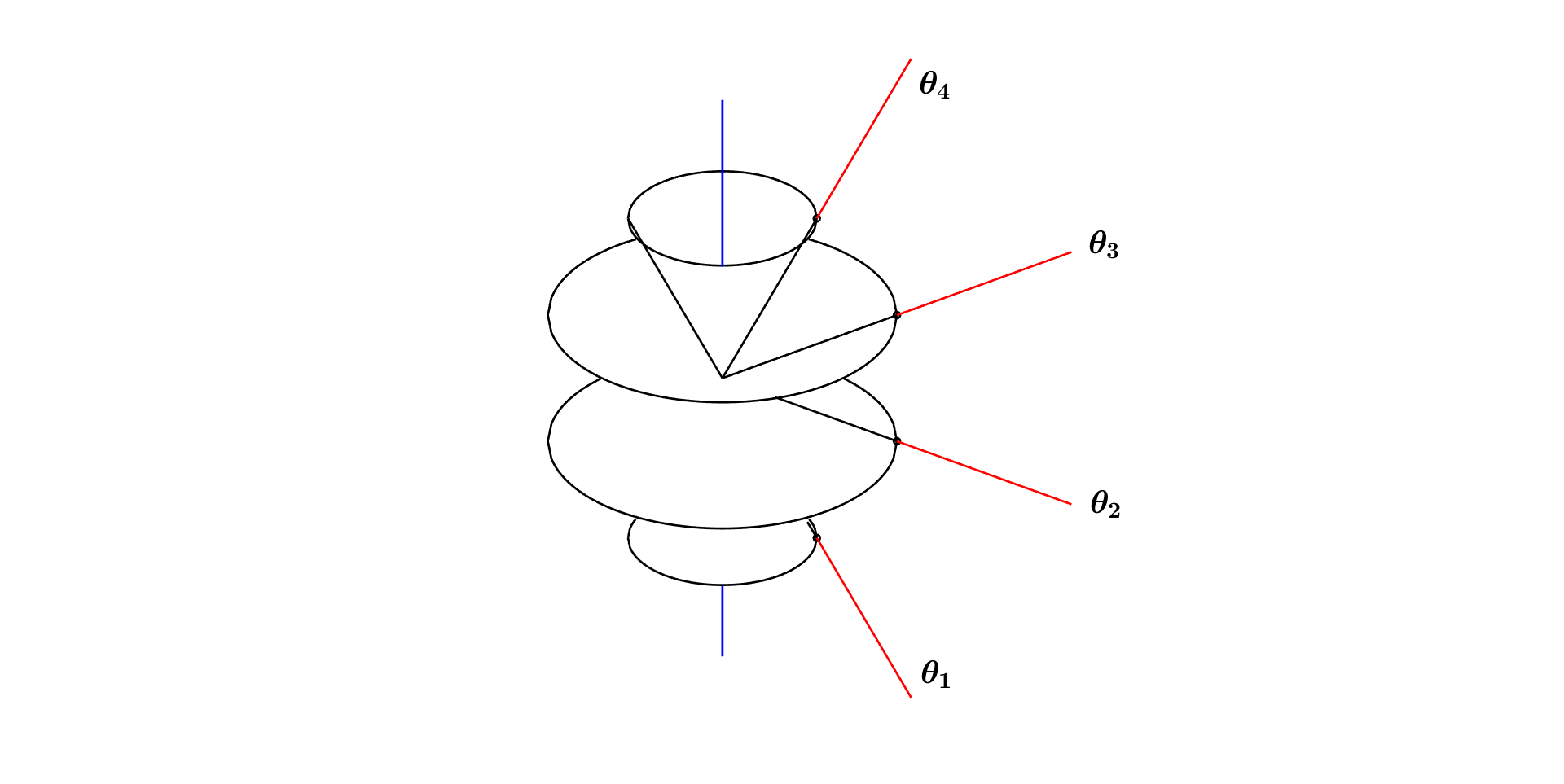}% Here is how to import EPS art
\caption {Sample directions of $2n = 4$ streams of axially symmetric radiation. The streams are centered on conical surfaces with opening angles $\theta_i$ to the zenith. In accordance  with (\ref{int2}), the direction cosines of the streams, $\mu_i = \cos \theta_i$, are the zeros of the Legendre polynomial $P_{2n}$, in this example, $P_{2n}(\mu_i)=P_4(\mu_i)=0$.}
\label{streams}
\end{figure}
\begin{figure}[t]
%\postscriptscale{streams2}{1.1}
\includegraphics[height=100mm,width=1\columnwidth]{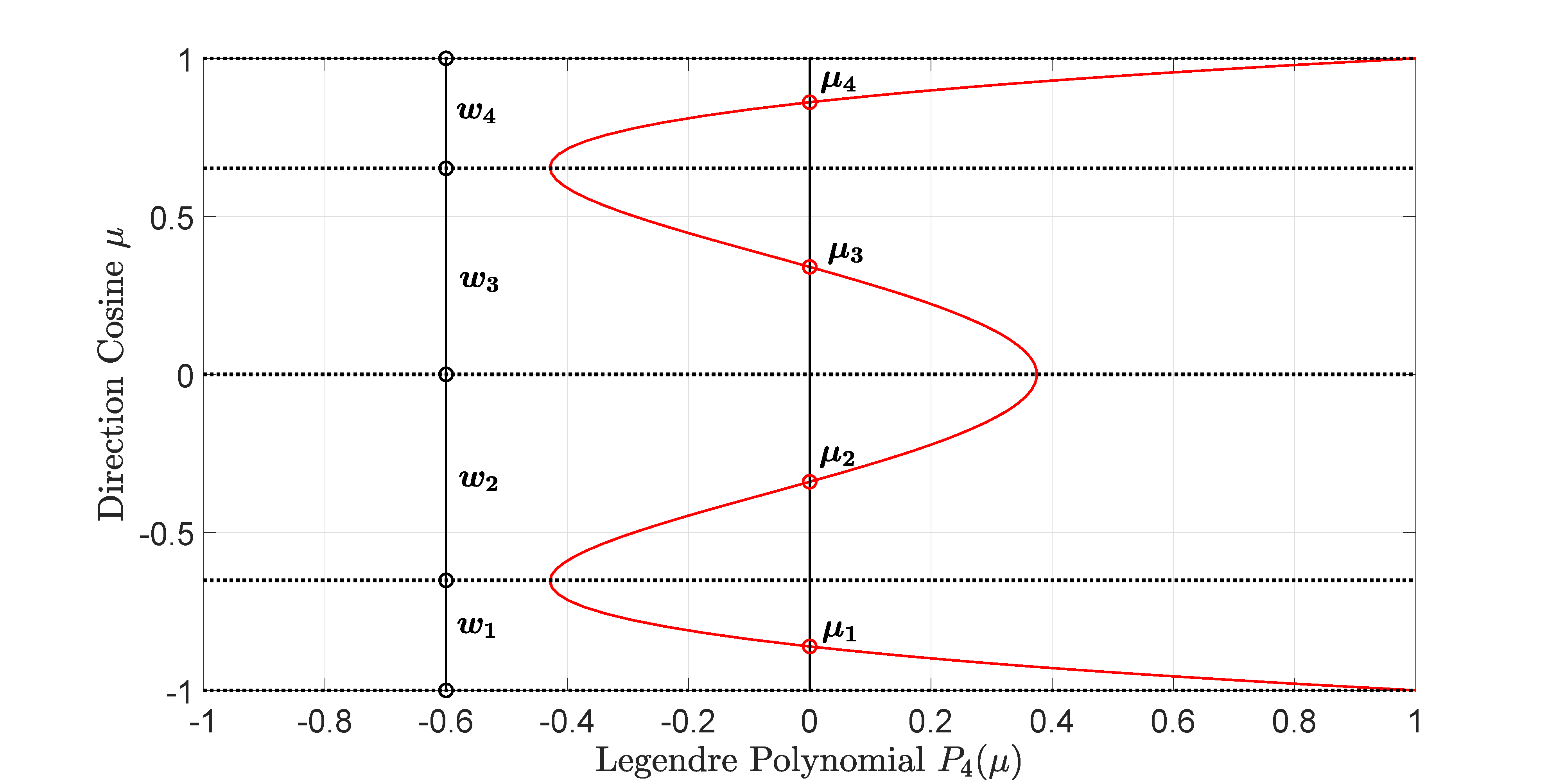}% Here is how to import EPS art
\caption {Representative parameters for Gauss-Legendre quadratures and $2n = 4$ streams.  In accordance with (\ref{int2}) the sample direction cosines,  $\mu_i=\cos\theta_i$, of the streams of Fig. \ref{streams} are the zeros of the Legendre polynomial, $P_{2n}(\mu)=P_4(\mu)$, which is shown as the continuous red curve. The zeros $\mu_i$, where $P_{2n}(\mu_i)=0$, are marked with small red circles.  For this example the values are $\mu_1, \mu_2, \mu_3, \mu_4 = -0.8611, -0.3400, \,0.3400,\,0.8611$.  The stream weights,  calculated with (\ref{int12}), are $w_1, w_2, w_3, w_4  = 0.3479,\,0.6521,\,0.6521,\, 0.3479$. See the text for more discussion.}
\label{streams2}
\end{figure}

\subsection{The $2n$-stream basis\label{int}}
The $2n$-stream method of reference\cite{WH1} allows one to solve the integro-differential equation of transfer (\ref{in10}) accurately and efficiently with modern computer software. The angular dependence of  the  intensity is characterized with $2n$ sample values,
$I(\mu_i,\tau)\ge 0$, along the directions of the streams $i=1,2,3,\ldots, 2n$. As sketched in Fig. \ref{streams}, the $i$th stream makes an angle $\theta_i=\cos^{-1}\mu_i$ to the zenith.  
The {\it Gauss-Legendre}\,\cite{Gauss} direction cosines, $\mu_i$,  are the zeros of the Legendre polynomial $P_{2n}$ of degree $2n$,  
\begin{equation}
P_{2n}(\mu_i)=0.
\label{int2}
\end{equation}
We will choose the indices $i=1,2,3,\cdots,2n$ such that
\begin{equation}
\mu_1<\mu_2<\mu_2<\cdots<\mu_{2n}.
\label{int4}
\end{equation}
Because the Legendre polynomial $P_{2n}$ is even, with $P_{2n}(\mu)=P_{2n}(-\mu)$,
the values of $\mu_i$ occur as equal and opposite pairs,
\begin{equation}
\mu_i=-\mu_{r(i)}.
\label{int6}
\end{equation}
The index reflection function is
\begin{equation}
r(i)=2n+1-i.
\label{int8}
\end{equation}
For the ordering convention (\ref{int4})
\begin{eqnarray}
\mu_j&<0&\quad\hbox{for}\quad j=1,2,3,\ldots, n,\label{int9a}\\
\mu_k&>0&\quad\hbox{for}\quad k=n+1,n+2,n+3,\ldots, 2n.\label{int9b}
\end{eqnarray}

The weighted  sample values of the intensity, $w_i I(\mu_i,\tau)$, will be denoted with the symbol
 $\lvec\mu_i|I(\tau)\}$,
\begin{equation}
\lvec\mu_i|I(\tau)\}=w_i I(\mu_i,\tau)\ge 0.
\label{int10}
\end{equation}
A formula for the Gauss-Legendre weights, $w_i>0$, was given by Eq. (72) of reference\,\cite{WH1} as
\begin{eqnarray}
\frac{1}{w_i}=\sum_{l=0}^{2n-1}\frac{2l+1}{2}P_l^2(\mu_i)>0.
\label{int12}
\end{eqnarray}
The weights $w_i$ sum to 2,
\begin{eqnarray}
\sum_{i=1}^{2n}w_i=2.
\label{int13}
\end{eqnarray}
The weights $w_i$ of (\ref{int12}) and the sample direction cosines $\mu_i$ defined by (\ref{int2}) are illustrated in Fig. \ref{streams2}. The weights are almost, but not precisely, the distances between local maxima and minima of $P_{2n}(\mu)$ for values of $\mu$ in the interval $[-1,\,1]$.  For example, $dP_4(\mu)/d\mu = (5\mu/3)(7\mu^2-3)$. The solutions to $dP_4(\mu)/d\mu = 0$ are $\mu =0,\pm \sqrt{3/7}$. But $\sqrt{3/7}=0.6547 \ne w_3 = 0.6521$, the number calculated with (\ref{int12}).

To simplify equations, we denote the intensity
as an abstract vector $|I\}=|I(\tau)\}$. We can represent the abstract vector $|I\}$ with a $2n\times 1$ column  vector
\begin{equation}
|I\}=\sum_{i=1}^{2n}|\mu_i)\lvec \mu_i|I\}=\left[\begin{array}{c}\lvec\mu_1|I\}\\\lvec\mu_2|I\}\\ \vdots \\ \lvec\mu_{2n}|I\}\end{array}\right].
\label{int14}
\end{equation}
We will call the column vector on the right of  (\ref{int14}) the {\it $\mu$-space} representation of  the abstract vector $|I\}$.  We will use other $2n\times 1$ arrays of numbers to represent the same abstract vector $|I\}$ in other bases, for example, the multipole basis $|l)$, discussed in Section {\bf \ref{mm}}.

The  {\it stream basis} vectors $|\mu_i)$ of (\ref{int14}) can be represented with the unit column vectors
\begin{equation}
|\mu_1)=\left[\begin{array}{c}1\\0\\ \vdots \\ 0\end{array}\right],\quad|\mu_2)=\left[\begin{array}{c}0\\1\\ \vdots \\ 0\end{array}\right],\quad \cdots,\quad|\mu_{2n})=\left[\begin{array}{c}0\\0\\ \vdots \\ 1\end{array}\right].
\label{int16}
\end{equation}
Corresponding left basis vectors can be represented as the unit row vectors
\begin{eqnarray}
\lvec\mu_1|&=&[1\quad 0\quad\cdots\quad 0],\nonumber\\
\lvec\mu_2|&=&[0\quad 1\quad\cdots\quad 0],\nonumber\\
&\vdots&\nonumber\\
\lvec\mu_{2n}|&=&[0\quad 0\quad\cdots\quad 1].\nonumber\\
\label{int18}
\end{eqnarray}
We use a double left parenthesis, $\lvec\mu_i|$, as a reminder that the row (or left)  basis vectors need not be Hermitian conjugates of the column (or right) basis vectors. The row  basis vectors $\lvec\mu_i|$ are like reciprocal lattice vectors of a crystal \,\cite{Reciprocal}.  The column  basis vectors, $|\mu_i)$ are like direct lattice vectors. Just as low-symmetry crystals can have oblique, non-orthogonal lattice vectors, the right basis vectors $|\mu_i)$ need not be orthogonal  to each other, although  they are orthonormal to the left eigenvectors $\lvec\mu_i|$.

As discussed in connection with Eq. (105) of reference\,\cite{WH1}, the stream basis vectors 
are right and left eigenvectors of the direction cosine matrix $\hat\mu$ 
\begin{equation}
\hat\mu|\mu_i)=\mu_i|\mu_i),\quad\hbox{and}\quad \lvec \mu_i|\hat\mu=\lvec \mu_i|\mu_i.
\label{int20}
\end{equation}
The eigenvectors are chosen to have the orthonormality property
\begin{equation}
\lvec \mu_i|\mu_{i'})=\delta_{i i'}.
\label{int22}
\end{equation}
They have the completeness property 
\begin{equation}
\sum_{i=1}^{2n}|\mu_i)\lvec \mu_i|=\hat 1.
\label{int24}
\end{equation}
In (\ref{int24}) and elsewhere, we will use the symbol $\hat 1$ to denote a square identity matrix with ones along the main diagonal and zeros elsewhere. It has the same dimensions as any square matrices that are added to, subtracted from,  or equated to it.

Multiplying (\ref{int24}) on the left or right by $\hat\mu$ and using (\ref{int20}) we find an expression for the direction-cosine operator,
\begin{equation}
\hat\mu =\sum_{i=1}^{2n}\mu_i|\mu_i)\lvec \mu_i|.
\label{int26}
\end{equation}
The direction secant matrix $\hat\varsigma$ is the inverse of the direction cosine matrix $\hat \mu$. We can use (\ref{int26}) to write
\begin{equation}
\hat\varsigma=\hat\mu^{-1} =\sum_{i=1}^{2n}\varsigma_i|\mu_i)\lvec \mu_i|.
\label{int28}
\end{equation}
The eigenvalues of the direction secant matrix are the inverses of the eigenvalues 
$\mu_i$ of the direction cosine matrix
\begin{equation}
\varsigma_i=\frac{1}{\mu_i}.
\label{int30}
\end{equation}
The direction secant matrix $\hat\varsigma$ has the same left and right eigenvectors as the direction cosine matrix $\hat\mu$,
\begin{equation}
\hat\varsigma|\mu_i)=\varsigma_i|\mu_i),\quad\hbox{and}\quad \lvec \mu_i|\hat\varsigma=\lvec \mu_i|\varsigma_i.
\label{int31}
\end{equation}

In accordance with Eq. (88) of reference\,\cite{WH1} it is convenient to use the stream basis vectors to define a projection matrix $\mathcal{M}_{\bf d}$  for downward streams with indices $j\le n$ and $\mu_j<0$, and a projection matrix $\mathcal{M}_{\bf u}$ for upward streams with indices $k>n$ and $\mu_k>0$.
\begin{equation}
\mathcal{M}_{\bf d }=\sum_{j=1}^{n}|\mu_j)\lvec \mu_j|\quad\hbox{and}\quad
\mathcal{M}_{\bf u} =\sum_{k=n+1}^{2n}|\mu_k)\lvec \mu_k|.
\label{int38}
\end{equation}
The projection matrices  of (\ref{int38}) have the simple algebra
\begin{eqnarray}
\mathcal{M}_{\bf d }+\mathcal{M}_{\bf u} &=&\hat 1\label{int40}\\
\mathcal{M}^2_{\bf d }=\mathcal{M}_{\bf d}\quad&\hbox{and}&\quad
\mathcal{M}^2_{\bf u}=\mathcal{M}_{\bf u}\label{int42}\\
\mathcal{M}_{\bf d }\mathcal{M}_{\bf u}=\breve 0\quad&\hbox{and}&\quad
\mathcal{M}_{\bf u}\mathcal{M}_{\bf d} =\breve 0.\label{int44}
\end{eqnarray}
Here and elsewhere, the symbol $\breve 0$ denotes a matrix, not necessarily square, for which all the elements are zero. The dimensions of $\breve 0$ are the same as the dimensions of other matrices to which it is added, subtracted or equated. 

In accordance with Eqs.  (109) and (110) of reference\, \cite{WH1}, we write the direction-cosine matrix 
$\hat\mu$ as the sum of a downward part $\hat\mu_{\bf d}$ and an upward part $\hat\mu_{\bf u}$ 
\begin{equation}
\hat \mu =\hat\mu_{\bf d} +\hat\mu_{\bf u}.
\label{int52}
\end{equation}
Expressions for the downward and upward parts are
\begin{eqnarray}
\hat\mu_{\bf d} &=&\mathcal{M}_{\bf d}\hat\mu =\hat\mu\mathcal{M}_{\bf d} =\sum_{j=1}^n\mu_j|\mu_j)\lvec \mu_j|,\label{int54}\\
 \hat\mu_{\bf u} &=&\mathcal{M}_{\bf u}\hat\mu =\hat\mu\mathcal{M}_{\bf u} =\sum_{k=n+1}^{2n}\mu_k|\mu_k)\lvec \mu_k|.
\label{int56}
\end{eqnarray}
In like manner, we write the direction-secant matrix 
$\hat\varsigma$ as the sum of a downward part $\hat\varsigma_{\bf d}$ and an upward part $\hat\varsigma_{\bf u}$ 
\begin{equation}
\hat \varsigma =\hat\varsigma_{\bf d} +\hat\varsigma_{\bf u}.
\label{int60}
\end{equation}
Expressions for the downward and upward parts are
\begin{eqnarray}
\hat\varsigma_{\bf d} &=&\mathcal{M}_{\bf d}\hat\varsigma =\hat\varsigma\mathcal{M}_{\bf d} =\sum_{j=1}^n\varsigma_j|\mu_j)\lvec \mu_j|,\label{int62}\\
 \hat\varsigma_{\bf u} &=&\mathcal{M}_{\bf u}\hat\varsigma =\hat\varsigma\mathcal{M}_{\bf u} =\sum_{k=n+1}^{2n}\varsigma_k|\mu_k)\lvec \mu_k|.
\label{int64}
\end{eqnarray}

It will often be convenient to write intensity vectors, flux vectors, source vectors, {\it etc.} as sums of downward and upward parts, for example,
\begin{equation}
|I \}=|I_{\bf d}\}+| I_{\bf u} \}=\left[\begin{array}{c}|I_{\bf d}\}\\ |I_{\bf u}\}\end{array}\right],
\label{int88}
\end{equation}
where
\begin{equation}
|I_{\bf d}\}=\mathcal{M}_{\bf d}| I\}\quad\hbox{and}\quad|I_{\bf u}\}=\mathcal{M}_{\bf u}| I\}.
\label{int90}
\end{equation}

\subsection{The multipole basis \label{mm}}
Describing the angular distribution of the axially symmetric intensity $I(\mu, \tau)$ with the $2n$ sample
 values,  $I(\mu_i, \tau)$, at the  Gauss-Legendre direction cosines $\mu_i$ of  (\ref{int2}), is equivalent to approximating the intensity as a superposition of the first $2n$ Legendre polynomials,
\begin{equation}
I(\mu,\tau)= \sum_{l=0}^{2n-1}(2l+1)P_{l}(\mu)I_l(\tau).
\label {mm2}
\end{equation}
The intensity multipoles  are
\begin{eqnarray}
I_l(\tau)&=&\lvec l|I(\tau)\}\nonumber\\
&=&\sum_{i=1}^{2n}\lvec l|\mu_i)\lvec\mu_i|I(\tau)\}.
\label {mm4}
\end{eqnarray}
Projections $\lvec l|\mu_i)$ of  the left multipole basis  $\lvec l|$ onto the right stream basis $|\mu_i)$, and vice versa, were given by Eqs.  (84) and (85) of reference\,\cite{WH1} in terms of Legendre polynomials $P_l$, and weights $w_i$ of (\ref{int12}) as
\begin{equation}
 \lvec l|\mu_i) =\frac{1}{2}P_l(\mu_i),
\label {mm6}
\end{equation}
and
\begin{equation}
\lvec\mu_i|l) = w_i(2l+1)P_{l}(\mu_i).
\label {mm8}
\end{equation}
Substituting (\ref{mm6}) and (\ref{int10}) into (\ref{mm4}), and noting from (\ref{mm2}) that $I(\mu,\tau)$ can be written as a superpositon of the first $2n$ Legendre polynomials,  we can use the Gauss-Legendre quadrature\,\cite{Gauss} to write the $l$-th multipole moment of the intensity as
\begin{eqnarray}
I_l(\tau)
&=&\frac{1}{2}\sum_{i=1}^{2n}w_i P_l(\mu_i)I(\mu_i,\tau)\nonumber\\
&\to&\frac{1}{2}\int_{-1}^1 d\mu P_l(\mu)I(\mu,\tau)\quad\hbox{as}\quad n\to\infty.
\label {mm9a}
\end{eqnarray}
In analogy to (\ref{int22}) the multipole basis vectors $|l')$ and $\lvec l|$ have been chosen to have the orthonormality property
\begin{equation}
\lvec l|l')=\delta_{l l'}.
\label{mm9b}
\end{equation}
In analogy to (\ref{int24}), they  have the completeness property 
\begin{equation}
\sum_{l=0}^{2n-1}|l)\lvec l|=\hat 1.
\label{mm9c}
\end{equation}
Using (\ref{mm9b}) we can write the intensity vector as
\begin{equation}
|I(\tau)\}=\sum_{l=0}^{2n-1}|l)\lvec l|I(\tau)\}=\sum_{l=0}^{2n-1}|l)I_l(\tau).
\label{mm9d}
\end{equation}
The expansion coefficients $\lvec l|I(\tau)\}=I_l(\tau)$ were given by (\ref{mm9a}).

From (\ref{mm8}) we see that the elements $\lvec \mu_i|0)$ of the right monopole basis vector  $|0)$ are the weights $w_i$ of (\ref{int12})
\begin{equation}
|0)=\sum_{i=1}^{2n}|\mu_i)\lvec \mu_i|0)=\sum_{i=1}^{2n}|\mu_i)w_i=\left[\begin{array}{c}w_1\\ w_2\\ \vdots \\ w_{2n}\end{array}\right],
\label{mm10}
\end{equation}
From (\ref{mm6}) we see that the elements $\lvec 0|\mu_i)$ of the left monopole basis vector $\lvec 0|$  are all equal to 1/2,
\begin{equation}
\lvec 0|=\sum_{i=1}^{2n}\lvec 0|\mu_i)\lvec \mu_i|=\frac{1}{2}\sum_{i=1}^{2n}\lvec \mu_i|=\bigg[\frac{1}{2}\quad \frac{1}{2}\quad\cdots \quad \frac{1}{2}\bigg].
\label{mm12}
\end{equation}
An identity from Eq. (53) of reference\,\cite{WH3} that will be useful subsequently is
\begin{equation}
\lvec 0|\hat \mu=\lvec 1| \quad\hbox{and}\quad \hat \mu|0)=\frac{1}{3} |1).
\label{mm13}
\end{equation}

To facilitate subsequent discussions, we note the identity from Eq. (217) of reference\,\cite{WH3}, 
\begin{equation}
\lvec 0|\hat \mu_{\bf u}^qe^{-\hat\varsigma_{\bf u}\tau}|0)=(-1)^q\lvec 0|\hat \mu_{\bf d}^qe^{\hat\varsigma_{\bf d}\tau}|0)=\frac{1}{2}E^{\{n\}}_{q+2}(\tau).
\label{mm20}
\end{equation}
Here $q$ is an integer, most often  $q=0$ or $q=1$ in our work. The $n$-stream exponential integral functions,
\begin{equation}
E^{\{n\}}_{q}(\tau)=\sum_{k=n+1}^{2n}w_k\mu_k^{q-2}e^{-\tau/\mu_k},
\label{mm22}
\end{equation}
can be obtained by evaluating the exact exponential integral functions,
\begin{equation}
E_{q}(\tau)=\int_0^1d\mu\,  \mu^{q-2} e^{-\tau/\mu}, 
\label{mm24}
\end{equation}
with Gauss-Legendre quadratures.  The exponential integral functions (\ref{mm24}) are discussed in Appendix I of Chandrasekhar's book\,\cite{Chandrasekhar}.  They account for the contributions to radiation transfer of intensity propagating at various slant angles with respect to the vertical. Graphical plots of the functions (\ref{mm22}) and (\ref{mm24}) for $n\ge 5$ can hardly be distinguished on a linear scale, as shown by Fig. 9 of reference\,\cite{WH3}. Values of $E_{q}^{\{n\}}(0)$ are given in Table 1 of reference\,\cite{WH3}.

The first few multipole moments $I_l$ of the intensity have useful physical interpretations. As shown by Eq. (19) of reference\,\cite{WH1}, the monopole moment ($l=0$) is proportional to the volume energy density $u$ of the radiation,
\begin{equation}
\lvec 0|I\}=I_0 = \frac{c u}{4\pi},
\label{mm26}
\end{equation}
where $c$ is the speed of light.
As shown by Eq. (20) of reference\,\cite{WH1}, the dipole  moment ($l=1$) is proportional to the
vertical energy flux $Z$ of the radiation,
\begin{equation}
\lvec 1|I\}=I_1 = \frac{Z}{4\pi}.
\label{mm28}
\end{equation}
Under many conditions of practical importance one finds  that the intensity is nearly isotropic and its quadrupole and higher moments  are very small compared to the monopole  moment
\begin{equation}
\lvec l|I\}=I_l\ll I_0,\quad\hbox{for}\quad l=2,3,4,\ldots
\label{mm30}
\end{equation}
When radiation-transfer conditions are such that (\ref{mm30}) is valid, it is common to use the Eddington approximation
\begin{equation}
\lvec l|I\}=I_l=0\quad\hbox{for}\quad l=2,3,4,\ldots\quad\hbox{(the Eddington approximation).}
\label{mm32}
\end{equation}
As shown in Fig. 1 and Fig. 2 of reference\,\cite{WH2}, the Eddington approximation can be very good deep inside an optically thick cloud. But the approximation is not good just inside the top and bottom surfaces. We do not use the Eddington approximation in this paper.
\subsection{Equation of radiative transfer\label{et}}
According to Eq.  (62) of reference\,\cite{WH1}, for a $2n$-stream model of radiative transfer the integro-differential equation of transfer (\ref{in10}) simplifies to the first-order, linear differential equation for the intensity vector $|I\}$,
\begin{equation}
\frac{d}{d\tau}|I\}=\hat \kappa\left(|B\}- |I\}\right).
\label{et2}
\end{equation}
The Planck intensity vector $|B\}=|B(\tau)\}$ of (\ref{et2}) is
\begin{equation}
|B\}=|0)B.
\label{et4}
\end{equation}
Here the right monopole basis vector 
$|0)$ was given by  (\ref{mm10}), and the scalar Planck intensity $B=B(\tau)$ was given by  (\ref{et10}).
The exponentiation-rate matrix $\hat \kappa$ of (\ref{et2}) was given by Eq.  (63) of reference\,\cite{WH1} as
\begin{equation}
\hat \kappa = \hat\varsigma\hat\eta,
\label{et12}
\end{equation}
the product of the direction-secant matrix $\hat\varsigma$ of  (\ref{int28}), and 
the efficiency matrix $\hat \eta$. As discussed in Section {\bf 3.2.1} of reference \cite{WH1}, the eigenvalues $\eta_l$ of the efficiency matrix are the fraction of the $l$th multipole moment  that remains after each generation of scattering. As shown in (54) of reference \cite{WH1}, the efficiency matrix can be written as 
\begin{equation}
\hat \eta =\hat 1-\frac{1}{2}\tilde\omega \hat p.
\label{et16}
\end{equation}
Here the single-scattering albedo $\tilde \omega$, which we mentioned in connection with (\ref{in10}),
is the probability that a photon that collides with a cloud particulate is scattered, rather than being absorbed and converted to heat. Probabilities must be nonnegative and  no larger than 1. So $\tilde\omega$ must be  bounded by
\begin{equation}
0\le \tilde\omega \le 1.
\label{et18}
\end{equation}
The continuous phase function $p(\mu,\mu')$ of (\ref{in14}) is represented by the scattering phase matrix $\hat p$ in (\ref{et16}).
The matrix elements $\lvec \mu_i|\hat p|\mu_{i'})=w_i p(\mu_i,\mu_{i'})$  give the representation of $\hat p$ in $\mu$-space as a $2n\times 2n$ array of numbers. The matrix $\hat p$ is defined such  that 
a photon in the stream  $i'$ that is not absorbed in a collision with a cloud particulate 
has a probability
$\lvec \mu_i|\hat p|\mu_{i'})/2$ to be scattered into the stream $i$. Therefore, in analogy to (\ref{in16}), we must have
\begin{equation}
\sum_{i=1}^{2n}\frac{1}{2}\lvec \mu_i|\hat p|\mu_{i'})=1\quad\hbox{and}\quad 
0\le\frac{1}{2}\lvec \mu_i|\hat p|\mu_{i'})\le 1.
\label{et20}
\end{equation}
In accordance with Eq. (40) of reference\,\cite{WH1}, for a cloud of randomly oriented scattering particulates or gas molecules
the scattering matrix $\hat p$ can be written in terms of  the right and left multipole basis vectors, $|l)$ and $\lvec l|$  of (\ref{mm6}) and (\ref{mm8}), as
\begin{equation}
\hat p = 2\sum_{l=0}^{2n-1}p_l|l)\lvec l|.
\label{et22}
\end{equation}
where the multipole phase coefficients are $p_l$. The monopole coefficient is always unity
\begin{equation}
p_0=1.
\label{et23}
\end{equation}
The coefficients $p_l$ of higher multipolarity, $1\le l\le 2n-1$, can be used to represent any physically permissible (nonnegative) phase function for a $2n$-stream model, for example: isotropic scattering, Rayleigh scattering, strongly peaked forward or backward scattering, etc.  

For {\it isotropic} scattering, the phase matrix is simply
\begin{eqnarray}
\hat p=2|0)\lvec 0|.
\label{et24}
\end{eqnarray}
According to Eq. (133) of reference\,\cite{WH1}, the only non-zero multipole phase coefficients $p_l$ of a Rayleigh-scattering phase function are $p_0=1$ and $p_2=1/10$. So the phase matrix of (\ref{et22}) for {\it Rayleigh} scattering is
\begin{eqnarray}
\hat p=2|0)\lvec 0|+\frac{1}{5}|2)\lvec 2|.
\label{et26}
\end{eqnarray}
For a $2n$ model, the possible multipole indices are $l=0,1,2,\ldots,2n-1$. Therefore, to represent the Rayleigh scattering operator (\ref{et26}), which includes terms with $l=2$, we must have $2n-1\ge 2$ or $n\ge 3/2$.  Since $n$ must be an integer, Raleigh scattering phase operators (\ref{et26}) can only be represented in models with $n\ge 2$.

As  in Section {\bf 5.2} of reference\,\cite{WH1}, we let $p(\mu,1)=\varpi^{\{p\}}(\mu)$  denote the phase function $p(\mu,1)$ that can be constructed from the first $2p$ Legendre polynomials $P_0(\mu), P_1(\mu),\ldots, P_{2p-1}(\mu)$, and which gives the maximum possible forward scattering $p(1,1)$, subject to the constraint that $p(\mu,1)\ge 0$ for any direction cosine, $-1\le \mu\le 1$. The multipole coefficients $p_l$ of $\varpi^{\{p\}}(\mu)$ are denoted by $p_l = \varpi^{\{p\}}_l$ and are  listed in  Table 1 of reference \,\cite{WH1}.  Then the phase operator (\ref{et22}) for {\it maximum forward scattering} of radiation modeled with $2n$ streams is 
\begin{equation}
\hat p = 2\sum_{l=0}^{2n-1}\varpi^{\{n\}}_l|l)\lvec l|.
\label{et27}
\end{equation}
For an illustative model with $2n= 10$, the phase operator (\ref{et27}) for maximum forward scattering becomes
\begin{eqnarray}
\hat p 
&=& 2\bigg(|0)\lvec 0|+.8182\,|1)\lvec 1|+.7273\,|2)\lvec 2|+.5967\,|3)\lvec 3|+.4988\,|4)\lvec 4|\nonumber\\
&&+.3869\,|5)\lvec 5|+.2937\,|6)\lvec 6|+.2016\,|7)\lvec 7|+.1209\,|8)\lvec 8|+.0573\,|9)\lvec 9|\bigg).
\label{et28}
\end{eqnarray}
The  phase function for {\it maximum backward scattering} differs from (\ref{et27}) for maximum forward scattering by
having alternating signs for the multipole expansion coefficients,
\begin{equation}
\hat p = 2\sum_{l=0}^{2n-1}(-1)^l\varpi^{\{n\}}_l|l)\lvec l|.
\label{et30}
\end{equation}

As shown in Eq. (138) of reference\,\cite{WH1}, the phase function $p(\mu,1)=\varpi^{\{n\}}(\mu)$ modeled by (\ref{et28})  is strongly  peaked in the forward direction, $\varpi^{\{n\}}(1)= n(n+1)$.
More detailed discussions of the scattering-phase matrix $\hat p$ and its multipole coefficients $p_l$ can be found in Section {\bf 5} of reference\, \cite{WH1}.

\subsection{Vertical flux\label{flx}}
For quantitative studies of vertical energy transfer,  the vertical flux vector
\begin{equation}
|Z \}=4\pi\hat\mu|I \},
\label{flx2}
\end{equation}
given by Eq. (210)  of reference\,\cite{WH1} and the corresponding scalar flux 
\begin{eqnarray}
Z&=&\lvec 0|Z \}\nonumber\\
&=&4\pi\lvec 0|\hat\mu|I \}\nonumber\\
&=&4\pi(1|I\}\nonumber\\
&=&4\pi I_1,
\label{flx4}
\end{eqnarray}
are more directly useful than the intensity vector $|I\}$ of (\ref{int14}). 

According to (\ref{int10}), the elements of the intensity vector are always nonnegative  $\lvec\mu_i|I\}\ge 0$. But according to (\ref{int9a}) and (\ref{int9b}),  the elements $\lvec\mu_i|Z\}$  of the vertical flux vector can have either positive or negative signs, with
\begin{eqnarray}
\lvec\mu_j|Z \}&=&4\pi\mu_j\lvec\mu_j|I \}\le 0\quad\hbox{for}\quad j=1,2,3,\ldots,n,
\label{flx5a}\\
\lvec\mu_k|Z \}&=&4\pi\mu_k\lvec\mu_k|I \}\ge 0\quad\hbox{for}\quad k=n+1,n+2,n+3,\ldots,2n.
\label{flx5b}
\end{eqnarray}

We can use (\ref{int24}),  (\ref{mm12}), (\ref{mm2}), (\ref{mm4}), (\ref{int20}) and (\ref{int10}) to write the scalar vertical flux (\ref{flx4}) as
\begin{eqnarray}
Z&=&4\pi\sum_{i=1}^{2n}\lvec 0|\mu_i)\lvec \mu_i|\hat\mu|I \}\nonumber\\
&=&2\pi\sum_{i=1}^{2n} w_i \mu_i I(\mu_i) \nonumber\\
&\to& 2\pi \int_{-1}^1d\mu\,\mu I(\mu)=\int_{4\pi}d\Omega \mu I(\mu).
\label{flx6}
\end{eqnarray}
We noted that the second line of (\ref{flx6}) is proportional to a Gauss-Legendre quadrature of the function $2\pi \mu\, I(\mu)$, which converges to the continuous integral of the third line when $n\to \infty$.  The unit of solid angle for axially symmetric radiation is $d\Omega = 2\pi d\mu$.

Using (\ref{flx5a}) and (\ref{flx5b}) with the definitions (\ref{int88}) and (\ref{int90}) of downward and upward parts of radiation vectors, we write
the downward and upward parts of the scalar flux as
\begin{eqnarray}
Z_{\bf d}&=&\lvec 0|Z_{\bf d}\}=\lvec 0|\mathcal{M}_{\bf d}|Z\}\le 0,\label{flx8}\\
Z_{\bf u}&=&\lvec 0|Z_{\bf u}\}=\lvec 0|\mathcal{M}_{\bf u}|Z\}\ge 0,
\label{flx10}
\end{eqnarray}
with 
\begin{equation}
Z=Z_{\bf d}+Z_{\bf u}.
\label{flx12}
\end{equation}
\subsection{Outgoing and incoming radiation \label{oir}}

The radiation coming into and going out of a stack of $m$ clouds  can be characterized  with the intensity vector $|I^{(m)}\}$ just above the top of the stack,   and by  the intensity vector  $|I^{(0)}\}$ just below the  bottom.   Alternatively, one can characterize the intensity with the incoming intensity $|I^{(\rm in)}\}$ and outgoing intensity $|I^{(\rm out)}\}$.  As discussed in Eq. (174) and (175) of reference\,\cite{WH1},  we can write  the incoming intensity vector as
\begin{eqnarray}
|I^{(\rm in)} \}&=&|I^{(\rm in)}_{\bf d} \}+|I^{(\rm in)}_{\bf u} \}\nonumber\\
&=&|I^{(m)}_{\bf d}\}+|I^{(0)}_{\bf u}\}.
\label{oir2}
\end{eqnarray}
where
\begin{equation}
|I^{(\rm in)}_{\bf d} \} =|I^{(m)}_{\bf d}\}\quad \hbox{and}\quad |I^{(\rm in)}_{\bf u} \} 
=|I^{(0)}_{\bf u}\}.
\label{oir4}
\end{equation}
The upward and downward parts of the intensity vectors of (\ref{oir2}) and  (\ref{oir4}) are defined in accordance with (\ref{int90}).
In like manner,
the outgoing intensity vector can be written as
\begin{eqnarray}
|I^{(\rm out)} \}&=&|I^{(\rm out)}_{\bf d} \}+|I^{(\rm out)}_{\bf u} \}\nonumber\\
&=&|I^{(0)}_{\bf d}\}+|I^{(m)}_{\bf u}\}.
\label{oir6}
\end{eqnarray}
where
\begin{equation}
|I^{(\rm out)}_{\bf d} \} =|I^{(0)}_{\bf d}\}\quad \hbox{and}\quad |I^{(\rm out)}_{\bf u} \}
 =|I^{(m)}_{\bf u}\}.
\label{oir8}
\end{equation}
The inverses of (\ref{oir2}) and  (\ref{oir6}) are
\begin{eqnarray}
|I^{(0)}\}&=&|I^{(0)}_{\bf d}\}+|I^{(0)}_{\bf u}\}\nonumber\\
 &=&
|I^{(\rm out)}_{\bf d} \}+|I^{(\rm in)}_{\bf u}\}, \label{oir10}
\end{eqnarray}
and
\begin{eqnarray}
|I^{(m)}\} &=&|I^{(m)}_{\bf d}\}+|I^{(m)}_{\bf u}\}\nonumber\\
&=& |I^{(\rm in)}_{\bf d}\}+|I^{(\rm out)}_{\bf u} \}.
\label{oir12}
\end{eqnarray}

In accordance with (\ref{flx2}) we write the flux vectors above and below a cloud as 
\begin{eqnarray}
|Z^{(0)}\}&=&4\pi\hat\mu|I^{(0)}\},\label{oir13}\\
|Z^{(m)}\}&=&4\pi\hat\mu|I^{(m)}\}.
\label{oir14}
\end{eqnarray}
Substituting (\ref{oir10}) into (\ref{oir13}) we find
\begin{equation}
|Z^{(0)}\}=|Z^{(\rm in)}_{\bf u}\}-|Z^{(\rm out)}_{\bf d}\},
\label{oir16}
\end{equation}
where we define  the upward part $|Z^{(\rm in)}_{\bf u}\}$ of the incoming flux and the downward part $|Z^{(\rm out)}_{\bf d}\}$ of the outgoing flux by
\begin{eqnarray}
|Z^{(\rm in)}_{\bf u}\}&=&4\pi\mu_{\bf u}|I^{(\rm in)}\},\label{oi13}\\
|Z^{(\rm out)}_{\bf d}\}&=&-4\pi\mu_{\bf d}|I^{(\rm out)}\}.
\label{oir18}
\end{eqnarray}
By convention, we use a negative sign on the right of (\ref{oir18}) to ensure that elements of the vector 
$|Z^{(\rm out)}_{\bf d}\}$ are nonnegative in $\mu$ space.
Multiplying (\ref{oir16}) on the left by $\lvec 0|$ we find the that the scalar flux defined by (\ref{flx4}) is
\begin{equation}
Z^{(0)}=Z^{(\rm in)}_{\bf u}-Z^{(\rm out)}_{\bf d}.
\label{oir20}
\end{equation}
Substituting (\ref{oir12}) into (\ref{oir14}) we find
\begin{equation}
|Z^{(m)}\}=-|Z^{(\rm in)}_{\bf d}\}+|Z^{(\rm out)}_{\bf u}\},
\label{oir22}
\end{equation}
where
\begin{eqnarray}
|Z^{(\rm in)}_{\bf d}\}&=&-4\pi\mu_{\bf d}|I^{(\rm in)}\},\label{oi17}\\
|Z^{(\rm out)}_{\bf u}\}&=&4\pi\mu_{\bf u}|I^{(\rm out)}\}.
\label{oir24}
\end{eqnarray}
Multiplying (\ref{oir22}) on the left by $\lvec 0|$ we find the that the scalar flux defined by (\ref{flx4}) is
\begin{equation}
Z^{(m)}=-Z^{(\rm in)}_{\bf d}+Z^{(\rm out)}_{\bf u}.
\label{oir26}
\end{equation}

We can use (\ref{oir16}) -- (\ref{oir24}) to write the 
outgoing and incoming flux vectors as
\begin{eqnarray}
|Z^{(\rm out)} \} 
&=&4\pi(\hat\mu_{\bf u}-\hat\mu_{\bf d})|I^{(\rm out)}\}=4\pi\sum_{\bf q}s_{\bf q}\hat\mu_{\bf q}|I^{(\rm out)}\}\label{oir27}\\
|Z^{(\rm in)} \} &=&4\pi(\hat\mu_{\bf u}-\hat\mu_{\bf d})|I^{(\rm in)} \}=4\pi\sum_{\bf q}s_{\bf q}\hat\mu_{\bf q}|I^{(\rm in)}\}
\label{oir28}
\end{eqnarray}
Here the sign factors are
\begin{equation}
s_{\bf u}=1\quad\hbox{and}\quad s_{\bf d}=-1.
\label{oir29}
\end{equation}
The negative signs in the equations (\ref{oir27}) and (\ref{oir28}) ensure that the elements of the outgoing and incoming flux vectors are nonnegative,
\begin{equation}
 \lvec\mu_i|Z^{(\rm out)}\}\ge 0\quad \hbox{and} \quad\lvec\mu_i|Z^{(\rm in)}\}\ge 0 \quad\hbox{for}\quad i=1,2,3,\ldots, 2n.
\label{oir30}
\end{equation}
From (\ref{mm12}) and (\ref{oir30}) we see that the scalar fluxes corresponding to (\ref{oir27}) and (\ref{oir28}) are nonnegative
\begin{eqnarray}
Z^{(\rm out)}&=&\lvec 0 |Z^{(\rm out)}\}=\frac{1}{2}\sum_{i=1}^{2n}\lvec\mu_i|Z^{(\rm out)}\} \ge 0, \label{oi24}\\
Z^{(\rm in)}&=&\lvec 0 |Z^{(\rm in)}\}=\frac{1}{2}\sum_{i=1}^{2n}\lvec\mu_i|Z^{(\rm in)}\} \ge 0. \label{oir32}
\end{eqnarray}

Subtracting (\ref{oir22}) from (\ref{oir16}) and using the identity (\ref{int40}) we find
\begin{equation}
|Z^{(0)}\}-|Z^{(m)}\} = |Z^{(\rm in)}\}-|Z^{(\rm out)} \}.
\label{oir34}
\end{equation}
Multiplying (\ref{oir34}) on the left by  $\lvec 0|$ we find
\begin{equation}
Z^{(0)}-Z^{(m)} = Z^{(\rm in)}-Z^{(\rm out)}.
\label{oir36}
\end{equation}
According to (\ref{oir36}),
the difference between the scalar fluxes $Z^{(0)}$ and $Z^{(m)}$, each of which can be positive, negative or zero,  is equal to the difference between the nonnegative flux $Z^{(\rm in)}\ge 0$ flowing into the cloud through the top and bottom, and the nonnegative flux $Z^{(\rm out)}\ge 0$ flowing out. Positive values of the flux differences of (\ref{oir36}) correspond to radiative heating of the cloud; negative values correspond to radiative cooling.
\subsection{Incident and thermal radiation \label{exin}}
As shown in Secton {\bf 2.6} of reference\, \cite{WH3}, the intensity at an optical depth $\tau$ above the bottom of a cloud can be written as the sum of a part $|\dot I\}=|\dot I(\tau)\} $ from thermal emission of particulates and gas molecules  inside the cloud and a part $|\ddot I\}=|\ddot I(\tau)\}$ from the transmission, absorption and scattering of incoming (incident) radiation, 
\begin{equation}
|I\} =|\dot I\}+|\ddot I\}.
\label{exin2}
\end{equation}
The single dots denote quantities originating from internally generated thermal radiation.  The double dots denote quantities originating from external incoming radiation.                                                                                                                                                                                                                                                                                                                                                                                                                                                                                                                                                                                                                                                                                                                                                                                                                                                                                                   We do not use single and double dot to represent first and second time derivatives, a convention that goes back to Isaac Newton.  Most of the work of this paper is focussed on steady-state radiation transfer for which there is no time dependence.

We write the intensity (\ref{oir2}) that is incident on the top and bottom of a stack of $m\ge 1$ clouds, or onto a single isolated cloud, with $m=1$ as
\begin{eqnarray}
|I^{(\rm in)} \}&=&|\dot I^{(\rm in)} \}+|\ddot I^{(\rm in)} \}\nonumber\\
&=&|\ddot I^{(\rm in)} \}
\label{exin6}
\end{eqnarray}
Thermal emission of particulates and gas molecules in the clouds can generate outgoing intensity, $|\dot I^{(\rm out)} \}>\breve 0$, but it cannot generate incoming intensity. Therefore,
\begin{equation}
|\dot I^{(\rm in)} \}=\breve 0,\quad\hbox{and}\quad|\dot Z^{(\rm in)} \}=\breve 0.
\label{exin8}
\end{equation}
The second equation of (\ref{exin8}) follows from (\ref{oir28}).

In accordance with (\ref{exin2}) we write the outgoing radiation (\ref{oir6}) 
\begin{equation}
|I^{(\rm out)} \}=|\dot I^{(\rm out)} \}+|\ddot I^{(\rm out)} \},
\label{exin4}
\end{equation}
As shown by Eq. (111) of reference\,\cite{WH3},  the outgoing intensity vector  $|\ddot I^{\{\rm out\}} \}$ is proportional to the incoming intensity vector $|\ddot I^{\{\rm in\}} \}$. The coefficient of proportionality is the scattering matrix $\mathcal{S}$,
\begin{equation}
|\ddot I^{(\rm out)} \}=\mathcal{S}|\ddot I^{(\rm in)} \}.
\label{exin10}
\end{equation}
We will frequently write the scattering matrix of a single cloud as the $2\times 2$ block matrix
\begin{equation}
\mathcal{S}=\left[\begin{array}{ll}\mathcal{S}_{\bf dd}&\mathcal{S}_{\bf du}\\\mathcal{S}_{\bf ud}&\mathcal{S}_{\bf uu}\end{array}\right]\quad\hbox{where}\quad\mathcal{S}_{\bf q q'}=\mathcal{M}_{\bf q}\mathcal{S}\mathcal{M}_{\bf q'}.
\label{exin11}
\end{equation}
Possible values of the stream-direction indices ${\bf q}$ and ${\bf q'}$ of (\ref{exin11})  are ${\bf d}$ and ${\bf u}$. In Section {\bf \ref{sm}} we review how to calculate the scattering matrix $\mathcal{S}$ for a homogeneous cloud.

Using (\ref{oir28}) and (\ref{exin10})  we write (\ref{oir27}) as
\begin{equation}
|\ddot Z^{(\rm out)} \}
=\Omega|\ddot Z^{(\rm in)}\}.
\label{exin12}
\end{equation}
In accordance with Eq.  (216) of reference\,\cite{WH1}, the cloud albedo matrix $\Omega$
of (\ref{exin12}) is a similarity transformation of the scattering matrix,
\begin{equation}
\Omega =(\hat\mu_{\bf u}-\hat \mu_{\bf d})\mathcal{S}(\hat\mu_{\bf u}-\hat \mu_{\bf d})^{-1}
=(\hat\mu_{\bf u}-\hat \mu_{\bf d})\mathcal{S}(\hat\varsigma_{\bf u}-\hat \varsigma_{\bf d}).
\label{exin14}
\end{equation}

The  part of the outgoing intensity (\ref{exin4}) that comes from thermal emission of cloud particulates and gas molecules can be written as
\begin{equation}
|\dot I^{(\rm out)}\}=|\dot J\}=\int_0^{\tau} d\tau' G(\tau')|0)B(\tau').
\label{exin16}
\end{equation}
In (\ref{exin16}) $\tau$  is the vertical optical thickness of the cloud,  $G(\tau')$ is the $2n\times 2n$ {\it continuous Green's  matrix} for thermally generated outgoing radiation, the monopole basis vector of (\ref{mm10}) is $|0)$, and $B(\tau')$, given by (\ref{et10}), is the Planck intensity at the source optical depth $\tau'$ above the bottom of the cloud. We review how to construct  $G(\tau')$ for homogeneous clouds in Section {\bf \ref{gf}}. The analogous {\it discrete Green's matrix} $G^{[m c\}}$ of (\ref{fs8}) depends on the discrete index $c=1,2,3,\ldots,m$ of individual clouds in a stack of $m$ clouds, rather than on the continuous optical depth $\tau'$ above the bottom of a single cloud.

For the special case of a single  isothermal cloud of constant Planck intensity $B(\tau')=B$,  (\ref{exin16}) simplifies to
\begin{equation}
|\dot I^{(\rm out)}\}=|\dot J\}=\mathcal{E}|0)B.
\label{exin20}
\end{equation}
The emissivity matrix $\mathcal{E}$ of the isothermal cloud is the integral of the continuous Green's matrix over the entire optical depth $\tau$ of the cloud,
\begin{equation}
\mathcal{E}=\int_0^{\tau} d\tau' G(\tau').
\label{exin22}
\end{equation}
According to Kirchhoff's law of radiation, Eq. (279) of reference\,\cite{WH1}, the emissivity matrix $\mathcal{E}$ of (\ref{exin22}) is related to the scattering matrix $\mathcal{S}$ by
\begin{equation}
\mathcal{E}=\hat 1-\mathcal{S}.
\label{exin24}
\end{equation}
An explicit proof of (\ref{exin24}) for homogeneous clouds is given by (\ref{gf8}) below.
In analogy to (\ref{exin11}) we will frequently write the emissivity matrix of an isothermal cloud as the $2\times 2$ block matrix
\begin{equation}
\mathcal{E}=\left[\begin{array}{ll}\mathcal{E}_{\bf dd}&\mathcal{E}_{\bf du}\\\mathcal{E}_{\bf ud}&\mathcal{E}_{\bf uu}\end{array}\right]\quad\hbox{where}\quad \mathcal{E}_{\bf q q'}=\mathcal{M}_{\bf q}\mathcal{E}\mathcal{M}_{\bf q'}.
\label{exin25}
\end{equation}
In summary,  for a single cloud we can write (\ref{exin4}) as
\begin{eqnarray}
|I^{(\rm out)} \}
=|\dot J \}+\mathcal{S}|\ddot I^{(\rm in)} \}.
\label{exin25a}
\end{eqnarray}
The output intensity vector $|I^{(\rm out)} \}$ of (\ref{exin25a}) is a  linear combination of  intensity  $|\dot J \}$, thermally emitted by cloud particulates and gas molecules, and transmitted and scattered incoming intensity $\mathcal{S}|\ddot I^{(\rm in)} \}$. For long wave thermal radiation, both cloud particulates and greenhouse gas molecules absorb and emit radiation, but only only cloud particulates contribute significantly to scattering. Short wave sunlight is efficiently scattered by cloud particulates. Especially for blue and ultraviolet sunlight, there is also significant Rayleigh scattering by atmospheric gases. Both particulates and gases absorb small fractions of sunlight, but Earth's atmosphere is normally too cool to emit short wave radiation.
\subsection{Identities for scattering matrices $\mathcal{S}$}
Some important fundamental properties of scattering  matrices $\mathcal{S}$ of (\ref{exin10}) are summarized here. Formal proofs will be given in a subsequent paper. 

For conservatively scattering clouds, with unit single-scattering albedo, $\tilde \omega = 1$, no thermal radiation can be  emitted.  That is, we must have
\begin{equation}
|\dot I^{(\rm out)}\}=\breve 0\quad\hbox{when}\quad\tilde \omega = 1.
\label{csc2}
\end{equation}
Comparing (\ref{csc2}) with (\ref{exin20}) we see that for a conservatively scattering cloud, with $\tilde \omega = 1$,
\begin{equation}
\mathcal{E}|0)=\breve 0.
\label{csc4}
\end{equation}
Using (\ref{csc4}) with Kirchhoff's law (\ref{exin24}) we find for a conservatively scattering cloud
\begin{equation}
\mathcal{S}|0)=|0).
\label{csc6}
\end{equation}
Eq. (\ref{csc6}) is the conservative-scattering limit of the more general inequality
\begin{equation}
0\le \lvec\mu_i|\mathcal{S}|0)\le\lvec\mu_i |0)=w_i\quad\hbox{for}\quad 0\le \tilde \omega \le 1.
\label{csc10}
\end{equation}
The elements of any scattering matrix, for a homogeneous or inhomogeneous cloud,  must satisfy the {\it Helmholtz-reciprocity} symmetry
\begin{equation}
\frac{\lvec\mu_o|\mathcal{S}|\mu_i)}{w_o|\mu_i|}=\frac{\lvec\mu_{r(i)}|\mathcal{S}|\mu_{r(o)})}{w_i|\mu_o|}.
\label{csc12}
\end{equation}
The index reflection function $r(i)$ was defined by (\ref{int8}).  The reciprocity theorem (\ref{csc12}) quantifies the plausible fact that for clouds described by (\ref{et2}), the rate of scattering from an input stream with index $i$ to an output stream with index $o$ is proportional to the rate of scattering of the time-reversed output stream with index $r(o)$ into the time-reversed input stream with index $r(i)$.

The elements of scattering matrices of homogeneous clouds also satisfy the simpler {\it reflection} symmetry 
\begin{equation}
\lvec\mu_o|\mathcal{S}|\mu_i)=\lvec\mu_{r(o)}|\mathcal{S}|\mu_{r(i)}).
\label{csc14}
\end{equation}
Scattering matrix elements of more general and realistic inhomogeneous clouds do not satisfy the reflection symmetry (\ref{csc14}) but they do satisfy the Helmholtz reciprocity symmetry (\ref{csc12}).

Helmholtz reciprocity  symmetries were discussed by Chandrasekhar\,\cite{Chandrasekhar} in his Sections \S {\bf 13} and \S {\bf 52}. Chandrasekhar uses a  normalization of the scattering matrix $\mathcal{S}$ that eliminates the factors $|\mu_i|$ and $|\mu_o|$ from (\ref{csc12}).  
We use a slightly different normalization   to simplify the forms of equations like  (\ref{csc6}), (\ref{csc10}) and (\ref{csc14}).  So formulas involving the $\mathcal{S}$ matrix are slightly different in our work from those  in Chandrasekhar's book.

\subsection{Scattering matrix $\mathcal{S}$ for a homogeneous cloud\label{sm}}
According to Eq. (206) of reference\,\cite{WH1}, the scattering matrix $\mathcal{S}$ of a homogeneous cloud  is given by the simple formula
\begin{equation}
\mathcal{S}=\mathcal{O}\mathcal{I}^{-1}.
\label{sm0}
\end{equation}
Here we review how to calculate the outgoing matrix $\mathcal{O}$ and incoming matrix $\mathcal{I}$ of (\ref{sm0}).
Let $\lvec \lambda_i|$ and  $|\lambda_i)$ be the left and right eigenvectors, and  let $\lambda_i$ be the corresponding eigenvalue of the penetration-length matrix, 
\begin{equation}
\hat \lambda =\hat\kappa^{-1} =\sum_{i=1}^{2n}\lambda_i|\lambda_i)\lvec\lambda_i|,
\label{sm2}
\end{equation}
the inverse of the exponentiation rate matrix $\hat \kappa$ of (\ref{et12}).
As in (\ref{int4})
the real eigenvalues or {\it penetration lengths} are ordered such that 
\begin{equation}
\lambda_1<\lambda_2<\lambda_3<\cdots\lambda_{2n}.
\label{sm4}
\end{equation}
As in (\ref{int6}) the penetration lengths for a homogenous cloud have the reflection symmetry
\begin{equation}
\lambda_i=-\lambda_{r(i)}.
\label{sm6}
\end{equation}
The index reflection function $r(i)$ was defined by (\ref{int8}). The {\it $\lambda$-space} basis vectors $\lvec \lambda_j|$ and $|\lambda_i)$ are chosen to have orthonormality and completeness relations analogous to (\ref{int24}) and (\ref{int26}),
\begin{equation}
\lvec \lambda_i|\lambda_j)=\delta_{ij},
\label{sm8}
\end{equation}
and 
\begin{eqnarray}
\hat 1 &=&\sum_{i=1}^{2n}|\lambda_i)\lvec \lambda_i|\nonumber\\
&=&\mathcal{L}_{\bf d}+\mathcal{L}_{\bf u}.
\label{sm9a}
\end{eqnarray}
For the limit of vanishing single scattering albedos, $\tilde\omega\to0$, $\lambda$-space quantities are chosen to approach the corresponding $\mu$-space quantities.
\begin{equation}
\left[\begin{array}{c}\lambda_i \\ \lvec\lambda_i|\\ |\lambda_i)\end{array}\right]\to
\left[\begin{array}{c}\mu_i \\ \lvec\mu_i|\\ |\mu_i)\end{array}\right]\quad\hbox{as}\quad \tilde \omega\to 0.
\label{sm9b}
\end{equation}
In analogy to (\ref{int38}) we define the downward and upward projection operators in $\lambda$ space by
\begin{equation}
\mathcal{L}_{\bf d}=\sum_{j=1}^{n}|\lambda_j)\lvec \lambda_j|\quad\hbox{and}\quad
\mathcal{L}_{\bf u} =\sum_{k=n+1}^{2n}|\lambda_k)\lvec \lambda_k|.
\label{sm10}
\end{equation}
The exponentiation-rate operator (\ref{et12}) can be written as the sum of downward and upward parts in $\lambda$ space
\begin{equation}
\hat\kappa=\hat\kappa_{\bf d}+\hat\kappa_{\bf u}. 
\label{sm12}
\end{equation}
where
\begin{equation}
\hat\kappa_{\bf d}=\sum_{j=1}^{n}\kappa_j|\lambda_j)\lvec \lambda_j|\quad\hbox{and}\quad
\hat\kappa_{\bf u} =\sum_{k=n+1}^{2n}\kappa_k|\lambda_k)\lvec \lambda_k|.
\label{sm14}
\end{equation}
The eigenvalues $\kappa_i$ of $\hat\kappa$ are the inverses of the eigenvalues $\lambda_i$ of $\hat\lambda$, and the eigenvectors are the same as those of $\hat\lambda$,
\begin{equation}
\kappa_i=\frac{1}{\lambda_i},\quad \lvec\kappa_i|= \lvec\lambda_i|,\quad\hbox{and}\quad |\kappa_i)=|\lambda_i).
\label{sm16}
\end{equation}
 The {\it overlap} matrix $\mathcal{C}$ between $\mu$ space and $\lambda$ space is defined by Eq. (162) of reference\, \cite{WH1} as
\begin{equation}
\mathcal{C}=\left[\begin{array}{ll}\mathcal{C}_{\bf d d}&\mathcal{C}_{\bf d u}\\ \mathcal{C}_{\bf u d}&\mathcal{C}_{\bf u u}\end{array}\right]=\left[\begin{array}{ll}\mathcal{M}_{\bf d}\mathcal{L}_{\bf d}&\mathcal{M}_{\bf d}\mathcal{L}_{\bf u}\\ \mathcal{M}_{\bf u}\mathcal{L}_{\bf d}&\mathcal{M}_{\bf u}\mathcal{L}_{\bf u}\end{array}\right].
\label{sm18}
\end{equation}
Note that the left directional index, ${\bf q=d}$ or ${\bf u}$, of $\mathcal{C}_{\bf q q'}$ refers to $\mu$ space, and the right directional index, ${\bf q'=d}$ or ${\bf u}$, refers to $\lambda$ space.
A useful identity for the overlap matrices was already given as Eq. (163) of reference\,\cite{WH1}
\begin{eqnarray}
\mathcal{C}_{\bf d d}+\mathcal{C}_{\bf u d}+\mathcal{C}_{\bf d u}+\mathcal{C}_{\bf u u}&=&\mathcal{M}_{\bf d}(\mathcal{L}_{\bf d}+\mathcal{L}_{\bf u})+\mathcal{M}_{\bf u}(\mathcal{L}_{\bf d}+\mathcal{L}_{\bf u})\nonumber\\
&=&\mathcal{M}_{\bf d}+\mathcal{M}_{\bf u}\nonumber\\
&=&\hat 1.
\label{sm22}
\end{eqnarray}
Here we noted the sum rules (\ref{int40}) and (\ref{sm9a}) of the upward and downward projection operators 
$\mathcal{M}_{\bf q}$ in $\mu$ space and $\mathcal{L}_{\bf q}$ in $\lambda$ space.  Each of the matrices $\mathcal{C}_{\bf q q'}$ can be represented as a $2n\times 2n$ matrix so it does not matter if we write the $\mathcal{C}_{\bf q q'}$ as an element of a $2\times 2$ block matrix or simply add them as in (\ref{sm22}).

A slight modification of  the overlap matrix gives the {\it incoming} matrix, defined by Eq. (199) of reference\,\cite{WH1} as
\begin{equation}
\mathcal{I}=\left[\begin{array}{ll}\mathcal{C}_{\bf d d}&\mathcal{C}_{\bf d u} e^{-\hat\kappa_{\bf u}\tau}\\ \mathcal{C}_{\bf u d} e^{\hat\kappa_{\bf d}\tau}&\mathcal{C}_{\bf u u}\end{array}\right].
\label{sm24}
\end{equation}
The {\it outgoing} matrix is defined by Eq. (201) of reference\,\cite{WH1} by
\begin{equation}
\mathcal{O}=\left[\begin{array}{ll}\mathcal{C}_{\bf d d}e^{\hat\kappa_{\bf d}\tau}&\mathcal{C}_{\bf d u} \\ \mathcal{C}_{\bf u d}& \mathcal{C}_{\bf u u}e^{-\hat\kappa_{\bf u}\tau}\end{array}\right].
\label{sm26}
\end{equation}

A simple limiting case of (\ref{sm0}) is a purely absorbing cloud with $\tilde\omega \to 0$ and negligible scattering, like a layer of clear air with greenhouse-gas absorption. Then $\mathcal{L}_{\bf q}\to
\mathcal{M}_{\bf q}$,  $\hat\kappa\to\hat\varsigma$ and
\begin{equation}
\mathcal{I}\to \hat 1,\quad
\mathcal{O}\to\left[\begin{array}{ll}e^{\hat\varsigma_{\bf d}\tau}&\breve 0 \\ \breve 0 &e^{-\hat\varsigma_{\bf u}\tau}\end{array}\right],\quad\hbox{and}\quad
\mathcal{S}=\mathcal{O}\mathcal{I}^{-1}\to\left[\begin{array}{ll}e^{\hat\varsigma_{\bf d}\tau}&\breve 0 \\ \breve 0 &e^{-\hat\varsigma_{\bf u}\tau}\end{array}\right].
\label{sm28}
\end{equation}
\subsection{Continuous Green's matrix $G(\tau')$ \label{gf}}
In reference\,\cite{WH3} we used a continuous Green's vector $|G(\tau')\}=G(\tau')|0)$ instead of the  continuous Green's matrix $G(\tau')$ of (\ref{exin22}).
In Eq. (151) of that reference we showed that $G (\tau')$ for a homogeneous cloud can be written as
\begin{eqnarray}
G (\tau') &=&\left[\mathcal{Q}(\tau')+\mathcal{S} \mathcal{R}(\tau')\right]\hat\kappa \nonumber\\
&=&\frac{\partial}{\partial \tau'}\left[\mathcal{Q}(\tau')+\mathcal{S} \mathcal{R}(\tau')\right] . \label{gf2}
\end{eqnarray}
In (\ref{gf2}) the scattering matrix $S=S(\tau)$ depends on the total optical thickness $\tau$ of the cloud, but is independent of the variable optical thickness $\tau'$ above the bottom of the cloud.
The {\it retro matrix} $\mathcal{R}(\tau')$ was given by Eq. (127) of reference\,\cite{WH3} as
\begin{eqnarray}
\mathcal{R}(\tau')
&=&\mathcal{C}_{\bf u d}e^{\hat\kappa_{\bf d}\tau'}-\mathcal{C}_{\bf d u}e^{-\hat\kappa_{\bf u}(\tau-\tau')}.
\label{gf4}
\end{eqnarray}
The matrix $\mathcal{Q}(\tau')$ was given by Eq. (149) of reference\,\cite{WH3} as 
\begin{eqnarray}
\mathcal{Q}(\tau')
&=&\mathcal{C}_{\bf u u}e^{-\hat\kappa_{\bf u}(\tau-\tau')} -\mathcal{C}_{\bf d d} e^{\hat\kappa_{\bf d}\tau'}.
\label{gf6}
\end{eqnarray}
To verify (\ref{exin22}) we integrate (\ref{gf2}) to find
\begin{eqnarray}
\int_0^{\tau}d\tau' G (\tau')&=&\mathcal{Q}(\tau)-\mathcal{Q}(0)+\mathcal{S} \left[\mathcal{R}(\tau)-\mathcal{R}(0) \right]\nonumber\\
&=&\mathcal{C}_{\bf u u} -\mathcal{C}_{\bf d d} e^{\hat\kappa_{\bf d}\tau}
-\mathcal{C}_{\bf u u}e^{-\hat\kappa_{\bf u}\tau} +\mathcal{C}_{\bf d d}\nonumber\\
&&+\mathcal{S} \left[\mathcal{C}_{\bf u d}e^{\hat\kappa_{\bf d}\tau}-\mathcal{C}_{\bf d u}
-\mathcal{C}_{\bf u d}+\mathcal{C}_{\bf d u}e^{-\hat\kappa_{\bf u}\tau} \right]\nonumber\\
&=&\hat 1-\mathcal{O}+\mathcal{S} \left[\mathcal{I}-\hat 1\right]\nonumber\\
&=&\hat 1-\mathcal{S}\nonumber\\
&=&\mathcal{E}.
 \label{gf8}
\end{eqnarray}
To derive the expression to the right of the third equal sign of (\ref{gf8}) from the previous line, we used (\ref{sm22}), (\ref{sm26}) and (\ref{sm24}). To derive  fourth  line we used (\ref{sm0}). The last line follows from (\ref{exin24}). We will call (\ref{gf8}) the {\it Kirchhoff identity}.  A discretized version (\ref{gf8}) for cloud stacks is given by (\ref{fs40}).

\subsection{Black clouds\label{bc}}
A single black cloud absorbs all incident intensity and scatters none so the scattering matrix $\mathcal{S}^{\{1\}}$ is the null matrix,
\begin{equation}
\mathcal{S}=\breve 0.
\label{bc2}
\end{equation}
A purely absorbing cloud with the scattering matrix (\ref{sm28}) becomes black as the optical thickness  approaches infinity, $\tau\to\infty$.
From Kirchhoff's law  (\ref{exin24}) and from (\ref{bc2}) we see that the emissivity matrix of a black cloud is the identity matrix
\begin{equation}
\mathcal{E}=\hat 1.
\label{bc4}
\end{equation}
Then according to (\ref{exin20}) and (\ref{bc4})  the intensity emitted by an isothermal black cloud with the Planck intensity $B$ is blackbody radiation
\begin{equation}
|\dot I^{(\rm out)}\}=|\dot J\}=|0)B.
\label{bc6}
\end{equation}
The output flux vector (\ref{oir27}) corresponding to (\ref{bc6}) is
\begin{equation}
|\dot Z^{(\rm out)} \}
 =4\pi(\hat\mu_{\bf u}-\hat\mu_{\bf d})|0)B.
\label{bc8}
\end{equation}
Using (\ref{bc8}) we  write the scalar output flux (\ref{flx4}) of a black cloud as
\begin{eqnarray}
\dot Z^{(\rm out)}=\lvec 0|\dot Z^{(\rm out)} \}=
2\dot Z^{(\rm bb)}.
\label{bc10}
\end{eqnarray}
Here the scalar blackbody flux $\dot Z^{(\rm bb)}$, emitted upward from the top of the black cloud, or downward from the bottom, is
\begin{eqnarray}
\dot Z^{(\rm bb)}&=&4\pi\lvec 0|\hat \mu_{\bf u}|0)B\nonumber\\
&=&-4\pi\lvec 0|\hat \mu_{\bf d}|0)B\nonumber\\
&=&2\pi E^{\{n\}}_3(0)\,B\nonumber\\
&\to& \pi B\quad\hbox{as}\quad n\to\infty.
\label{bc12}
\end{eqnarray}
To write the third line of (\ref{bc12}) we used the expresssion (\ref{mm20}) for the $n$-stream analog $E^{\{n\}}_3(\tau)$ of the exponential integral function $E_3(\tau)$.
To write the last line of (\ref{bc12}) we recalled from (222) of reference\, \cite{WH3} that $ E^{\{n\}}_3(0)$ converges to  $1/2$ as $n\to \infty$. The convergence is rapid. For $n=5$, Table 1 of reference\, \cite{WH3} gives  $ E^{\{5\}}_3(0)= 0.5038$.

The frequency-integrated value of the  flux (\ref{bc12}) is
\begin{eqnarray}
\overline{\dot Z^{(\rm bb)}}&=&\int_0^{\infty}d\nu\,\dot Z^{(\rm bb)}\nonumber\\
&=&2 E^{\{n\}}_3(0)\, \sigma_{\rm SB}T^4\nonumber\\
&\to& \sigma_{\rm SB}T^4\quad\hbox{as}\quad n\to\infty.
\label{bc14}
\end{eqnarray}
Here we noted that the integral of the  Planck intensity (\ref{et10}) over all upward solid angle increments, 
$2\pi\mu d\mu$, and frequency increments, $d\nu$, is
\begin{equation}
2\pi\int_0^1 d\mu\,\mu\int_0^{\infty}d\nu B= \pi \int_0^{\infty}d\nu B=\sigma_{\rm SB}T^4.
\label{bc16}
\end{equation}
The Stefan-Boltzmann constant is
\begin{eqnarray}
\sigma_{\rm SB}&=&\frac{2\pi k_{\rm B}^4}{c^2h_{\rm P}^3}\int_0^{\infty}dx \frac {x^3}{e^x-1}\nonumber\\
&=&\frac{2\pi^5 k_{\rm B}^4}{15 c^2h_{\rm P}^3}.
\label{bc18}
\end{eqnarray}
According to (\ref{bc10}),
an isothermal black cloud emits equal upward and downward fluxes $\dot Z^{(\rm bb)}$ through its top and bottom surfaces. The frequency-integrated fluxes  (\ref{bc14})  are very nearly $\overline{\dot Z^{(\rm bb)}}=\sigma_{\rm SB}T^4$.
\subsection{Planck emissivities\label{ter}}
For a non-black, isothermal cloud with Planck intensity $B$, we can use (\ref{exin20}) with (\ref{oir27}) to write the outgoing  thermal flux vector as
\begin{eqnarray}
|\dot Z^{(\rm out)}\}&=&4\pi(\hat\mu_{\bf u}-\hat\mu_{\bf d})|\dot I^{(\rm out)}\}\nonumber\\
&=&4\pi(\hat\mu_{\bf u}-\hat\mu_{\bf d})\mathcal{E}|0)B.
\label{ter2}
\end{eqnarray}
In accordance with (\ref{oir13}) the scalar outgoing flux that corresponds to (\ref{ter2}) is
\begin{equation}
\dot Z^{(\rm out)}=\lvec 0|\dot Z^{(\rm out)}\}=\dot Z^{(\rm out)}_{\bf u}+\dot Z^{(\rm out)}_{\bf d}.
\label{ter4}
\end{equation}
The upward flux from the top of the cloud is
\begin{equation}
\dot Z^{(\rm out)}_{\bf u}=4\pi\lvec 0| \hat\mu_{\bf u}\mathcal{E}|0)B
=\varepsilon_{\bf u}\dot Z^{(\rm bb)},
\label{ter6}
\end{equation}
where the blackbody flux $\dot Z^{(\rm bb)}$ was given by  (\ref{bc12}).
The Planck emissivity $\varepsilon_{\bf u}$ for the cloud top is
\begin{equation}
\varepsilon_{\bf u}=\frac{\dot Z^{(\rm out)}_{\bf u}}{\dot Z^{(\rm bb)}}=\frac{\lvec 0| \hat\mu_{\bf u}\mathcal{E}|0)}{\lvec 0| \hat\mu_{\bf u}|0)}=1-\frac{\lvec 0| \hat\mu_{\bf u}\mathcal{S}|0)}{\lvec 0| \hat\mu_{\bf u}|0)}.
\label{ter8}
\end{equation}
In like manner, we write
the downward flux from the bottom of the cloud as
\begin{equation}
\dot Z^{(\rm out)}_{\bf d}=-4\pi\lvec 0| \hat\mu_{\bf d}\mathcal{E}|0)B^{\{1\}}
=\varepsilon_{\bf d}\dot Z^{(\rm bb)}.
\label{ter10}
\end{equation}
The Planck emissivity $\varepsilon_{\bf d}$ for the cloud bottom is
\begin{equation}
\varepsilon_{\bf d}=\frac{\dot Z^{(\rm out)}_{\bf d}}{\dot Z^{(\rm bb)}}=\frac{\lvec 0| \hat\mu_{\bf d}\mathcal{E}|0)}{\lvec 0| \hat\mu_{\bf d}|0)}=1-\frac{\lvec 0| \hat\mu_{\bf d}\mathcal{S}|0)}{\lvec 0| \hat\mu_{\bf d}|0)}.
\label{ter12}
\end{equation}
The total flux from the top and bottom of an isothermal cloud is
\begin{equation}
\dot Z^{(\rm out)}
=(\varepsilon_{\bf u}+\varepsilon_{\bf d})\dot Z^{(\rm bb)},
\label{ter13}
\end{equation}
From (\ref{csc10}), (\ref{ter8}) and (\ref{ter12}) we see that the emissivities are bounded by
\begin{equation}
0\le \varepsilon_{\bf u}\le 1\quad\hbox{and}\quad 0\le \varepsilon_{\bf d}\le 1
\label{ter14}
\end{equation}
By symmetry, a homogeneous isothermal cloud emits equal upward and downward scalar fluxes, so $\varepsilon_{\bf u}=\varepsilon _{\bf d}$.  
For an inhomogeneous cloud, $\varepsilon_{\bf u}$ and $\varepsilon_{\bf d}$ may not be equal because particulates near the top of the cloud may have different single-scattering albedos $\tilde\omega$ and scattering-phase matrices $\hat p$ from those  near the bottom. 

\begin{figure}[t]
%\postscriptscale{emis2}{1.1}
\includegraphics[height=100mm,width=1\columnwidth]{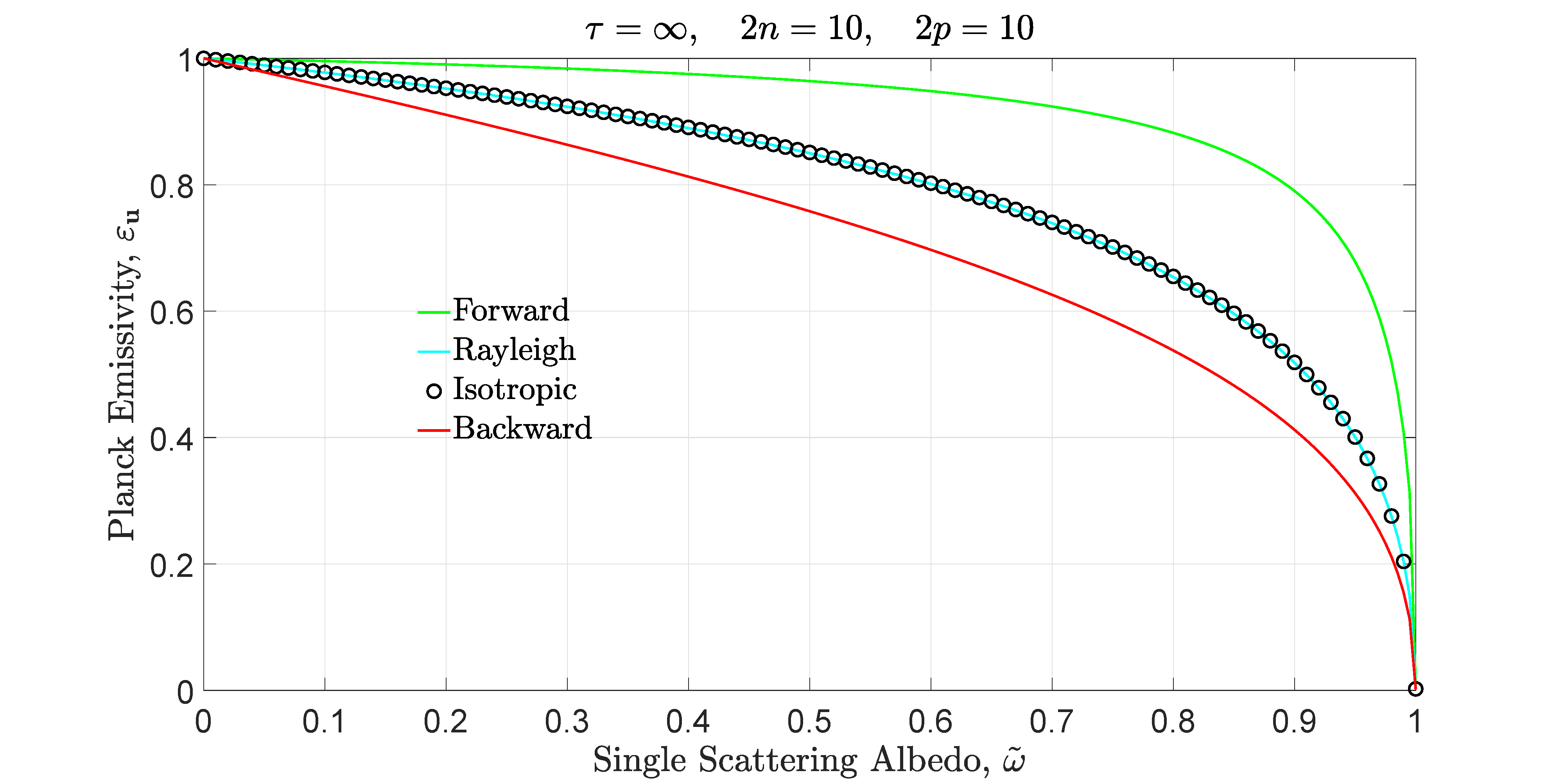}% Here is how to import EPS art
\caption {Planck emissivities $\varepsilon_{\bf u}$ of (\ref{ter8}) for homogeneous, optically thick clouds with the scattering-phase matrices $\hat p$ of (\ref{et24}) -- (\ref{et30}) as functions of the single-scattering albedos $\tilde\omega$. See the text for more detail.}
\label{emis2}
\end{figure}

Examples of Planck emissivities $\varepsilon_{\bf u}(\tau)$ for  homogeneous clouds of finite thickness $\tau$ were shown in Fig. 10 of reference \cite{WH3}, where the expression (\ref{ter8}) for the emissivity was given as Eq. (229).  The emissivities increase with increasing optical depths $\tau$ of the clouds.  In the limit $\tau\to\infty$, the emissivity saturates at a value, $\varepsilon_{\bf u}(\infty)\le 1$, that depends on the single-scattering albedo $\tilde \omega$ and the scattering-phase matrix $\hat p$ of (\ref{et22}). For purely absorbing clouds with  $\tilde\omega =0$, the optically-thick emissivity is  $\varepsilon_{\bf u}(\infty)=1$.

Some representative examples of the Planck emissivities $\varepsilon_{\bf u}$ of homogeneous, optically thick clouds, calculated with (\ref{ter8}) as functions of the single-scattering albedo $\tilde\omega$  are shown in 
Fig. \ref{emis2}. The curves are for fixed values of the four scattering-phase matrices $\hat p$ of (\ref{et24}) --(\ref{et30}).  Independent of $\hat p$,  for black clouds with no scattering or transmission and only absorption ($\tilde\omega =0$) we have $\varepsilon_{\bf u}=1$. For clouds with 100\% scattering or transmission and no absorption ($\tilde\omega = 1$) we have   $\varepsilon_{\bf u}=0$.  Only for intermediate values of the single scattering albedo, $0<\tilde \omega<1$, does the emissivity depend on the scattering-phase matrix $\hat p$ of (\ref{et22}).

\subsection{Planck albedos\label{pa}}
Suppose that the cloud discussed in the preceding section is illuminated from above and below with isotropic radiation of Planck intensity $B$, so the incoming intensity is
\begin{equation}
|\ddot I^{(\rm in)}\}= |0)B 
\label{pa2}
\end{equation}
According to (\ref{exin10}), the cloud will scatter the incoming intensity into outgoing intensity
\begin{equation}
|\ddot I^{(\rm out)}\}=\mathcal{S}|\ddot I^{(\rm in)}\}=\mathcal{S} |0)B. 
\label{pa4}
\end{equation}
The scattering matrix $\mathcal{S}$ of the cloud is related to the emissivity matrix  $\mathcal{E}$ in accordance with Kirchhoff's law (\ref{exin24}).  We use (\ref{oir13}) to write the scattered scalar flux  as
\begin{equation}
\ddot Z^{(\rm out)}=4\pi \lvec 0|(\hat\mu_{\bf u}-\hat\mu_{\bf d})|\ddot I^{(\rm out)}\}=\ddot Z^{(\rm out)}_{\bf u}+\ddot Z^{(\rm out)}_{\bf d}.
\label{pa6}
\end{equation}
We can write the upward part of (\ref{pa6}) as
\begin{equation}
\ddot Z^{(\rm out)}_{\bf u}=4\pi \lvec 0|\hat\mu_{\bf u}\mathcal{S} |0)B=\omega_{\bf u}\dot Z^{(\rm bb)}.
\label{pa8}
\end{equation}
where we can use the expression (\ref{bc12}) for the blackbody flux $\dot Z^{(\rm bb)}$, and Kirchhoff's law (\ref{exin24}) to write the  upward part of the {\it Planck albedo} $\omega_{\bf u}$  as
\begin{equation}
\omega_{\bf u}=\frac{\ddot Z^{(\rm out)}_{\bf u}}{\dot Z^{(\rm bb)}}=\frac{\lvec 0|\hat\mu_{\bf u}\mathcal{S} |0)}{\lvec 0|\hat\mu_{\bf u} |0)}
=1-\varepsilon_{\bf u}.
\label{pa10}
\end{equation}

In like manner we can write the downward part of (\ref{pa6}) as
\begin{equation}
\ddot Z^{(\rm out)}_{\bf d}=-4\pi \lvec 0|\hat\mu_{\bf d}\mathcal{S} |0)B=\omega_{\bf d}\dot Z^{(\rm bb)}.
\label{pa12}
\end{equation}
where
\begin{equation}
\omega_{\bf d}=\frac{\ddot Z^{(\rm out)}_{\bf d}}{\dot Z^{(\rm bb)}}=\frac{\lvec 0|\hat\mu_{\bf d}\mathcal{S} |0)}{\lvec 0|\hat\mu_{\bf u} |0)}
=1-\varepsilon_{\bf d}.
\label{pa14}
\end{equation}
 The Planck albedos $\omega_{\bf u}$  and $\omega_{\bf d}$ of (\ref{pa10}) and (\ref{pa14})
should not be confused with the single-scattering albedo $\tilde\omega$ of a cloud particulate or gas molecule, which we distinguish with a tilde.  A cloud with purely absorbing particulates  and gas molecules, and therefore with a vanishing single-scattering albedo, $\tilde\omega = 0$, can have a Planck albedo $\omega_{\bf q}$  close to 1 if the cloud is optically thin enough that most of the incoming  radiation can be transmitted through the cloud without absorption.  As one would intuitively expect, and as follows formally from (\ref{csc10})
the Planck albedos are bounded by
\begin{equation}
0\le\omega_{\bf u}\le 1\quad\hbox{and}\quad 0\le\omega_{\bf d}\le 1
\label{pa16}
\end{equation}

The isotropic input intensity (\ref{pa2}) corresponds to the input flux

\begin{eqnarray}
\ddot Z^{(\rm in)}&=&4\pi\lvec 0|(\hat \mu_{\bf u}-\hat \mu_{\bf d})|\ddot I^{(\rm in)}\}\nonumber\\
&=&4\pi\lvec 0|(\hat \mu_{\bf u}-\hat \mu_{\bf d}) |0)B \nonumber\\
&=&2\dot Z^{(\rm bb)},
\label{pa18}
\end{eqnarray}
where the blackbody flux $\dot Z^{(\rm bb)}$ was given by (\ref{bc12}).
The ratio of the output flux $\ddot Z^{(\rm out)}$ of (\ref{pa6}) to the input flux $\ddot Z^{(\rm in)}$ of (\ref{pa18}) is therefore
\begin{equation}
\frac{\ddot Z^{(\rm out)}}{\ddot Z^{(\rm in)}}=\frac{\omega_{\bf u}+\omega_{\bf d}}{2}.
\label{pa20}
\end{equation}

The mean  Planck albedo, $(\omega_{\bf u}+\omega_{\bf d})/2$ of (\ref{pa20}), is  much like the Bond albedo of a planet\,\cite{Bond},  the fraction of the intercepted solar energy flux that the planet transmits or scatters back to space without absorption. But the mean Planck albedo
$(\omega_{\bf u}+\omega_{\bf d})/2$  is the fraction of isotropic, monochromatic incoming radiation per unit area of the top and bottom of a cloud that is reflected or transmitted, rather than being absorbed and converted to heat.  The Bond albedo for planets in our solar system is defined for nearly collimated illumination  of the entire planet by the full frequency spectrum of the Sun.

Using (\ref{oir20}) and (\ref{oir26})   we write the scalar flux $Z(\tau')$ as $\dot Z^{(0)}=\dot Z(0)$ at the bottom of the cloud, where $\tau' = 0$ and $\dot Z^{(1)}=\dot Z(\tau)$ at the top, where $\tau' = \tau$.
\begin{eqnarray}
\dot Z^{(0)}&=&\dot Z^{(\rm in)}_{\bf  u}-\dot Z^{(\rm out)}_{\bf  d}=-\dot Z^{(\rm out)}_{\bf  d}=4\pi\lvec 0|\hat\mu_{\bf d}\mathcal{E}|0)B\label{ter30}\\
\dot Z^{(1)}&=&-\dot Z^{(\rm in)}_{\bf  d}+\dot Z^{(\rm out)}_{\bf  u}=\dot Z^{(\rm out)}_{\bf  u}=4\pi\lvec 0|\hat\mu_{\bf u}\mathcal{E}|0)B.\label{ter32}
\end{eqnarray}
Here we noted that a single cloud has no incoming thermal flux, $|\dot Z^{(\rm in)}\}=\breve 0$. 

To facilitate subsequent discussions of stacks of more than one cloud, we write (\ref{ter30}) and (\ref{ter32}) as the single vector equation
\begin{equation}
|\dot Z)=4\pi \dot M |B]\quad\hbox{or}\quad Z^{(g)}=4\pi \sum_{c=1}^1\dot M^{(gc\}}B^{\{c\}},
\label{ter34}
\end{equation}
where  $|B]=B^{\{c\}}=B^{\{1\}}$ is the Planck intensity of cloud $c=1$ in a stack of $m=1$ clouds, and $Z^{(g)}$ is the flux in the space $g=c$ above the  cloud $c$ or in the space $g=c-1$ below the cloud c. We will call the spaces below or above clouds {\it gaps}.  For a stack of $m$ clouds the cloud indices $c$ and gap indices $g$ can take on the values
\begin{equation}
c=1,2,3,\ldots,m\quad\hbox{and}\quad g=0,1,2,3,\ldots m.
\label{ter35}
\end{equation}
On the left of (\ref{ter34}) the abstract vector and matrix symbols mean
\begin{equation}
|\dot Z)=\left[\begin{array}{c}\dot Z^{(0)}\\ \dot Z^{(1)}\end{array}\right]\quad\hbox{and}\quad
\dot M=\left[\begin{array}{c}\dot M^{(01\}}\\ \dot M^{(11\}}\end{array}\right]
=\left[\begin{array}{c}\lvec 0|\hat\mu_{\bf d}\mathcal{E}|0)\\
\lvec 0|\hat\mu_{\bf u}\mathcal{E}|0)\end{array}\right].
\label{ter36}
\end{equation}
Since the abstract matrix $M$ converts a cloud property $|B]$  to a gap quantity $|\dot Z)$ we write its 
elements as $M^{(gc\}}$ with a left parenthesis and a right curly bracket as delimiters for the gap and cloud indices.
\subsection{Half isotropic incoming intensity\label{qir}}
To fully specify the intensity vector $|\ddot I^{(\rm in)}\}$ of incoming radiation onto the bottom and top of a cloud stack,  one needs $2n$ numbers, for example, the stream amplitudes $\lvec\mu_i|\ddot I^{(\rm in)}\}$ for $i=1,2,3,\ldots,2n$.   But  for this expository work it is convenient to model input radiation as {\it half-isotropic} Planck radiation, which would be generated by an external black cloud with a
Planck intensity $B^{\{\rm in\}}_{\bf d}$, located above the cloud stack, and a second external black cloud  of  Planck intensity $B^{\{\rm in\}}_{\bf u}$, located below the cloud  stack.
The two nonnegative numbers, $B^{\{\rm in\}}_{\bf d}$ and $B^{\{\rm in\}}_{\bf u}$,
are sufficient to specify half-isotropic incoming  radiation with the incoming  intensity vector  
\begin{equation}
|\ddot I^{(\rm in)}\}=\sum_{\bf q=d}^{\bf u}\mathcal{M}_{\bf q}|0)B^{\{\rm in\}}_{\bf q}.
\label{hi2}
\end{equation}
The input flux vector (\ref{oir28}) corresponding to (\ref{hi2}) is
\begin{equation}
|\ddot Z^{(\rm in)}\}
=4\pi\sum_{\bf q=d}^{\bf u}s_{\bf q}\hat\mu_{\bf q}B^{\{\rm in\}}_{\bf q}|0).
\label{hi4}
\end{equation}
The sign factors $s_{\bf q}$ were defined by (\ref{oir29}).
The upward and downward scalar incoming fluxes corresponding to (\ref{hi4}) are
\begin{eqnarray}
\ddot Z^{(\rm in)}_{\bf u}&=&\lvec 0| \mathcal{M}_{\bf u}|\ddot Z^{(\rm in)}\}=4\pi \lvec 0|\hat\mu_{\bf u}|0)B^{\{\rm in\}}_{\bf u}, \label{hi6a}\\
\ddot Z^{(\rm in)}_{\bf d}&=&\lvec 0| \mathcal{M}_{\bf d}|\ddot Z^{(\rm in)}\}=-4\pi \lvec 0|\hat\mu_{\bf d}|0)B^{\{\rm in\}}_{\bf d}. \label{hi6b}
\end{eqnarray}
From  (\ref{exin12}) we see that the outgoing flux corresponding to the incoming flux (\ref{hi4}) is
\begin{equation}
|\ddot Z^{(\rm out)}\}
=4\pi\sum_{\bf q=d}^{\bf u}\Omega \,s_{\bf q}\,\hat\mu_{\bf q}|0)B^{\{\rm in\}}_{\bf q}.
\label{hi10}
\end{equation}
The cloud albedo matrix $\Omega$ in  (\ref{hi10}) was  given by (\ref{exin14}).
The upward and downward scalar outgoing fluxes corresponding to the vector flux (\ref{hi10}) are
\begin{eqnarray}
\ddot Z^{(\rm out)}_{\bf u}&=&\lvec 0| \mathcal{M}_{\bf u}|\ddot Z^{(\rm out)}\}=4\pi\sum_{\bf q}\lvec 0|\mathcal{M}_{\bf u}\Omega \,s_{\bf q}\,\hat\mu_{\bf q}|0)B^{\{\rm in\}}_{\bf q}, \label{hi10a}\\
\ddot Z^{(\rm out)}_{\bf d}&=&\lvec 0| \mathcal{M}_{\bf d}|\ddot Z^{(\rm out)}\} =4\pi\sum_{\bf q}\lvec 0|\mathcal{M}_{\bf d}\Omega \,s_{\bf q}\,\hat\mu_{\bf q}|0)B^{\{\rm in\}}_{\bf q},\label{hi10b}
\end{eqnarray}

We can use (\ref{hi6a}), (\ref{hi6b}), (\ref{hi10a}) and (\ref{hi10b}) with (\ref{oir20}) to write the scalar flux $\ddot Z^{(0)}$  below  the cloud as a linear combination of the input Planck intensities $B^{\{\rm in\}}_{\bf d}$ and $B^{\{\rm in\}}_{\bf u}$,
\begin{eqnarray}
\ddot Z^{(0)}&=&\ddot Z^{(\rm in)}_{\bf u}-\ddot Z^{(\rm out)}_{\bf d}\nonumber\\
&=&4\pi \lvec 0|\hat\mu_{\bf u}-\mathcal{M}_{\bf d}\Omega\hat\mu_{\bf u}|0)B^{\{\rm in\}}_{\bf u}
+4\pi\lvec 0|\mathcal{M}_{\bf d}\Omega\hat\mu_{\bf d}|0) B^{\{\rm in\}}_{\bf d}\nonumber\\
&=&\ddot N^{(0)}_{\bf u}B^{\{\rm in\}}_{\bf u}+\ddot N^{(0)}_{\bf d}B^{\{\rm in\}}_{\bf d}.\label{hi12}
\end{eqnarray}
In like manner, can use (\ref{hi6a}), (\ref{hi6b}), (\ref{hi10a}) and (\ref{hi10b}) with (\ref{oir26})  to write the scalar flux $\ddot Z^{(1)}$  above  the cloud as
\begin{eqnarray}
\ddot Z^{(1)}&=&-\ddot Z^{(\rm in)}_{\bf d}+\ddot Z^{(\rm out)}_{\bf u}\nonumber\\
&=&4\pi \lvec 0|\hat\mu_{\bf d}-\mathcal{M}_{\bf u}\Omega\hat\mu_{\bf d}|0)B^{\{\rm in\}}_{\bf d}
+4\pi\lvec 0|\mathcal{M}_{\bf u}\Omega\hat\mu_{\bf u}|0) B^{\{\rm in\}}_{\bf u}\nonumber\\
&=&\ddot N^{(1)}_{\bf d}B^{\{\rm in\}}_{\bf d}+\ddot N^{(1)}_{\bf u}B^{\{\rm in\}}_{\bf u}.\label{hi14}
\end{eqnarray}

To facilitate subsequent discussions of stacks of more than one cloud, we write (\ref{hi12}) and (\ref{hi14}) as the single vector equation
\begin{equation}
|\ddot Z)=4\pi \ddot N|B^{\{\rm in\}}\rangle\quad\hbox{or}\quad \ddot Z^{(g)}=4\pi \sum_{\bf q=d}^{\bf u}\ddot N^{(g)}_{\bf q}B^{\{\rm in\}}_{\bf q},
\label{hi16}
\end{equation}
where
\begin{equation}
|\ddot Z)=\left[\begin{array}{c}\ddot Z^{(0)}\\ \ddot Z^{(1)}\end{array}\right],
\label{hi18}
\end{equation}
and
\begin{equation}
|B^{\{\rm in\}}\rangle=\left[\begin{array}{c}B^{\{\rm in\}}_{\bf d}\\B^{\{\rm in\}}_{\bf u} \end{array}\right].
\label{hi20}
\end{equation}
The elements of the $(m+1)\times 2 = 2\times 2$ matrix $\ddot N$  of (\ref{hi16}) follow from (\ref{hi12}) and (\ref{hi14}) and are
\begin{equation}
 \ddot N = \left[\begin{array}{cc}\ddot N^{(0)}_{\bf d}&\ddot N^{(0)}_{\bf u}\\ \ddot N^{(1)}_{\bf d}&\ddot N^{(1)}_{\bf u}\end{array}\right]
=   \left[\begin{array}{lr}\lvec 0|\mathcal{M}_{\bf d}\Omega\hat\mu_{\bf d}|0)    &\lvec 0|\hat\mu_{\bf u}-\mathcal{M}_{\bf d}\Omega\hat\mu_{\bf u}|0)
\\ \lvec 0|\hat\mu_{\bf d}-\mathcal{M}_{\bf u}\Omega\hat\mu_{\bf d}|0)&  \lvec 0|\mathcal{M}_{\bf u}\Omega\hat\mu_{\bf u}|0) \end{array}\right].
\label{hi22}
\end{equation}
Summing the flux $|\dot Z)$ of (\ref{ter34}) from thermal emission and the flux $|\ddot Z)$ of (\ref{hi16}) from scattering of incoming radiation we find the total fluxes
\begin{eqnarray}
|Z)&=&\left[\begin{array}{c}Z^{(0)}\\ Z^{(1)}\end{array}\right]=|\dot Z)+|\ddot Z)=4\pi\dot M |B]+4\pi \ddot N |B^{\{\rm in\}}\rangle.
\label{hi24}
\end{eqnarray}

\subsection{Radiative heating and cooling of an isolated cloud\label{hc}}
It is instructive to consider a 1-cloud stack consisting of a single {\it isolated} cloud. The number of clouds in the stack is $m=1$ and the cloud index can only be $c=1$.
Using (\ref{hi24}) we write the net radiative absorption rate $R^{\{c\}}$ per unit area of the isolated cloud  as the difference between the  vertical flux $Z^{(0)}$ at the bottom of the cloud and the vertical flux $Z^{(1)}$ at the top,
\begin{eqnarray}
R^{\{c\}}
&=& Z^{(0)}-Z^{(1)}\nonumber\\
&=&-\sum_{g=0}^1 \Delta^{\{cg)}Z^{(g)}.
\label{hc2}
\end{eqnarray}
The elements of the $ m\times (m+1) =1\times 2$ differencing matrix $\Delta$ are
\begin{eqnarray}
\Delta &=&\left[\begin{array}{rr}\Delta^{\{10)}& \Delta^{\{11)}\end{array}\right]\nonumber\\
&=&\left[\begin{array}{rr}-1& 1\end{array}\right].
\label{hc4}
\end{eqnarray}
In preparation for discussions of stacks of more than one cloud we have written (\ref{hc2}) as the element of an abstract vector equation
\begin{eqnarray}
|R]&=&-\Delta |Z)\nonumber\\
&=& -4\pi \Delta \dot M|B]-4\pi\Delta\ddot N|B^{\{\rm in\}}\rangle. 
\label{hc6}
\end{eqnarray}
The rate $R^{\{c\}}$ of (\ref{hc2}) is the diabatic heating rate (or cooling rate if $R^{\{c\}}<0$) due to absorption and  emission of radiation by cloud particulates and gas molecules. One definition of diabatic heating is: {\it A process that occurs with the addition or loss of heat. The opposite of adiabatic. Meteorological examples include air parcels warming due to the absorption of infrared radiation or release of latent heat}  \cite{diabatic}. This definition refers to the enthalpy the non-condensible (non-water) molecules of an air parcel.  If one were to include the enthalpy of water and water vapor in the total enthalpy of the air parcel,  condensation or evaporation would generate no diabatic heating or cooling. For example, the heat  (enthalpy) added by condensation of water vapor to the dry air of an expanding air parcel would be equal and opposite to the heat (enthalpy) lost from the condensing vapor.

We can write the net heating rate of (\ref{hc2}) as the difference between the heating rate $\ddot H^{\{c\}}$ due to absorption of external radiation and the cooling rate $\dot C^{\{c\}}$ due to thermal emission by cloud particulates and gas molecules. 
\begin{eqnarray}
R^{\{c\}} = \ddot H^{\{c\}}-\dot C^{\{c\}}.
\label{h2}
\end{eqnarray}
From inspection of (\ref{hc2}) we see that
\begin{eqnarray}
\ddot H^{\{c\}}&=&\ddot Z^{(0)}-\ddot Z^{(1)}\nonumber\\
&=&\ddot Z^{(\rm in)}-\ddot Z^{(\rm out)}\nonumber\\
&=&\lvec 0|\ddot Z^{(\rm in)}\}-\lvec 0|\ddot Z^{(\rm out)}\}\nonumber\\
&=&\lvec 0|(\hat 1 -\Omega^{\{c\}})|\ddot Z^{(\rm in)}\}\nonumber\\
&=&\lvec 0|\mathcal{A}^{\{c\}}|\ddot Z^{(\rm in)}\}.
\label{h4}
\end{eqnarray}
We used (\ref{oir36}) to write the second line of (\ref{h4}) and we used  (\ref{exin12}) to write the fourth line.  In the last line we have introduced the {\it absorptivity matrix}  $\mathcal{A}^{\{c\}}$ which we define as the complement of  the albedo matrix $\Omega^{\{c\}}$,
\begin{eqnarray}
\mathcal{A}^{\{c\}}&=&\hat 1-\Omega^{\{c\}}\nonumber\\
&=&(\hat\mu_{\bf u}-\hat\mu_{\bf d})\mathcal{E}^{\{c\}}(\hat\varsigma_{\bf u}-\hat\varsigma_{\bf d}).
\label{h6}
\end{eqnarray}
The second line of (\ref{h6}) comes from (\ref{exin14}) and (\ref{exin24}). From (\ref{h6}) we see that the same similarity transformation  (\ref{exin14}) that converts the scattering matrix   $\mathcal{S}^{\{c\}}$ to the albedo matrix $\Omega^{\{c\}}$, converts the emissivity matrix $\mathcal{E}^{\{c\}}$ to the absorptivity matrix  $\mathcal{A}^{\{c\}}$. 
For future reference, we use (\ref{hi4}) to write the last line of (\ref{h4}) as
\begin{eqnarray}
\ddot H^{\{c\}}
&=&\lvec 0|\mathcal{A}^{\{c\}}|\ddot Z^{(\rm in)}\}\nonumber\\
&=&\lvec 0|\mathcal{A}^{\{c\}}\left(\hat\mu_{\bf  u}B^{(\rm in)}_{\bf u}-\hat\mu_{\bf  d}B^{(\rm in)}_{\bf d}\right)|0)
\label{h7}
\end{eqnarray}

From  inspection of (\ref{hc2}) we see that
the cooling  rate due to thermal emission by particulates and gas molecules of a single isothermal cloud of Planck intensity $B^{\{c\}}$ is
\begin{eqnarray}
\dot C^{\{c\}}
&=&\dot Z^{(1)}- \dot Z^{(0)}\nonumber\\
&=&\Delta |\dot Z)\nonumber\\
&=&4\pi \Delta\dot M B^{\{c\}}\nonumber\\
&=&4\pi\lvec 0|(\hat\mu_{\bf u}-\hat\mu_{\bf d})\mathcal{E}^{\{c\}}|0)B^{\{c\}}\nonumber\\
&=&(\varepsilon_{\bf u}+\varepsilon_{\bf d})\dot Z^{\{\rm bb\}}.
\label{h8}
\end{eqnarray}
We used the $1\times 2$ matrix $\Delta$ of (\ref{hc4}) and the $2\times 1$ vector $|\dot Z)$ of (\ref{ter36})  to write the second line of (\ref{h8}); we used the expression (\ref{hc4}) for $\Delta$ and the expression (\ref{ter34}) for $|\dot Z)$ to write the third line; we used the matrix elements of (\ref{ter36}) to write the fourth line;  for the fifth line we used the definitions (\ref{ter8}) and (\ref{ter12}) of the Planck emissivities $\varepsilon_{\bf u}$ and $\varepsilon_{\bf d}$, along with the definition (\ref{bc12}) of the blackbody flux $\dot Z^{\{\rm bb\}}$.
Not surprisingly, the cooling rate (\ref{h8}) of a single, isolated, isothermal cloud is
the sum of the thermal flux $\varepsilon_{\bf u}\dot Z^{\{\rm bb\}}$ emitted upward from the cloud top and the thermal flux $\varepsilon_{\bf d}\dot Z^{\{\rm bb\}}$ emitted downward from the cloud bottom.

We can use the second line of (\ref{h6}) to write the fourth line of (\ref{h8}) as the isolated cloud cooling rate
\begin{equation}
\dot C^{\{c\}}=4\pi\lvec 0|\mathcal{A}^{\{c\}}(\hat\mu_{\bf u}-\hat\mu_{\bf d})|0)B^{\{c\}}.
\label{h12}
\end{equation}
Substituting (\ref{h7}) and (\ref{h12}) into (\ref{h2}) we find that the net heating rate of the cloud is 
\begin{eqnarray}
R^{\{c\}} &=&\ddot H^{\{c\}}-\dot  C^{\{c\}}\nonumber\\
&=&4\pi \lvec 0|\mathcal{A}^{\{c\}}
([B^{\{\rm in\}}_{\bf u}-B^{\{c\}}]\hat\mu_{\bf u}-[B^{\{\rm in\}}_{\bf d}-B^{\{c\}}]\hat\mu_{\bf d})|0).
\label{h14}
\end{eqnarray}

\subsection{Radiative equilibrium\label{re}}
The net radiative heating rate for an  isolated cloud will vanish, $R^{\{c\}}=0$. Then the cloud thermally emits just as much radiation out of its top and bottom surfaces as it absorbs from incoming radiation. Suppose that convection and other heat transfer mechanism are negligibly small compared to radiative heat transfer. Then  for fixed half-isotropic incoming radiation, described by the $2\times 1$ vector of upward and downward Planck intensities, $|B^{(\rm in)}\rangle$ of (\ref{hi20}), the temperature $T^{\{c\}}$ of the cloud will rise if $R^{\{c\}}>0$  or drop if $R^{\{c\}}<0$ until the Planck intensity $B^{\{c\}}$ of the isothermal cloud makes the expression (\ref{h14}) equal to zero.  Setting $R^{\{c\}}=0$ in (\ref{h14}) we see that for given values,  $B^{\{\rm in\}}_{\bf d}$ and $ B^{\{\rm in\}}_{\bf u}$, of half-isotropic incoming radiation the heating rate will vanish if the  Planck intensity of a isolated, isothermal cloud is
\begin{equation}
B^{\{c\}}=\frac{\lvec 0|\mathcal{A}^{\{c\}}
(B^{\{\rm in\}}_{\bf u}\hat\mu_{\bf u}-B^{\{\rm in\}}_{\bf d}\hat\mu_{\bf d})|0)}
{\lvec 0|\mathcal{A}^{\{c\}}(\hat\mu_{\bf u}-\hat\mu_{\bf d})|0)}.
\label{re4}
\end{equation}
The cloud Planck intensity $B^{\{c\}}$ for radiative equilibrium will be  somewhere between $B^{\{\rm in\}}_{\bf d}$ and  $B^{\{\rm in\}}_{\bf u}$, and the cloud temperature $T^{\{c\}}$ will be somewhere between the temperatures $T^{\{\rm in\}}_{\bf d}$ and  $T^{\{\rm in\}}_{\bf u}$ of the downward and upward incoming radiation.

The fundamental symmetries of a homogeneous cloud, with the same single-scattering albedo $\tilde\omega$ and the same scattering-phase matrix $\hat p$ of (\ref{et22}) from top  to bottom, are such that
\begin{equation}
\lvec 0|\mathcal{A}^{\{c\}}\hat\mu_{\bf u}|0)=-\lvec 0|\mathcal{A}^{\{c\}}\hat\mu_{\bf d}|0).
\label{re6}
\end{equation}
Then (\ref{re4}) simplifies to 
\begin{equation}
B^{\{1\}}=\frac{B^{\{\rm in\}}_{\bf u}+B^{\{\rm in\}}_{\bf d}}{2}.
\label{re8}
\end{equation}
In radiative equilibrium the Planck intensity $B^{\{1\}}$ of a homogeneous cloud is the average of the Planck intensities $B^{\{\rm in\}}_{\bf u}$  and $B^{\{\rm in\}}_{\bf d}$ of the half isotropic incoming radiation.
\subsection{1-cloud examples \label{nex}}
To graph numerical results it is convenient to introduce a reference Planck intensity $B_0$ and a reference flux $Z_0$. In accordance with (\ref{bc12}), these are related as the flux of a blackbody is related to its Planck intensity
\begin{eqnarray}
Z_0&=&4\pi\lvec 0|\hat\mu_{\bf u}|0)B_0\nonumber\\
&=&2\pi E^{\{n\}}_3(0)B_0\nonumber\\
&\to &\pi B_0\quad\hbox{as}\quad n\to\infty.
\label{nex2}
\end{eqnarray}
We use caligraphic fonts to denote  fluxes measured in units of $Z_0$ or Planck intensities measured in units of $B_0$.  For example, for half-isotropic incoming radiation incident on a homogeneous cloud we can use (\ref{nex2}) and (\ref{hi4}) to write
\begin{eqnarray}
\ddot {\mathcal{Z}}^{(\rm in)}_{\bf q}=\frac{\ddot Z^{(\rm in)}_{\bf q}}{Z_0}=\frac{B^{\{\rm in\}}_{\bf q}}{B_0}=\mathcal{B}^{\{\rm in\}}_{\bf q}.
\label{nex4}
\end{eqnarray}
\begin{figure}[t]
%\postscriptscale{one1}{1}
\includegraphics[height=90mm,width=1\columnwidth]{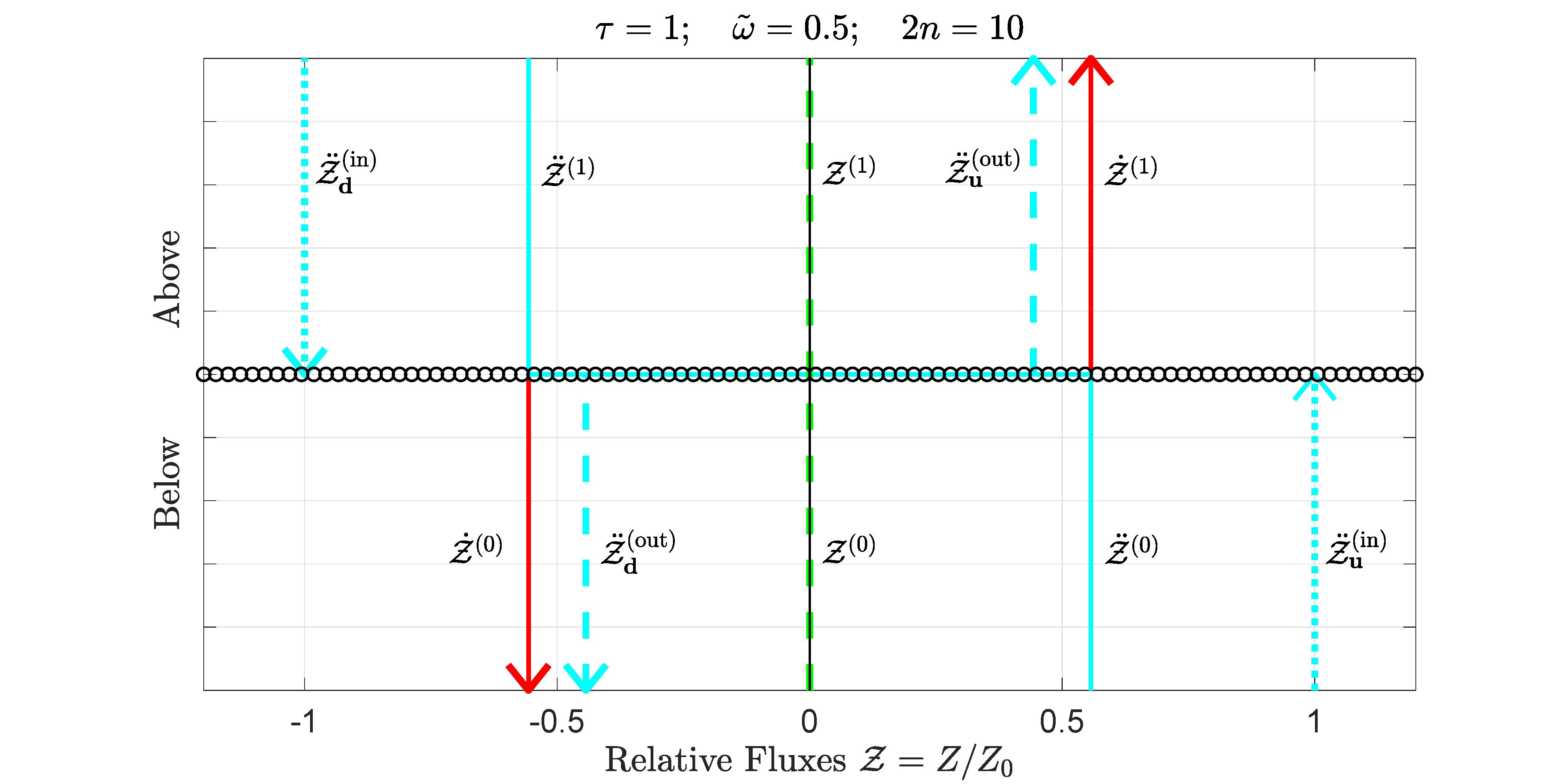}% Here is how to import EPS art
\caption{ A Rayleigh scattering cloud of optical depth $\tau=1$ with the same temperature as incoming radiation from above and below.  
The downward and upward relative fluxes of incoming radiation are denoted by  $\ddot{\mathcal{Z}}^{(\rm in)}_{\bf d}$ and $\ddot{\mathcal{Z}}^{(\rm in)}_{\bf u}$, and are shown as the dotted cyan lines. The downward and upward parts of  transmitted and reflected radiation are denoted by $\ddot {\mathcal{Z}}^{(\rm out)}_{\bf d}$ and 
$\ddot {\mathcal{Z}}^{(\rm out)}_{\bf u}$, and are shown as the dashed cyan lines.  The net fluxes above and below the cloud from the incoming radiation
are denoted by  $\ddot{\mathcal{Z}}^{(1)}$ and   $\ddot {\mathcal{Z}}^{(0)}$, and are shown as the continuous cyan lines. For thermal emission by cloud particulates and gas molecules, the fluxes below and above the clouds are denoted by  $\dot {\mathcal{Z}}^{(0)}$ and   $\dot {\mathcal{Z}}^{(1)}$, and are shown as the continuous red lines.
The net fluxes are denoted by  $ \mathcal{Z}^{(0)}$ and $\mathcal{Z}^{(1)}$, and are shown as the dashed green lines. Fluxes with only upward or only downward radiation are distinguished with arrowheads from fluxes which are the net of both upward and downward radiation.  See the text for more detailed discussions.
\label{one1}}
\end{figure}

\begin{figure}[t]
%\postscriptscale{one2}{1}
\includegraphics[height=100mm,width=1\columnwidth]{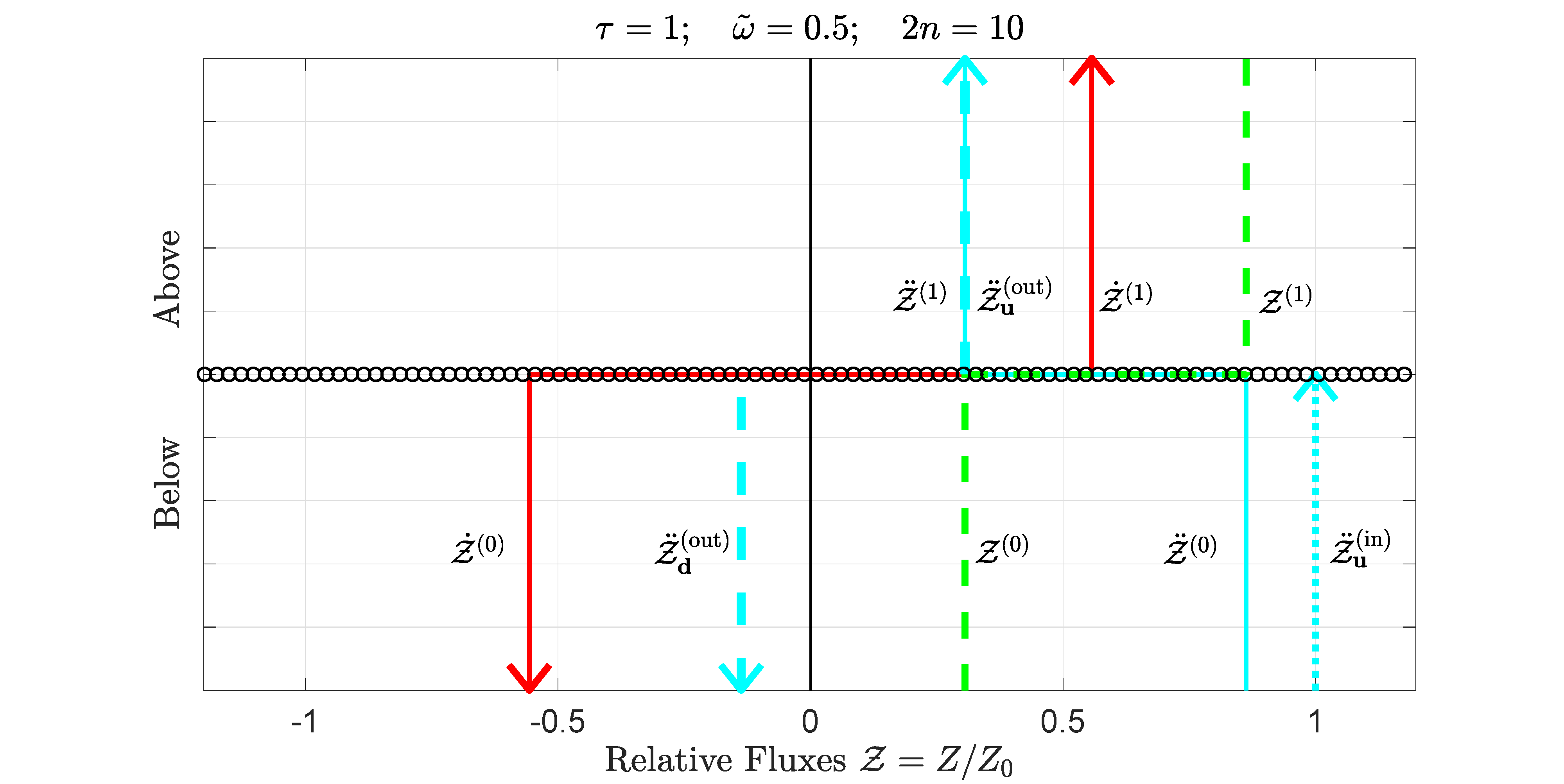}% Here is how to import EPS art
\caption{The same cloud as for Fig. \ref{one1}, but with only upward incoming radiation incident on the bottom of the cloud. See the text for more detailed discussions.
\label{one2}}
\end{figure}
\begin{figure}[t]
%\postscriptscale{one3}{1}
\includegraphics[height=100mm,width=1\columnwidth]{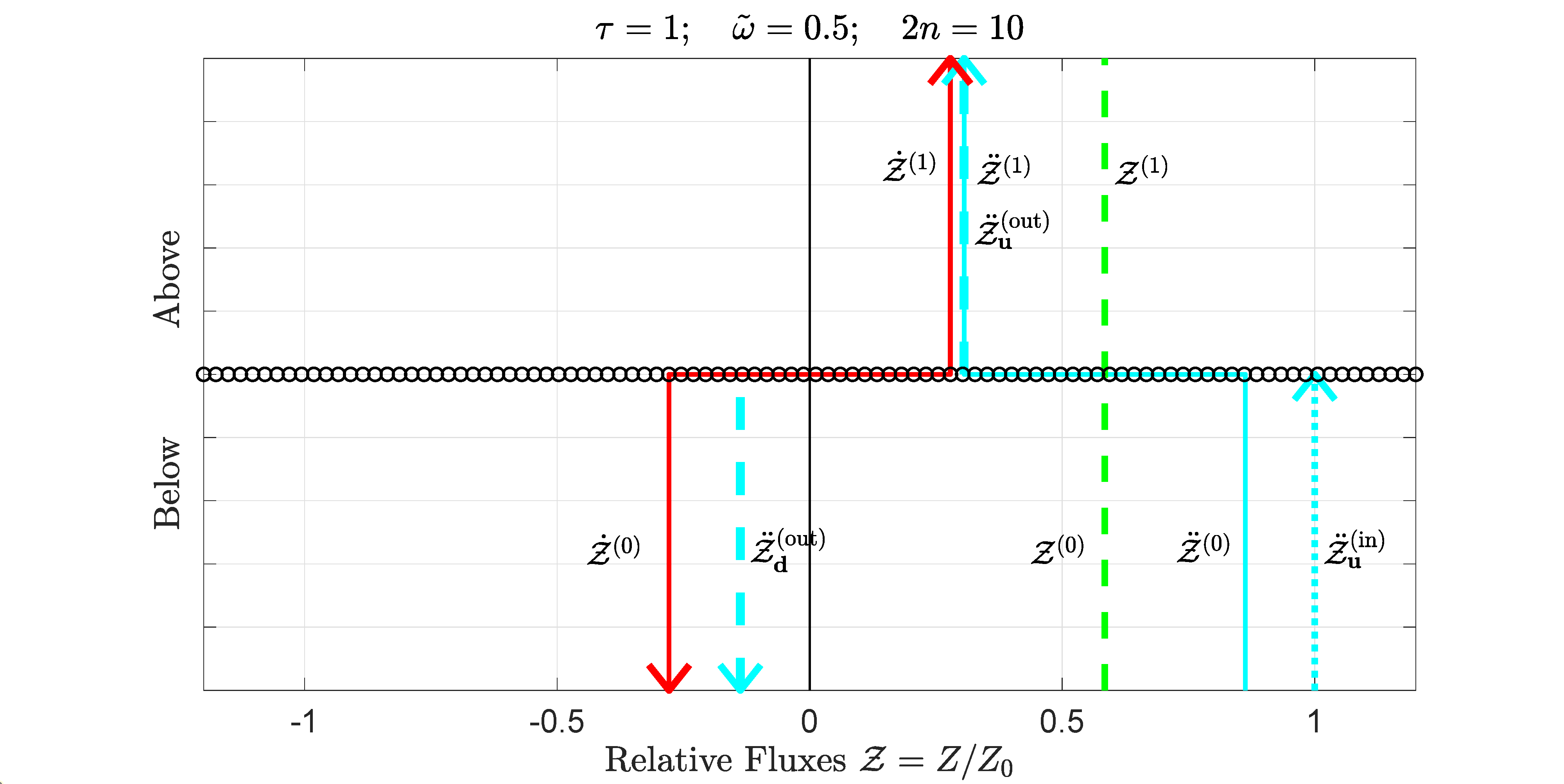}% Here is how to import EPS art
\caption{The same incoming radiation as for Fig. \ref{one2}, but after the cloud has cooled to radiative equilibrium. See the text for more detailed discussions.
\label{one3}}
\end{figure}

In Fig. \ref{one1} -- \ref{one3} we show  numerical examples of radiation transport by a single  homogeneous cloud of  optical thickness
\begin{eqnarray}
\tau= 1.
\label{nex6}
\end{eqnarray}
For the examples considered here, we assume that both the particulates and gas molecules have a Rayleigh scattering-phase matrix $\hat p$ of (\ref{et26}) with a single-scattering albedo
\begin{eqnarray}
\tilde\omega=\frac{1}{2}.
\label{nex8}
\end{eqnarray}
We can use the value  (\ref{nex8}) for the single-scattering albedo $\tilde \omega$, together with the Rayleigh-scattering-phase matrix $\hat p$ of (\ref{et26}) to evaluate the efficiency matrix  $\hat \eta$ of (\ref{et16}). Multiplying $\hat\eta$ on the left with the secant matrix $\hat \varsigma$ of (\ref{int60}) gives the  exponentiation matrix $\hat\kappa$ of (\ref{et12}).  As summarized in connection with (\ref{exin11}), 
 $\hat\kappa$, together with the optical depth $\tau$ of (\ref{nex6}), can be used to evaluate the scattering matrix $\mathcal{S}$ of the cloud.
The closely related albedo matrix $\Omega$ follows from
(\ref{exin14}), the emissivity matrix $\mathcal{E}$ follows from (\ref{exin24}) and the absorptivity matrix $\mathcal{A}$ follows from (\ref{h6}).  
We use $2n=10$ streams so $\mathcal{S}$, $\mathcal{E}$, $\Omega$ and $\mathcal{A}$ are $10\times 10$ matrices, too large for convenient display in this paper, but readily used by modern mathematical software packages like Matlab.
Using $\mathcal{E}$ to evaluate the upward Planck emissivity $\varepsilon_{\bf u}$ of   (\ref{ter8}), and using $\mathcal{S}$ to evaluate  the upward
 Planck albedo $\omega_{\bf u}$ of (\ref{pa10}), we find 
\begin{eqnarray}
\varepsilon_{\bf u} = 0.5568\quad\hbox{and}\quad\omega_{\bf u} = 0.4432 .
\label{nex12}
\end{eqnarray}
Doubling the number of streams to $2n=20$ decreases the Planck albedo by about $0.15\%$, from 
$\omega_{\bf u} = 0.4432$ to   $\omega_{\bf u} = 0.4426$, too small to be displayed in Fig. \ref{one1}.
So $2n=10$ streams gives good modeling accuracy.  

Using $\mathcal{E}$ to evaluate the matrix $\dot M$ of (\ref{ter36}) we find
\begin{equation}
4\pi \dot M=\left[\begin{array}{r}-0.5568\\ 0.5568\end{array}\right]\frac{Z_0}{B_0}.
\label{nex13a}
\end{equation}
Here $Z_0$ is the reference flux and $B_0$ is the reference Planck intensity of (\ref{nex2}).
Using $\Omega$ to evaluate the matrix $\ddot N$ of (\ref{hi22}) we find
\begin{equation}
4\pi\ddot N=\left[\begin{array}{rr}-0.3058&0.8627\\ -0.8627&0.3058\end{array}\right]\frac{Z_0}{B_0}.
\label{nex13b}
\end{equation}

Let the Planck intensity $B^{\{c\}}$ of the cloud and the Planck intensities   $B^{\{\rm in\}}_{\bf d}$ and $B^{\{\rm in \}}_{\bf u}$ of the half-isotropic  incoming radiation be 
\begin{equation}
 B^{\{c\}}=B_0\quad\hbox{and}\quad|B^{\{\rm in\}}\rangle=\left[\begin{array}{c} B^{\{\rm in\}}_{\bf d}\\ B^{\{\rm in \}}_{\bf u}\end{array}\right]=
 \left[\begin{array}{c} 1\\ 1\end{array}\right]B_0.
\label{nex14}
\end{equation}
Evaluating (\ref{ter34}) with $B^{\{c\}}$ from (\ref{nex14}) and $\dot M$ from (\ref{nex13a}) we find
\begin{equation}
|\dot Z)=\left[\begin{array}{c}\dot Z^{(0)}\\ \dot Z^{(1)}\end{array}\right]=4\pi \dot M B^{\{c\}}=\left[\begin{array}{r}-0.5568\\ 0.5568\end{array}\right]Z_0.
\label{nex16}
\end{equation}
Evaluating (\ref{hi16}) with $|B^{\{\rm in\}}\rangle$ from (\ref{nex14}) and $\ddot N$ from (\ref{nex13b}) we find (after adjusting the last numerical digit for roundoff error)
\begin{equation}
|\ddot Z)=\left[\begin{array}{c}\ddot Z^{(0)}\\ \ddot Z^{(1)}\end{array}\right]=4\pi \ddot N|B^{\{\rm in\}}\rangle = \left[\begin{array}{r}0.5568\\ -0.5568\end{array}\right]Z_0.
\label{nex18}
\end{equation}
We see that the flux (\ref{nex16})   due to thermal emission of cloud particulates and gas molecules is equal and opposite the flux (\ref{nex18}) due to transmission and scattering of incoming radiation. Summing (\ref{nex16}) and (\ref{nex18}) we find that the total flux vanishes,
\begin{equation}
|Z)=\left[\begin{array}{c}Z^{(0)}\\ Z^{(1)}\end{array}\right]=|\dot Z)+|\ddot Z)=\left[\begin{array}{r}0\\ 0\end{array}\right].
\label{nex20}
\end{equation}
The cloud heating rate (\ref{hc2}) also vanishes
\begin{equation}
R^{\{c\}}=Z^{(0)}-Z^{(1)}=0.
\label{nex22}
\end{equation}
The cloud of Fig. \ref{one1} is in full  thermal equilibium with the incoming radiation, and it neither heats nor  cools.

In Fig. \ref{one2} we show what happens if the downward incoming radiation is removed but everything else remains the same.
The Planck intensity $|B^{\{\rm in\}}\rangle$   of (\ref{nex14}) is changed to
\begin{equation}
|B^{\{\rm in\}}\rangle=\left[\begin{array}{c} B^{\{\rm in\}}_{\bf d}\\ B^{\{\rm in \}}_{\bf u}\end{array}\right]=
 \left[\begin{array}{c} 0\\ 1\end{array}\right]B_0 .
\label{nex24}
\end{equation}
Since the cloud retains the same Planck intensity $B^{\{1\}}=B_0$ as for (\ref{nex14}),
the thermally emitted flux $|\dot Z)$ stays the same as for (\ref{nex16}).
Evaluating (\ref{hi16}) with $|B^{\{\rm in\}}\rangle$ from (\ref{nex24}) and $\ddot N$ from (\ref{nex13b}) we find values different from (\ref{nex18})
\begin{equation}
|\ddot Z)=\left[\begin{array}{c}\ddot Z^{(0)}\\ \ddot Z^{(1)}\end{array}\right]=\left[\begin{array}{r}0.8627\\ 0.3058\end{array}\right]Z_0.
\label{nex28}
\end{equation}
 Summing (\ref{nex16}) and (\ref{nex28}) we find that the net flux is positive both  below and above the cloud
\begin{equation}
|Z)=\left[\begin{array}{c}Z^{(0)}\\ Z^{(1)}\end{array}\right]=|\dot Z)+|\ddot Z)=\left[\begin{array}{r}0.3058\\  0.8627\end{array}\right]Z_0.
\label{nex30}
\end{equation}
The cloud heating rate (\ref{hc2}) is negative,  that is, the cloud cools by releasing more vertical flux $Z^{(1)} = 0.8627\,Z_0$ out of the top than the flux $Z^{(0)} = 0.3058\,Z_0$ that comes into the bottom
\begin{equation}
R^{\{c\}}=Z^{(0)}-Z^{(1)}=-0.5568 Z_0.
\label{nex32}
\end{equation}

In Fig. \ref{one3} we show what happens to the cloud of Fig. \ref{one2} if it is allowed to cool to radiative equilibrium.
The incoming  radiation $|B^{\{\rm in\}}\rangle$ retains the value of (\ref{nex24}).  In accordance with (\ref{re8}),
the Planck intensity of the cloud   is halved from the value of (\ref{nex14}).
\begin{equation}
B^{\{c\}}=\frac{1}{2}B_0.
\label{nex34}
\end{equation}
Evaluating (\ref{ter34})  with $B^{\{c\}}$ from (\ref{nex34})  and $\dot M$ from (\ref{nex13a}) we find that the thermally emitted fluxes $|\dot Z)$ are half as large as those of (\ref{nex16})
\begin{equation}
|\dot Z)=\left[\begin{array}{c}\dot Z^{(0)}\\ \dot Z^{(1)}\end{array}\right]=\left[\begin{array}{r}-0.2784\\ 0.2784\end{array}\right]Z_0.
\label{nex36}
\end{equation}
The flux $|\ddot Z)$ from scattered incoming radiation remains the same as (\ref{nex28}).
Summing the thermally emitted flux (\ref{nex36}) and scattered flux $|\ddot Z)$ of  (\ref{nex28}) we find that the total flux for radiative equilibrium is
\begin{equation}
|Z)=\left[\begin{array}{c}Z^{(0)}\\ Z^{(1)}\end{array}\right]=|\dot Z)+|\ddot Z)=\left[\begin{array}{r}0.5842\\ 0.5842\end{array}\right]Z_0.
\label{nex40}
\end{equation}

The cloud heating rate (\ref{hc2}) is zero, that is, the cloud thermally emits the same amount of radiation as it absorbs from incoming radiation
\begin{equation}
R^{\{c\}}=Z^{(0)}-Z^{(1)}=0.
\label{nex42}
\end{equation}
But the cloud is colder than the source of external radiation. The Planck intensity $B^{\{c\}} = B_0/2$ of the cloud is only half of the Planck intensity of $B^{\{\rm in\}}_{\bf u}=B_0$ of the  radiation coming up from below.
\begin{figure}[t]
%\postscriptscale{seriesclouds6}{1}
\includegraphics[height=100mm,width=1\columnwidth]{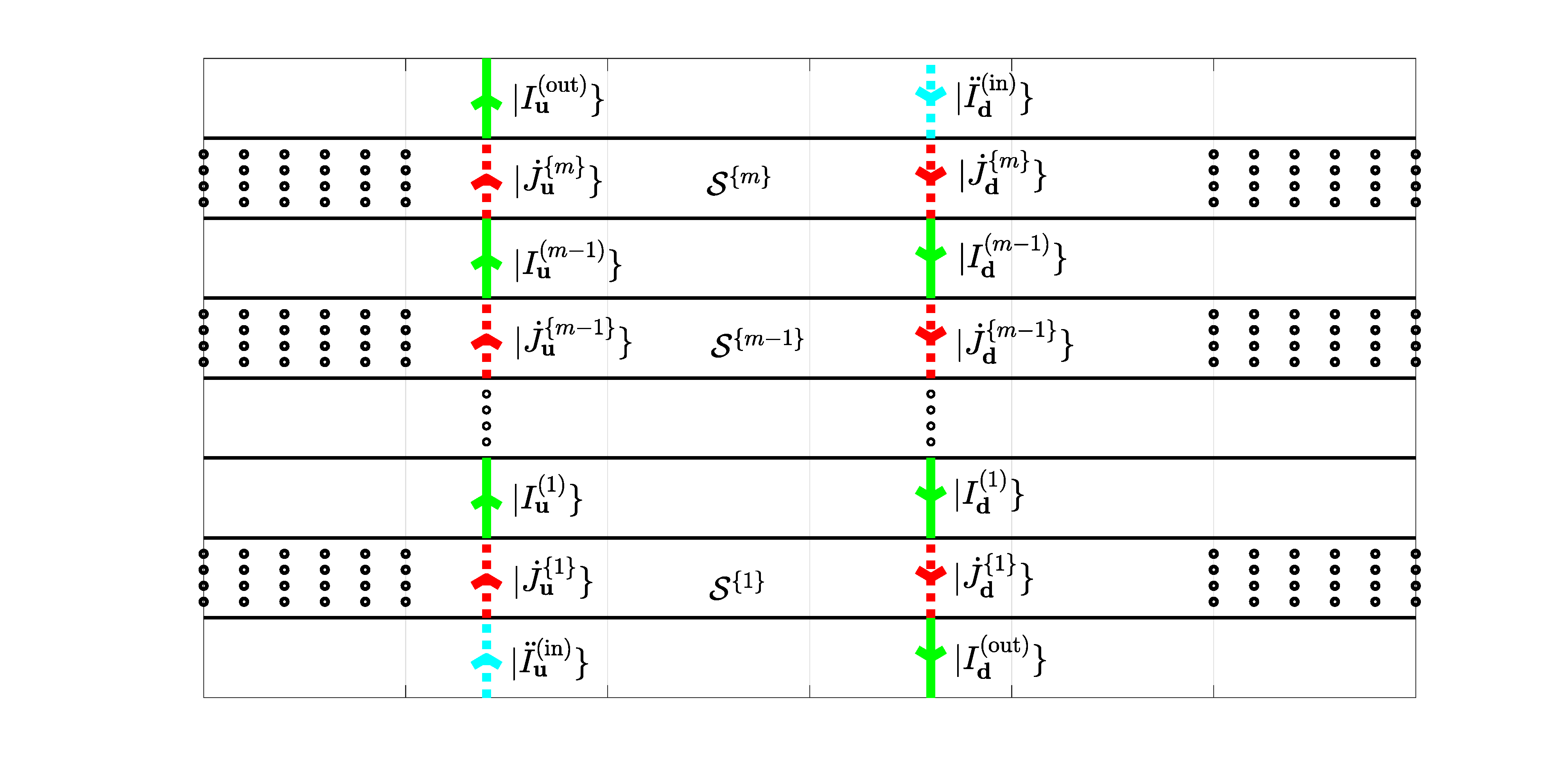}% Here is how to import EPS art
\caption{ A stack of $m$ clouds, labeled by the indices $c=1,2,3,\dots, m$. As described in the text,  the figure illustrates the fundamental equations (\ref{ic2}) -- (\ref{ic18}) for radiation transfer in a cloud stack.
\label{seriesclouds6}}
\end{figure}
\section{Cloud Stacks\label{cs}}
Having discussed transmission, scattering  and absorption of radiation, along with radiative heating and cooling of single clouds, we turn to stacks of $m$ clouds, like those shown schematically by Fig. \ref{seriesclouds6}.  Each cloud is identified by a superscript  $\{c\}$ where $c=1,2,\ldots, m$. The cloud $c$ will have a scattering matrix $\mathcal{S}^{\{c\}}$.  The thermal radiation emitted by cloud $c$ will be denoted by 
 $|\dot J^{\{c\}}\}$. The dotted red arrows of Fig. \ref{seriesclouds6} indicate the  upward and downward parts, $|\dot J^{\{c\}}_{\bf u}\}$ and $|\dot J^{\{c\}}_{\bf d}\}$. For a nonisothermal cloud,  $|\dot J^{\{c\}}\}$ is given by the (\ref{exin22}) as
\begin{equation}
|\dot J^{\{c\}}\}=\int_0^{\tau^{\{c\}}} d\tau' G^{\{c\}}(\tau')|0)B^{\{c\}}(\tau').
\label{rrs6}
\end{equation}
Here $B^{\{c\}}(\tau')$ is the Planck intensity at the optical depth $\tau'$ above the bottom of the cloud, and $ |G^{\{c\}}(\tau')\}$ is the Green's function vector for a cloud with a variable internal temperature, $T(\tau')$ at $\tau'$. The vertical optical depth between the bottom and top of the  cloud is $\tau^{\{c\}}$.
For an isothermal cloud of constant Planck intensity $B^{\{c\}}$ we can use the simpler expression (\ref{exin20}),
\begin{equation}
|\dot J^{\{c\}}\}= \mathcal{E}^{\{c\}}|0) B^{\{c\}},
\label{rrs8}
\end{equation}
where the emissivity matrix $\mathcal{E}^{\{c\}}$ of the cloud follows from (\ref{exin22}) and is
\begin{equation}
 \mathcal{E}^{\{c\}}=\int_0^{\tau^{\{c\}}} d\tau' G^{\{c\}}(\tau').
\label{rrs9}
\end{equation}
In accordance with Kirchhoff's law  (\ref{exin24}), the sum of 
the emissivity matrix $\mathcal{E}^{\{c\}}$ and the scattering matrix $\mathcal{S}^{\{c\}}$ is the $2n\times 2n$ identity matrix $\hat 1$.
\begin{equation}
\mathcal{E}^{\{c\}}+\mathcal{S}^{\{c\}}=\hat 1.
\label{rrs9}
\end{equation}

We characterize the incoming radiation by the intensity vector $|\ddot I^{(\rm in)}\}=|\ddot I^{(m)}_{\bf d}\}+|\ddot I^{(0)}_{\bf u}\}$  of (\ref{oir2}), the sum of the  downward part $|I^{(m)}_{\bf d}\}$ of the intensity vector in the $m$th ``gap," above the top cloud  ($c=m$), and the upward part $|I^{(0)}_{\bf u}\}$ of the intensity vector in the $0$th ``gap" below the bottom cloud ($c=1$). We recall from (\ref{exin6}) that $|\dot I^{(\rm in)}\}=\breve 0$ and
$| I^{(\rm in)}\}=|\ddot I^{(\rm in)}\}$. There is no internally generated thermal radiation coming into the bottom and top of the stack. In Fig. \ref{seriesclouds6} the  incoming intensity is shown as the dotted cyan arrows.

Just as for a single cloud there is an outgoing intensity vector $|I^{(\rm out)}\}$  like that of (\ref{oir6}), the sum of the  downward part $|I^{(\rm out)}_{\bf d}\}=|I^{(0)}_{\bf d}\}$ of the intensity vector in the $0$th gap below the cloud stack and the upward part $|I^{(\rm out)}_{\bf u}\}=|I^{(m)}_{\bf u}\}$ of the intensity vector in the $m$th gap above the cloud stack. The downward and upward parts of the outgoing intensity,  $|I^{(\rm out)}_{\bf d}\}$ and $|I^{(\rm out)}_{\bf u}\}$ are denoted by the green arrows above and below the cloud stack of Fig. \ref{seriesclouds6}.

The most convenient descriptors of radiation transfer in cloud stacks are the gap intensities
\begin{eqnarray}
|I^{(g)}\}&=&|\dot I^{(g)}\}+|\ddot I^{(g)}\}\nonumber\\
&=&|I^{(g)}_{\bf d}\}+| I^{(g)}_{\bf u}\},
\label{rrs9a}
\end{eqnarray}
The gap index can take on the values $g=0,1,2,\ldots,m$.  We already defined the gap intensity  $|I^{(0)}\}$ in the bottom gap, that is, the intensity below the stack in (\ref{oir10}). The gap intensity $|I^{(m)}\}$, that is, the intensity above the stack, was defined by (\ref{oir12}). In addition,
a stack of $m>1$ clouds has $m-1$ internal gaps.  For internal gaps $g=1,2,3,\ldots,m-1$ we
denote the intensity in the
 gap  between cloud $c=g$ and the next higher cloud $c=g+1$ by $|I^{(g)}\}$ .  The downward and upward parts of the internal gap intensities,  $|I^{(g)}_{\bf d}\}$ and $|I^{(g)}_{\bf u}\}$ for $g=1,2,3,\ldots,m-1$ are  denoted by the green arrows between adjacent clouds of Fig. \ref{seriesclouds6}.

We see that the radiation in a stack of clouds is determined by $2n(m+1)$ known, nonnegative numbers, the $2n$ stream projections $\lvec\mu_i|\ddot I^{(\rm in)}\}$ of the incoming intensity, and the $2nm$ stream projections $\lvec\mu_i|\dot J^{\{c\}}\}$ of the thermal source vectors. 
These known variables of the problem determine the values of $2nm$ unknown variables, the $2n$ nonnegative stream projections $\lvec\mu_i|I^{(\rm out)}\}$ of the outgoing radiation, and the $2n(m-1)$ nonnegative stream projections $\lvec\mu_i|I^{(g)}\}$  of the internal gap intensities.

From inspection of Fig. \ref{seriesclouds6} we see that the $2nm$ unknown intensities can be determined by $2nm$ linear equations, which we abbreviate as $2m$ equations of $n\times 1$ vectors
\begin{eqnarray}
|I^{(1)}_{\bf d} \}&=&\mathcal {S}^{\{2\}}_{\bf du}|I^{(1)}_{\bf u} \}+\mathcal {S}^{\{2\}}_{\bf dd}|I^{(2)}_{\bf d} \}+|\dot J^{\{2\}}_{\bf d}\}.\label{ic2}\\
|I^{(1)}_{\bf u} \}&=&\mathcal {S}^{\{1\}}_{\bf uu}|\ddot I^{(\rm in)}_{\bf u} \}+\mathcal {S}^{\{1\}}_{\bf ud}|I^{(1)}_{\bf d} \}+|\dot J^{\{1\}}_{\bf u}\}.\label{ic4}\\
|I^{(2)}_{\bf d} \}&=&\mathcal {S}^{\{3\}}_{\bf du}|I^{(2)}_{\bf u} \}+\mathcal {S}^{\{3\}}_{\bf dd}|I^{(3)}_{\bf d} \}+|\dot J^{\{3\}}_{\bf d}\}.\label{ic6}\\
|I^{(2)}_{\bf u} \}&=&\mathcal {S}^{\{2\}}_{\bf uu}|I^{(1)}_{\bf u} \}+\mathcal {S}^{\{2\}}_{\bf ud}|I^{(2)}_{\bf d} \}+|\dot J^{\{2\}}_{\bf u}\}.\label{ic8}\\
\vdots &=&\vdots\nonumber\\
|I^{(m-1)}_{\bf d} \}&=&\mathcal {S}^{\{m\}}_{\bf du}|I^{(m-1)}_{\bf u} \}+\mathcal {S}^{\{m\}}_{\bf dd}|\ddot I^{(\rm in)}_{\bf d}\}+|\dot J^{\{m\}}_{\bf d}\}.\label{ic10}\\
|I^{(m-1)}_{\bf u} \}&=&\mathcal {S}^{\{m-1\}}_{\bf uu}|I^{(m-2)}_{\bf u} \}+\mathcal {S}^{\{m-1\}}_{\bf ud}|I^{(m-1)}_{\bf d} \}+|\dot J^{\{m-1\}}_{\bf u}\}.\label{ic12}\\
 |I^{(\rm out)}_{\bf d} \}&=&\mathcal {S}^{\{1\}}_{\bf du}|\ddot I^{(\rm in)}_{\bf u} \}+\mathcal {S}^{\{1\}}_{\bf dd}|I^{(1)}_{\bf d} \} +|\dot J^{\{1\}}_{\bf d}\}.\label{ic16}\\
 |I^{(\rm out)}_{\bf u} \}&=&\mathcal {S}^{\{m\}}_{\bf uu}|I^{(m-1)}_{\bf u} \}+\mathcal {S}^{\{m\}}_{\bf ud}|\ddot I^{(\rm in)}_{\bf d} \} +|\dot J^{\{m\}}_{\bf u}\}.\label{ic18}
\end{eqnarray}
We see that  the downward radiation $|I^{(g)}_{\bf d}\}$ in the gap $g$ is the sum of three parts
\begin {enumerate}
 \item{ $\mathcal {S}^{\{g+1\}}_{\bf du}|I^{(g)}_{\bf u}\}$, the downward reflection of upward radiation $|I^{(g)}_{\bf u}\}$ in gap $g$ by the next higher cloud $c=g+1$.}
\item{$\mathcal {S}^{\{g+1\}}_{\bf dd}|I^{(g+1)}_{\bf d}\}$, the downward transmission of downward radiation $|I^{(g+1)}_{\bf d}\}$ in gap $g+1$ by the cloud $c=g+1$.}
\item{The downward thermal radiation $|\dot J^{\{g+1\}}_{\bf d}\}$ emitted by the cloud $c=g+1$.}
\end{enumerate}
The upward radiation $|I^{(g)}_{\bf u}\}$ in the gap $g$ is the sum of three analogous parts.  

To account for intensity below and above the cloud stack,  there are four exceptional cases in (\ref{ic2}) -- (\ref{ic18}). In (\ref{ic4}) and  (\ref{ic16}) the upward gap radiation on the right side of the equations is replaced by $|\ddot I^{(\rm in)}_{\bf u}\}$ of (\ref{oir2}) the upward incoming intensity incident on the bottom of the stack. In (\ref{ic10}) and  (\ref{ic18}) the downward gap radiation on the right side of the equations is replaced by $|\ddot I^{(\rm in)}_{\bf d}\}$, the downward incoming intensity incident on the top of the stack. The left sides of (\ref{ic16}) and (\ref{ic18}) show how the downward and upward parts of the unknown intensity $|I^{(\rm out)}\}$ are related to the other unknown intensities and to the thermal and external sources.
\subsection{Stack analysis\label{sa}}
To find the unknown intensities  on the left sides of equations (\ref{ic2}) -- (\ref{ic18})  it is helpful to write the equations as the formally simpler {\it stack equation}
\begin{equation}
| U]=X| U]+ P|\dot J]+Y|\ddot I^{\{\rm in\}}\}.
\label{sa2}
\end{equation}
The left side of (\ref{sa2}) is the {\it unknown intensity} stack vector, which we define as the concatenation
\begin{equation}
|U]=
\left[\begin{array}{l}|U^{[1]}\}\\  | U^{[2]}\} \\ 
\vdots\\|U^{[m-1]}\}\\ |U^{[m]}\} \end{array}\right]
=
\left[\begin{array}{l}|U^{[1]}_{\bf d}\}\\ |U^{[1]}_{\bf u}\}\\ | U^{[2]}_{\bf d}\} \\ | U^{[2]}_{\bf u}\}\\
\vdots\\|U^{[m-1]}_{\bf d}\}\\ |U^{[m-1]}_{\bf u}\}\\|U^{[m]}_{\bf d}\}\\| U^{[m]}_{\bf u}\} \end{array}\right].
\label{sa4}
\end{equation}
For $x=1,2,3,\ldots,m-1$ the unknown intensity vectors $|U^{[x]}\}$  are identical to the gap intensity vectors $|I^{(x)}\}$, that is
\begin{equation}
|U^{[x]}\}=|I^{(x)}\} \quad\hbox{for}\quad x=1,2,3,\ldots,m-1.
\label{sa6}
\end{equation}
For the special case of $x=m$, the unknown  intensity vector is the same as the outgoing intensity (\ref{oir6}) 
\begin{equation}
|U^{[m]}\}= |I^{(\rm out )}\}=|I^{(0)}_{\bf d}\}+|I^{(m)}_{\bf u}\}.
\label{sa8}
\end{equation}
We use square brackets $[$ and $]$ as delimiters for the superscript indices $x$ of the unknown intensity vectors $|U^{[x]}\}$.

The first term on the right of the stack equation (\ref{sa2}) is the product of the {\it generation} matrix $X$ with the unknown intensity stack vector $|U]$ of (\ref{sa4}),
\begin{equation}
X|U]=\sum_{xx' {\bf q q}'}X^{[x x']}_{\bf q q'}|U^{[x']}_{\bf q'}\}.
\label{sa10}
\end{equation}
The unknown  intensity indices, $x$ and $x'$,  can have the values $1,2,3,\ldots,m$. The directional indices ${\bf q}$ and ${\bf q}'$ can have the  values ${\bf d}$ and ${\bf u}$.

From inspection of (\ref{ic2}) --  (\ref{ic18}) and (\ref{sa2}) we see that  if $m$ is not so small that the pattern is distorted by the four exceptions, (\ref{ic4}), (\ref{ic10}), (\ref{ic16}) and (\ref{ic18}), the $n\times n$  block matrix elements in the upper left corner of $X$ are
\begin{equation}
X =
\left[\begin{array}{cccccccccc} 
\breve 0&\mathcal{S}^{\{2\}}_{\bf d u}&\mathcal{S}^{\{2\}}_{\bf d d}&\breve 0&\breve 0 &\breve 0&\breve 0&\cdots\\ 
\mathcal{S}^{\{1\}}_{\bf u d}&\breve 0&\breve0&\breve 0&\breve 0 &\breve 0&\breve 0&\cdots\\ 
\breve 0& \breve 0&\breve 0&\mathcal{S}^{\{3\}}_{\bf d u}&\mathcal{S}^{\{3\}}_{\bf d d} &\breve 0&\breve 0&\cdots\\
\breve 0&\mathcal{S}^{\{2\}}_{\bf u u}&\mathcal{S}^{\{2\}}_{\bf u d}&\breve 0&\breve 0&\breve 0&\breve 0&\cdots \\
\breve 0& \breve 0&\breve 0& \breve 0&\breve 0&\mathcal{S}^{\{4\}}_{\bf d u}&\mathcal{S}^{\{4\}}_{\bf d d}&\cdots \\
\breve 0& \breve 0&\breve 0&\mathcal{S}^{\{3\}}_{\bf u u}&\mathcal{S}^{\{3\}}_{\bf u d} &\breve 0&\breve 0&\cdots\\
\vdots&\vdots&\vdots&\vdots&\vdots&\vdots&\vdots&\ddots\\ 
\end{array}\right].
\label{sa12}
\end{equation}
The block matrix elements in the bottom right corner of $X$ are
\begin{equation}
X =
\left[\begin{array}{cccccccccc} 
\vdots&\ddots&\vdots &\vdots&\vdots&\vdots &\vdots &\vdots&\vdots\\
\breve 0&\cdots&\breve 0&\breve 0&\mathcal{S}^{\{m-1\}}_{\bf d u}&\mathcal{S}^{\{m-1\}}_{\bf dd }&\breve 0&\breve 0&\breve 0 \\
\breve 0&\cdots&\mathcal{S}^{\{m-2\}}_{\bf u u}&\mathcal{S}^{\{m-2\}}_{\bf u d}&\breve 0&\breve 0&\breve 0&\breve 0&\breve 0 \\
\breve 0&\cdots&\breve 0&\breve 0&\breve 0&\breve 0&\mathcal{S}^{\{m\}}_{\bf du}&\breve 0&\breve 0 \\
\breve 0&\cdots&\breve 0&\breve 0&\mathcal{S}^{\{m-1\}}_{\bf u u}&\mathcal{S}^{\{m-1\}}_{\bf ud }&\breve 0&\breve 0&\breve 0 \\
\mathcal{S}^{\{1\}}_{\bf dd}&\cdots&\breve 0&\breve 0&\breve 0&\breve 0&\breve 0&\breve 0&\breve 0 \\
\breve 0&\cdots&\breve 0&\breve 0&\breve 0&\breve 0&\mathcal{S}^{\{m\}}_{\bf uu}&\breve 0&\breve 0 \\
\end{array}\right].
\label{sa14}
\end{equation}
The last two columns of $X$ have only $n\times n$ null matrices, $\breve 0$, as elements.  For the simplest case of $m=2$ clouds,  $X$ is constructed entirely from the four exceptional equations, (\ref{ic4}), (\ref{ic10}), (\ref{ic16}) and (\ref{ic18}), and can be written as
\begin{equation}
X =
\left[\begin{array}{llll} 
X^{[1 1]}_{\bf d d}&X^{[1 1]}_{\bf d u}&X^{[1 2]}_{\bf d d}&X^{[1 2]}_{\bf d u}\\
X^{[1 1]}_{\bf u d}&X^{[1 1]}_{\bf u u}&X^{[1 2]}_{\bf u d}&X^{[1 2]}_{\bf u u}\\
X^{[2 1]}_{\bf d d}&X^{[2 1]}_{\bf d u}&X^{[2 2]}_{\bf d d}&X^{[2 2]}_{\bf d u}\\
X^{[2 1]}_{\bf u d}&X^{[2 1]}_{\bf u u}&X^{[2 2]}_{\bf u d}&X^{[2 2]}_{\bf u u}
\end{array}\right]
=
\left[\begin{array}{llll} 
\breve 0&\mathcal{S}^{\{2\}}_{\bf d u}&\breve 0 &\breve 0\\ 
\mathcal{S}^{\{1\}}_{\bf u d}&\breve 0&\breve0&\breve 0\\
\mathcal{S}^{\{1\}}_{\bf dd}& \breve 0&\breve 0&\breve 0\\ 
\breve 0&\mathcal{S}^{\{2\}}_{\bf u u}&\breve 0&\breve 0\\
\end{array}\right].
\label{sa16}
\end{equation}

The second term on the right of (\ref{sa2}) 
\begin{equation}
P|\dot J]=\sum_{xc'{\bf q q}'} P^{[x c'\}}_{\bf q q'}|J^{\{c'\}}_{\bf q'}\}.
\label{sa18}
\end{equation}
 is the product of the {\it permutation} matrix $P$ with the known thermal source stack matrix  $|\dot J]$, which we write as
\begin{equation}
|\dot J]=\left[\begin{array}{l}|\dot J^{\{1\}}\}\\  | \dot J^{\{2\}}\} \\ 
\vdots\\ |\dot J^{\{m-1\}}\}\\ |\dot J^{\{m\}}\} \end{array}\right]=
\left[\begin{array}{l}|\dot J^{\{1\}}_{\bf d}\}\\ |\dot J^{\{1\}}_{\bf u}\}\\ |\dot J^{\{2\}}_{\bf d}\}\\ |\dot J^{\{2\}}_{\bf u}\}\\
\vdots\\|\dot J^{\{m-1\}}_{\bf d}\}\\ |\dot J^{\{m-1\}}_{\bf u}\}\\ |\dot J^{\{m\}}_{\bf d}\}\\ |\dot J^{\{m\}}_{\bf u}\} \end{array}\right].
\label{sa20}
\end{equation}
Here the thermal source vectors  $| \dot J^{\{c\}}\}$ are given by (\ref{rrs6}) for a cloud $c$ with a variable internal temperature,  or by  (\ref{rrs8}) for an isothermal cloud. The downward and upward parts of the source vectors are $|\dot J^{\{c\}}_{\bf d}\}=\mathcal{M}_{\bf d}|\dot J^{\{c\}}\}$ and $|\dot J^{\{c\}}_{\bf u}\}=\mathcal{M}_{\bf d}|\dot J^{\{c\}}\}$.

From inspection of (\ref{ic2}) --  (\ref{ic18}) and (\ref{sa2}) we see that the matrix 
 $P$  is a block version of a permutation matrix\,\cite{permutation}
\begin{equation}
P=
\left[\begin{array}{cccccccccc} 
\breve 0&\breve 0&\hat 1&\breve 0&\breve 0&\cdots &\breve 0&\breve 0&\breve 0&\breve 0\\ 
\breve 0&\hat 1&\breve 0&\breve 0&\breve 0&\cdots &\breve 0&\breve  0&\breve 0&\breve 0\\ 
\breve 0&\breve 0&\breve 0&\breve 0&\hat 1&\cdots &\breve 0&\breve 0&\breve 0&\breve  0\\
\breve 0&\breve 0&\breve 0&
\hat 1&\breve 0&\cdots &\breve 0&\breve 0&\breve 0&\breve 0\\
\vdots&\vdots&\vdots&\vdots&\vdots&\ddots&\vdots&\vdots&\vdots&\vdots\\
\breve 0&\breve 0&\breve 0&\breve 0&\breve 0&\cdots &\breve 0&\breve 0&\hat 1&\breve 0\\
\breve 0&\breve 0&\breve 0&\breve 0&\breve 0&\cdots &\breve 0&\hat 1&\breve 0&\breve 0
\\  \hat 1&\breve 0&\breve 0&\breve 0&\breve 0&\cdots &\breve 0&\breve 0&\breve 0&\breve 0\\
\breve 0&\breve 0&\breve 0&\breve 0&\breve 0&\cdots &\breve 0&\breve 0&\breve 0&\hat 1
\end{array}\right].
\label{sa22}
\end{equation}
Here $\hat 1$ denotes an $n\times n$ identity matrix and $\breve 0$ is an $n\times n$ null matrix. For the simplest case of $m=2$,  the matrix $P$ becomes
\begin{equation}
P=
\left[\begin{array}{llll} P^{[1 1\}}_{\bf d d}&P^{[1 1\}}_{\bf d u}&P^{[1 2\}}_{\bf d d}&P^{[1 2\}}_{\bf d u}\\
P^{[1 1\}}_{\bf u d}&P^{[1 1\}}_{\bf u u}&P^{[1 2\}}_{\bf u d}&P^{[1 2\}}_{\bf u u}\\
P^{[2 1\}}_{\bf d d}&P^{[2 1\}}_{\bf d u}&P^{[2 2\}}_{\bf d d}&P^{[2 2\}}_{\bf d u}\\
P^{[2 1\}}_{\bf u d}&P^{[2 1\}}_{\bf u u}&P^{[2 2\}}_{\bf u d}&P^{[2 2\}}_{\bf u u}\\
\end{array}\right]\\
=
\left[\begin{array}{cccc} 
\breve 0&\breve 0&\hat 1&\breve 0\\ 
\breve 0&\hat 1&\breve 0&\breve 0\\ 
\hat 1&\breve 0&\breve 0&\breve 0\\
\breve 0&\breve 0&\breve 0&\hat 1\\
\end{array}\right].
\label{sa24}
\end{equation}

The last term on the right of (\ref{sa2})
\begin{equation}
Y|\ddot I^{(\rm in)}\}=\sum_{x{\bf q q'}}Y^{[x]}_{\bf q q'}|\ddot I^{(\rm in)}_{\bf q'}\}
\label{sa26}
\end{equation}
is the product of the {\it insertion} matrix $Y$ with the
incoming intensity vector  $|\ddot I^{(\rm in)}\}$ of  (\ref{oir2}).
From inspection of  (\ref{ic2}) --  (\ref{ic18}) and (\ref{sa2}) we see that $Y$ has  the general form
\begin{equation}
Y =
\left[\begin{array}{ll} 
Y^{[1]}_{\bf d d}&Y^{[1]}_{\bf d u}\\ 
Y^{[1]}_{\bf u d}&Y^{[1]}_{\bf u u}\\
\vdots& \vdots\\ 
Y^{[m-1]}_{\bf d d}&Y^{[m-1]}_{\bf d u}\\
Y^{[m-1]}_{\bf u d}&Y^{[m-1]}_{\bf u u}\\
Y^{[m]}_{\bf d d}&Y^{[m]}_{\bf d u}\\
Y^{[m]}_{\bf u d}&Y^{[m]}_{\bf u u}\\
\end{array}\right]
=
\left[\begin{array}{ll} 
\breve 0&\breve 0\\ 
\breve 0&\mathcal{S}^{\{1\}}_{\bf u u}\\
\vdots& \vdots\\ 
\mathcal{S}^{\{m\}}_{\bf d d}&\breve 0\\
\breve 0&\breve 0\\
\breve 0&\mathcal{S}^{\{1\}}_{\bf d u}\\
\mathcal{S}^{\{m\}}_{\bf u d}&\breve 0\\
\end{array}\right].
\label{sa28}
\end{equation}
The four non-zero elements, $Y^{[1]}_{\bf uu}$,   $Y^{[m-1]}_{\bf dd}$, $Y^{[m]}_{\bf du}$, and  $Y^{[m]}_{\bf ud}$, are parts of the scattering matrices 
$\mathcal{S}^{\{1\}}$ and $\mathcal{S}^{\{m\}}$ of the bottom and top clouds of the stack.
From inspection of (\ref{sa28}) we see that
 (\ref{sa26}) determines:
 \begin{enumerate}
  \item { how much the upward incoming intensity 
$|\ddot I^{(\rm in)}_{\bf u}\}$ contributes to the upward unknown intensity $|\ddot U^{[1]}_{\bf u}\}$ in the gap above the bottom  cloud with $c=1$}
\item{ how much the downward incoming intensity 
$|\ddot I^{(\rm in)}_{\bf d}\}$ contributes to the downward unknown intensity $|\ddot U^{[m-1]}\}$ in the gap below the top  cloud with $c=m-1$}
\item { how much the upward incoming intensity 
$|\ddot I^{(\rm in)}_{\bf u}\}$ contributes to the downward unknown intensity element $|\ddot U^{[m]}_{\bf d}\}$.  According to (\ref{sa8}), this is equal to the downward outgoing intensity $|\ddot I^{\{\rm out\}}_{\bf d}\}$ from the cloud stack, $|\ddot U^{[m]}_{\bf d}\}=|\ddot I^{\{\rm out\}}_{\bf d}\}$ }
\item { how much the downward incoming intensity 
$|\ddot I^{(\rm in)}_{\bf d}\}$ contributes to the upward unknown intensity element $|\ddot U^{[m]}_{\bf u}\}$.  According to (\ref{sa8}), this is equal to the upward outgoing intensity $|\ddot I^{\{\rm out\}}_{\bf u}\}$ from the cloud stack, $|\ddot U^{[m]}_{\bf u}\}=|\ddot I^{\{\rm out\}}_{\bf u}\}$ }
\end{enumerate}   
For the simplest case of $m=2$ we see from inspection of (\ref{ic2}) --  (\ref{ic18})  that the matrix $Y$ simpifies to
\begin{equation}
Y =
\left[\begin{array}{ll} 
Y^{[1]}_{\bf d d}&Y^{[1]}_{\bf d u}\\ 
Y^{[1]}_{\bf u d}&Y^{[1]}_{\bf u u}\\
Y^{[2]}_{\bf d d}&Y^{[2]}_{\bf d u}\\
Y^{[2]}_{\bf u d}&Y^{[2]}_{\bf u u}\\
\end{array}\right]
=
\left[\begin{array}{ll} 
\mathcal{S}^{\{2\}}_{\bf d d}&\breve 0\\
\breve 0&\mathcal{S}^{\{1\}}_{\bf u u}\\
\breve 0&\mathcal{S}^{\{1\}}_{\bf d u}\\
\mathcal{S}^{\{2\}}_{\bf u d}&\breve 0\\
\end{array}\right].
\label{sa30}
\end{equation}
\subsection{Thermal equilibrium\label{te}}
Suppose that all clouds, as well as the external radiation that is incident on them, have the same temperature and therefore the same Planck intensity $B$. Then the gap intensities must be given by 
\begin{equation}
|U]=|0]B,
\label{te2}
\end{equation}
The angular part of the thermal-equilibrium intensity (\ref{te2}) is a stack of $m$ copies of the monopole basis vector $|0)$ of (\ref{mm10})
\begin{equation}
|0]=\left[\begin{array}{c}|0)\\ | 0) \\ 
\vdots\\|0)\end{array}\right].
\label{te4}
\end{equation}

For isothermal clouds, each with the same Planck intensity $B$, we can use (\ref{rrs8}) to write the thermal source stack vector of (\ref{sa20}) as
\begin{equation}
|\dot J]=[\mathcal{E}]|0]B.
\label{te6}
\end{equation}
The stack matrix for emissivity  is block diagonal and given by
\begin{equation}
[\mathcal{E}]=\left[\begin{array}{llllll} 
\mathcal{E}^{\{1\}}&\breve 0 &\breve 0&\cdots&\breve 0&\breve 0\\ 
\breve 0&\mathcal{E}^{\{2\}}&\breve 0&\cdots&\breve 0&\breve0\\
\breve 0&\breve 0&\mathcal{E}^{\{3\}}&\cdots&\breve 0& \breve 0\\ 
\vdots&\vdots&\vdots&\ddots&\vdots&\vdots\\
\breve 0&\breve 0&\breve 0&\cdots&\mathcal{E}^{\{m-1\}}&\breve 0\\
\breve 0&\breve 0&\breve 0&\cdots&\breve 0&\mathcal{E}^{\{m\}}\\
\end{array}\right].
\label{te8}
\end{equation}
The emissivity matrices $\mathcal{E}^{\{c\}}$  of individual clouds were given by (\ref{exin24}).

External radiation in thermal equilibrium is simply Planck radiation, or blackbody radiation, like that of (\ref{bc6}),
\begin{equation}
|\ddot I^{(\rm in)}\}=|0)B.
\label{te10}
\end{equation}
Substituting (\ref{te2}), (\ref{te6}) and (\ref{te10})  into (\ref{sa2}) we find
\begin{equation}
\left(\hat 1-X- P[\mathcal{E}]\right)|0]=Y|0).
\label{te12}
\end{equation}
The identity (\ref{te12}) can be used as a consistency check of the numerical values of the matrices  $X$, $P$, 
$[\mathcal{E}]$ and $Y$.
\subsection{Formal solutions\label{fs}}
Rewriting (\ref{sa2}) as
\begin{equation}
(1-X)| U]= P|\dot J]+Y|\ddot I^{(\rm in)}\},
\label{fs2}
\end{equation}
we see that the formal solution is 
\begin{eqnarray}
| U]&=&(\hat 1 -X)^{-1}P|\dot J]+(\hat 1 -X)^{-1}Y|\ddot I^{(\rm in)}\}\nonumber\\
&=&G|\dot J]+W|\ddot I^{(\rm in)}\}\nonumber\\
&=&|\dot U]+|\ddot U].
\label{fs4}
\end{eqnarray}
According to the last line of (\ref{fs4}),
the unknown intensity stack vector $|U]$ is a linear combination of the known thermal source stack vector $|\dot J]$ of (\ref{sa20}) and the known incoming intensity vector $|\ddot I^{(\rm in)}\}$ of (\ref{oir2}).
We will discuss the proportionality matrices $G$ and $W$ below.

The contribution to (\ref{fs4}) of thermal emission by cloud particulates and gas molecules is
\begin{equation}
|\dot U]= G|\dot J]\quad\hbox{or}\quad |\dot U^{[x]}\}=\sum_{c=1}^{m}G^{[xc\}}|\dot J^{\{c\}}\}.
\label{fs6}
\end{equation}
Here the stack matrix  $G$ of (\ref{fs6}) can be written as the $m \times m$ block matrix 
\begin{equation}
G=(\hat 1-X)^{-1}P=
\left[\begin{array}{ccccccc} 
G^{[1 1\}}&G^{[1 2\}}&\cdots
&G^{[1, m-1\}}&G^{[1, m\}}\\ 
G^{[2 1\}}&G^{[2 2\}}&\cdots
&G^{[2, m-1\}}&G^{[2, m\}}\\
\vdots&\vdots&\ddots&\vdots&\vdots\\
G^{[m-1, 1\}}&G^{[m-1, 2\}}&\cdots
&G^{[m-1, m-1\}}&G^{[m-1, m\}}\\
G^{[m, 1\}}&G^{[m,2\}}&\cdots
&G^{[m, m-1\}}&G^{[m, m\}}\\
\end{array}\right].
\label{fs8}
\end{equation}
The $2n\times 2n$ {\it discrete Green's matrices} $G^{[x c\}}$, that are the elements of the block matrix (\ref{fs8}), are themselves $2\times 2$ block matrices,
\begin{equation}
G^{[x c\}}=
\left[\begin{array}{ccccccc} 
G^{[x c\}}_{\bf d d}&G^{[x c\}}_{\bf d u}\\
G^{[x c\}}_{\bf u d}&G^{[x c\}}_{\bf u u}\end{array}\right]\quad\hbox{where}\quad
\mathcal{M}_{\bf q}G^{[x c\}}\mathcal{M}_{\bf q'}.
\label{fs10}
\end{equation}
Both the  row index $x$, which labels unknown intensities, and the column  index $c$ of $G^{[x c\}}$,  which labels clouds, can have the values $1,2,3,\ldots,m$. 

We can use the last block element of the matrix equation  (\ref{fs6}) to write the equivalent thermal source vector $|\dot J^{\{\rm ev\}}\}$ of the entire stack as a linear combination of the thermal source vectors of the individual clouds,
\begin{eqnarray}
|\dot J^{\{\rm ev\}}\}&=&|\dot I^{(\rm out)}\}=|\dot U^{[m]}\}\nonumber\\
&=&\sum_{c=1}^m G^{[m c\}}|\dot J^{\{c\}}\}.
\label{fs12}
\end{eqnarray}
For the special case that each cloud is  isothermal with a Planck intensity $B^{\{c\}}$ and a corresponding thermal source vector $|\dot J^{\{c\}}\}= \mathcal {E}^{\{c\}}|0)B^{\{c\}}$ from (\ref{rrs8}), we can simplify   (\ref{fs12})  to
\begin{eqnarray}
|\dot J^{\{\rm ev\}}\}&=&|\dot I^{(\rm out)}\}=|\dot U^{[m]}\}\nonumber\\
&=&\sum_{c=1}^m G^{[m c\}}\mathcal {E}^{\{c\}}|0)B^{\{c\}}.
\label{fs14}
\end{eqnarray}
If all clouds have the same temperature and the same Planck intensities $B^{\{c\}}=B$, (\ref{fs14}) can be further simplified to 
\begin{eqnarray}
|\dot J^{\{\rm ev\}}\}&=&|\dot I^{(\rm out)}\}=|\dot U^{[m]}\}\nonumber\\
&=&\mathcal {E}^{\{\rm ev\}}|0)B.
\label{fs16}
\end{eqnarray}
The emissivity  of the isothermal  cloud stack is a weighted sum of the individual cloud emissivities,
\begin{equation}
\mathcal{E}^{\{\rm ev\}} = \sum_{c=1}^m G^{[m c\}}\mathcal{E}^{\{c\}}.
\label{fs18}
\end{equation}
From inspection (\ref{fs18}) and (\ref{gf8})  we see that in the limit of clouds $c$ of infinitesimal optical thickness, $\tau^{\{c\}}\to d\tau'\to 0$,
 the discrete Green's matrix $G^{[m c\}}$ and the emissivity $\mathcal{E}^{\{c\}}$ approach the limits
\begin{equation}
\frac{G^{[m c\}}}{\tau^{\{c\}}} \to G(\tau')\quad\hbox{and}\quad \mathcal{E}^{\{c\}}\tau^{\{c\}}\to
 \hat 1\, d\tau'\quad \hbox{as}\quad \tau^{\{c\}}\to d\tau'\to 0.
\label{fs19}
\end{equation}
Here $G^{[m c\}}$ is the discrete Green's matrix of (\ref{fs40}),   $G(\tau')$ is the continuous Green's matrix of (\ref{gf8}), and $\tau'$ is the total optical depth from the bottom of the cloud stack to the bottom of the cloud $c$.

The contribution to (\ref{fs4}) of incoming radiation is
\begin{equation}
|\ddot U]= W|\ddot I^{(\rm in)}\} \quad\hbox{or}\quad |\ddot U^{[x]}\}=W^{[x]}|\ddot I^{(\rm in)}\}.
\label{fs20}
\end{equation}
The scattering coefficients $W^{[x]}$ are generalized scattering matrices that give the contribution to the unknown gap intensity $|U^{[x]}\}$ from the incoming  intensity $|I^{\{\rm in\}}\}$  after transmissions and reflections by all of the clouds of the stack.
The $m\times 1$ block matrix $W$  of {\it scattering coefficients} is
\begin{equation}
W=(\hat 1-X)^{-1}Y =
\left[\begin{array}{cc} 
W^{[1]}\\
W^{[2 ]}\\
\vdots\\
W^{[m-1 ]}\\
W^{[m] }\\
\end{array}\right].
\label{fs22}
\end{equation}
The elements $W^{[x]}$ of the block matrix (\ref{fs22}) are themselves $2\times 2$ block matrices, with $n\times n$ elements $W^{[x]}_{\bf q q'}$,
\begin{equation}
W^{[x]}=
\left[\begin{array}{ccccccc} 
W^{[x]}_{\bf d d}&W^{[x]}_{\bf d u}\\
W^{[x]}_{\bf u d}&W^{[x]}_{\bf u u}\end{array}\right]\quad\hbox{where}\quad
\mathcal{M}_{\bf q}W^{[x ]}\mathcal{M}_{\bf q'}.
\label{fs24}
\end{equation}
We write the last element of the matrix equation (\ref{fs20}) as
\begin{equation}
|\ddot U^{[m]}\}=W^{[m]}|\ddot I^{(\rm in)}\}.
\label{fs26}
\end{equation}
or since $|\ddot U^{[m]}\}=|\ddot I^{(\rm out)}\}$ according to (\ref{sa8}),
\begin{equation}
|\ddot I^{(\rm out)}\}=\mathcal{S}^{\{\rm ev\}}|\ddot I^{(\rm in)}\}.
\label{fs28}
\end{equation}
From inspection of (\ref{fs26}) and (\ref{fs28}) we see that the last scattering coefficient $W^{[m]}$ is the equivalent scattering matrix of the entire cloud stack, 
\begin{equation}
\mathcal{S}^{\{\rm ev\}}=W^{[m]}.
\label{fs30}
\end{equation}
As we will discuss in more detail in Section {\bf\ref{dc}},
the equivalent scattering matrix $\mathcal{S}^{\{\rm ev\}}$   of a cloud stack is the Redheffer {\it star product}\,\cite{Redheffer} of the scattering matrices of the individual clouds,
\begin{equation}
\mathcal{S}^{\{\rm ev\}}=W^{[m]}=\mathcal{S}^{\{m\}}\rstar\mathcal{S}^{\{m-1\}}\rstar\cdots\rstar\mathcal{S}^{\{2\}}\rstar
\mathcal{S}^{\{1\}}.
\label{fs32}
\end{equation}
An explicit expression for the  star product of two clouds is given  in Section {\bf \ref{sc}} as Eq.  (\ref{sc8}). The general expression (\ref{fs32}) for a stack of $m\ge 2$ clouds follows by mathematical induction. 

In summary, we can use (\ref{fs12}) and (\ref{fs26})   to write  the total output intensity, the sum of a part due to thermal emission of cloud particulates and gas molecules, and a part due to transmission, scattering  and absorption of incoming  radiation, as
\begin{equation}
|I^{(\rm out)} \}=|\dot J^{\{\rm ev\}}\}+\mathcal{S}^{\{\rm ev\}}|\ddot I^{(\rm in)}\}.
\label{fs34}
\end{equation}
The thermal source vector $|\dot J^{\{\rm ev\}}\}$ of the equivalent cloud was given by (\ref{fs12}) for non-isothermal clouds and by the simpler expression (\ref{fs14}) for isothermal clouds.
For a stack of isothermal clouds, we can
substitute the emissivity matrix (\ref{fs18}), and the scattering matrix (\ref{fs30}) into 
Kirchhoff's law (\ref{exin24}), to find the identity,
\begin{equation}
\sum_{c=1}^m G^{[m c\}}\mathcal{E}^{\{c\}}+
\mathcal{S}^{\{m\}}\rstar\mathcal{S}^{\{m-1\}}\rstar\cdots\rstar\mathcal{S}^{\{2\}}\rstar
\mathcal{S}^{\{1\}}=\hat 1.
\label{fs40}
\end{equation}
Eq. (\ref{fs40}), a discretized version of the Kirchhoff identity (\ref{gf8}),  can provide a useful check of numerical calculations.  

Using the stack  Green's matrix $G$ to find $|\dot U]=G|\dot J]$ in (\ref{fs6}), and using the scattering coefficient matrix $W$ to find $|\ddot U]=W|\ddot I^{(\rm in)}\}$ in (\ref{fs20}),  have the advantage of physical clarity. But this is not the most efficient and accurate way to numerically solve (\ref{fs4}). For extensive numerical calculations, it is better to avoid calculating the inverse matrix $(1-X)^{-1}$ needed to evaluate $G$ and $W$, and to evaluate the contribution  $|\dot U]$ of internal thermal emission and  $|\ddot U]$ of incoming radiation with  {\it backslash fractions},
\begin{equation}
|\dot U]=(\hat 1-X)\backslash P|\dot J],
\label{fs42}
\end{equation}
and
\begin{equation}
|\ddot U]=(\hat 1-X)\backslash Y|\ddot I^{\{\rm in\}}\}.
\label{fs44}
\end{equation}
The algorithms used by modern mathematical software  like Matlab\,\cite{backslash} to evaluate backslash fractions like (\ref{fs42}) and (\ref{fs44}) are more efficient  and much more accurate than those used to evaluate the inverse matrix $(\hat 1-X)^{-1}$.

\subsection{Gap intensities and fluxes\label{gi}}
Having solved (\ref{fs4}) for the stack vector $|U]$ of unknown intensities, we can use the results to evaluate the {\it gap intensity stack vector}
\begin{equation}
|I)=
\left[\begin{array}{l}|I^{(0)}\}\\|I^{(1)}\}\\ |I^{(2)}\}\\ \vdots\\
|I^{(m-1)}\}\\|I^{(m)}\}\\  \end{array}\right].
\label{gi2}
\end{equation}
The gap intensity stack vector $|I)$ of (\ref{gi2}) and the unknown intensity vector $|U]$ of (\ref{sa4}) are almost the same, but the small differences are important. 
The  gap intensity vector $|I)$ of (\ref{gi2}) has $m+1$ block elements, $|I^{(g)}\}$ for $g=0,1,2,\ldots,m$, in contrast to the $m$ block elements  $|U^{[x]}\}$ for $x=1,2,3,\ldots,m$ of the unknown intensity vector $|U]$ of (\ref{sa4}). According to (\ref{sa6}) the  $m-1$ middle block elements $|I^{(g)}\}=|U^{[g]}\}$ for the  internal gap indices $g=1,2,3,\ldots,m-1$. 
The first block element $|I^{(0)}\}$ of the gap intensity vector $|I)$ of (\ref{gi2}) was given by  (\ref{oir10}), which we rewrite as
\begin{eqnarray}
|I^{(0)}\}&=&|I^{(\rm out)}_{\bf d}\}+|I^{(\rm in)}_{\bf u}\}\nonumber\\
&=&|U^{[m]}_{\bf d}\}+|\ddot          I^{(\rm in)}_{\bf u}\}.
\label{gi2a}
\end{eqnarray}
The second line of (\ref{gi2a}) comes from (\ref{exin6}) and (\ref{sa8}). 
The last block element $|I^{(m)}\}$ of the gap intensity vector $|I)$ of (\ref{gi2}) was given by  (\ref{oir12}), which we rewrite as
\begin{eqnarray}
|I^{(m)}\}&=&|I^{(\rm in)}_{\bf d}\}+|I^{(\rm out)}_{\bf u}\}\nonumber\\
&=&|\ddot I^{(\rm in)}_{\bf d}\}+|U^{[m]}_{\bf u}\}.
\label{gi2c}
\end{eqnarray}
The second line of (\ref{gi2c}) comes from (\ref{exin6}) and (\ref{sa8}). 

The gap intensity stack vector $|I)$ of (\ref{gi2}) is fully characterized by $2n(m+1)$ nonnegative numbers $\lvec\mu_i|I^{(g)}\}$. One can also fully characterize the intensity with the  $2n(m+1)$ multipole moments
\begin{equation}
I^{(g)}_l =\lvec l|I^{(g)}\},
\label{gi8}
\end{equation}
where the possible values of the multipole index are $l=0,1,2,...,2n-1$ and the possible values of the gap index are $g=0,1,2,\ldots,m$.
Expressions for the multipole amplitudes $I^{(g)}_l$ were given in Section {\bf \ref{mm}}.	
Of particular importance are the dipole moments $I^{(g)}_1$. According to  (\ref{mm28}) we can write the scalar gap fluxes $Z^{(g)}$ as
\begin{equation}
Z^{(g)}=4\pi I^{(g)}_1=4\pi\lvec 1|I^{(g)}\}.
\label{gi10}
\end{equation}
We will denote  the $(m+1)\times 1$ array of scalar gap fluxes by
\begin{equation}
|Z)=
\left[\begin{array}{l}Z^{(0)}\\ Z^{(1)}\\ Z^{(2)}\\ 
\vdots\\ Z^{(m-1)}\\Z^{(m)}\end{array}\right].
\label{gi12}
\end{equation}
\subsection{Model cloud stacks\label{bb}}
To gain insight into radiation transfer for cloud stacks we consider the simplified model where
each cloud is isothermal and has a Planck intensity $B^{\{c\}}$, and where the incoming radiation  $|\ddot I^{(\rm in)}\}$ is the half-isotropic radiation of (\ref{hi2}).  We assume that the scattering matrices, $\mathcal{S}^{\{c\}}$, and therefore the emissivity matrices, $\mathcal{E}^{\{c\}}=\hat 1-\mathcal{S}^{\{c\}}$, are known for all of the clouds. Then the radiation transfer is determined by only $m+2$  initial conditions: the $m$ Planck intensities $B^{\{c\}}$ of the isothermal clouds and the upward and downward Planck intensities, $B^{\{\rm in\}}_{\bf u}$ and $B^{\{\rm in\}}_{\bf d}$ of the half-isotropic incoming radiation. 

We first consider the scalar fluxes of internal gaps $g=1,2,3,\ldots, m-1$, when
$|I^{(g)}\}=|U^{[g]}\}$ in accordance with (\ref{sa6}).  Using (\ref{fs6}) with $|\dot J^{\{c\}}\}=\mathcal{E}^{\{c\}}|0)B^{\{c\}}$ in accordance with (\ref{rrs8}), we write the thermally generated part of (\ref{gi10}) as
\begin{equation}
\dot Z^{(g)}=4\pi \sum_{c=1}^m \lvec 1|G^{[gc\}}\mathcal{E}^{\{c\}}|0)B^{\{c\}}.
\label{bb2}
\end{equation}
We can use (\ref{fs20}) and (\ref{hi2}) to write the contribution of incoming radiation to the internal gap flux as
\begin{equation}
\ddot Z^{(g)}=4\pi\sum_{\bf q=d} ^{\bf u}\lvec 1|W^{[g]}\mathcal{M}_{\bf q}|0)B^{\{\rm in\}}_{\bf q}.
\label{bb4}
\end{equation}
Summing (\ref{bb2}) and (\ref{bb4}), we find that the total scalar flux for the internal gap $g$ is
\begin{eqnarray}
 Z^{(g)}&=&\dot Z^{(g)}+\ddot Z^{(g)}\nonumber\\
&=&4\pi \sum_{c=1}^m \lvec 1|G^{[gc\}}\mathcal{E}^{\{c\}}|0)B^{\{c\}}+4\pi\sum_{\bf q=d} ^{\bf u}\lvec 1|W^{[g]}\mathcal{M}_{\bf q}|0)B^{\{\rm in\}}_{\bf q}\nonumber\\
&=&4\pi\sum_{c=1}^{m}\dot M^{(g c\}}B^{\{c\}}+4\pi\sum_{\bf q=d}^{\bf u}\ddot N^{(g)}_{\bf q}B^{\{\rm in\}}_{\bf q}.
\label{bb5}
\end{eqnarray}
For the internal gaps with $g=1,2,3,\ldots,m-1$, the coupling matrices are
\begin{eqnarray}
\dot M^{(g c\}}=\lvec 1|G^{[gc\}}\mathcal{E}^{\{c\}}|0)\quad\hbox{for}\quad c=1,2,3,\ldots,m,\quad \hbox{and}\quad \ddot N^{(g)}_{\bf q}=\lvec 1|W^{[g]}\mathcal{M}_{\bf q}|0).
\label{bb5a}
\end{eqnarray}

The gap fluxes $Z^{(0)}$ below the stack, and $ Z^{(m)}$
above, require special attention.  Using (\ref{fs14}), (\ref{fs20}) and (\ref{fs30}) we write (\ref{gi2a})  as
\begin{eqnarray}
|I^{(0)}\}
&=&|U^{[m]}_{\bf d}\}+ |\ddot I^{(\rm in)}_{\bf u}\}\nonumber\\
&=&\mathcal{M}_{\bf d}\left(\sum_{c=1}^m G^{[m c\}}\mathcal{E}^{\{c\}}|0)B^{\{c\}}+W^{[m]}|\ddot I^{(\rm in)}\}\right)+\mathcal{M}_{\bf u}|\ddot I^{(\rm in)}\}\nonumber\\
&=&\sum_{c=1}^m \mathcal{M}_{\bf d} G^{[m c\}}\mathcal{E}^{\{c\}}|0)B^{\{c\}}
+\left(\mathcal{M}_{\bf u}+\mathcal{M}_{\bf d}W^{[m]}\right)|\ddot I^{(\rm in)}\}.
\label{bb6}
\end{eqnarray}
Multiplying the last line of (\ref{bb6}) on the left by $4\pi\lvec 1|$, and using (\ref{hi2}) for $|\ddot I^{(\rm in)}\}$  we write the scalar flux (\ref{gi10}) for the gap $g=0$ as
\begin{eqnarray}
Z^{(0)}
&=&4\pi\sum_{c=1}^m\lvec 1|\mathcal{M}_{\bf d} G^{[mc\}}\mathcal{E}^{\{c\}}|0)B^{\{c\}}+4\pi \sum_{\bf q}\lvec 1| (\mathcal{M}_{\bf u}+\mathcal{M}_{\bf d}W^{[m]})\mathcal{M} _{\bf q}|0)B^{\{\rm in\}}_{\bf q}\nonumber\\
&=&4\pi\sum_{c=1}^{m}\dot M^{(0 c\}}B^{\{c\}}+4\pi\sum_{\bf q=d}^{\bf u}\ddot N^{(0)}_{\bf q}B^{\{\rm in\}}_{\bf q}.
\label{bb8}
\end{eqnarray}
The coupling matrices are
\begin{eqnarray}
\dot M^{(0 c\}}=\lvec 1|\mathcal{M}_{\bf d} G^{[mc\}}\mathcal{E}^{\{c\}}|0)\quad\hbox{and}\quad
\ddot N^{(0)}_{\bf q}= \lvec 1|(\mathcal{M}_{\bf u}+\mathcal{M}_{\bf d}W^{[m]} )\mathcal{M}_{\bf q}|0).
\label{bb8a}
\end{eqnarray}

Using (\ref{fs14}), (\ref{fs20}) and (\ref{fs30})  we write (\ref{gi2c})  as
\begin{eqnarray}
|I^{(m)}\}&=&|U^{[m]}_{\bf u}\}+ |\ddot I^{(\rm in)}_{\bf d}\}\nonumber\\
&=&\mathcal{M}_{\bf u}\left(\sum_{c=1}^m G^{[m c\}}\mathcal{E}^{\{c\}}|0)B^{\{c\}}+W^{[m]}|\ddot I^{(\rm in)}\}\right)+\mathcal{M}_{\bf d}|\ddot I^{(\rm in)}\}\nonumber\\
&=&\sum_{c=1}^m \mathcal{M}_{\bf u} G^{[m c\}}\mathcal{E}^{\{c\}}|0)B^{\{c\}}
+\left(\mathcal{M}_{\bf d}+\mathcal{M}_{\bf u}W^{[m]}\right)|\ddot I^{(\rm in)}\}.
\label{bb10}
\end{eqnarray}
Multiplying the last line of (\ref{bb10}) on the left by $4\pi\lvec 1|$, and using (\ref{hi2}) for $|\ddot I^{(\rm in)}\}$  we write the scalar flux (\ref{gi10})  for the gap $g=m$ as
\begin{eqnarray}
Z^{(m)}
&=&4\pi\sum_{c=1}^m\lvec 1|\mathcal{M}_{\bf u} G^{[mc\}}\mathcal{E}^{\{c\}}|0)B^{\{c\}}+ 4\pi\sum_{\bf q}\lvec 1| (\mathcal{M}_{\bf d}+\mathcal{M}_{\bf u}W^{[m]})\mathcal{M}_{\bf q} |0)B^{\{\rm in\}}_{\bf q}\nonumber\\
&=&4\pi\sum_{c=1}^{m}\dot M^{(m c\}}B^{\{c\}}+4\pi\sum_{\bf q=d}^{\bf u}\ddot N^{(m)}_{\bf q}B^{\{\rm in\}}_{\bf q}.
\label{bb12}
\end{eqnarray}
The coupling matrices are
\begin{eqnarray}
\dot M^{(m c\}}=\lvec 1|\mathcal{M}_{\bf u} G^{[mc\}}\mathcal{E}^{\{c\}}|0)\quad\hbox{and}\quad
\ddot N^{(m)}_{\bf q}=
\lvec 1| (\mathcal{M}_{\bf d}+\mathcal{M}_{\bf u}W^{[m]})\mathcal{M}_{\bf q} |0)
\label{bb12a}
\end{eqnarray}

Using $|B^{\{\rm in\}}\rangle$, defined by (\ref{hi20}), and the $m\times 1$ array of the Planck intensities $B^{\{c\}}$ of individual clouds,
\begin{equation}
|B]=
\left[\begin{array}{l}B^{\{1\}}\\ B^{\{2\}}\\ 
\vdots\\B^{\{m-2\}}\\ B^{\{m-1\}}\\ B^{\{m\}}\end{array}\right],
\label{bb14}
\end{equation}
we can summarize (\ref{bb2}), (\ref{bb4}), (\ref{bb8}) and (\ref{bb12}) with the formally simpler equation, the analog of (\ref{hi24}) for a single cloud,
\begin{equation}
|Z)=4\pi \dot M|B]+4\pi \ddot N|B^{\{\rm in\}}\rangle,
\label{bb18}
\end{equation}
or more explcitly
\begin{equation}
Z^{(g)}=4\pi\sum_{c=1}^{m}\dot M^{(g c\}}B^{\{c\}}+4\pi\sum_{\bf q=d}^{\bf u}\ddot N^{(g)}_{\bf q}B^{\{\rm in\}}_{\bf q}.
\label{bb20}
\end{equation}
We note that matrix $\dot M$ of (\ref{bb18}), with the scalar elements $\dot M^{(g c\}}$ of (\ref{bb20}), is not square but has $m+1$ rows, one for each gap index $g$, and  $m$ columns, one for each cloud index $c$.  
The matrix $\ddot N$ of (\ref{bb18}), with the scalar elements $\ddot N^{(g)}_{\bf q}$ of (\ref{bb20}),  is also not square, but has $m+1$ rows, one for each gap index $g$, but only $2$ column indices, labeled by ${\bf q=d}$ or ${\bf q=u}$, for Planck intensities of the downward and upward parts of the half-isotropic incoming radiation.

\subsection{Radiative heating and cooling of cloud stacks\label{hcs}}
Guided by the first line of (\ref{hc2}), we write the stack vector for net radiative absorption as
\begin{equation}
|R]=\left[\begin{array}{c} R^{\{1\}}\\ R^{\{2\}}\\ R^{\{3\}}\\ 
\vdots\\ R^{\{m\}}\end{array}\right]=-\Delta |Z)=-4\pi\Delta |I_1)=
\left[\begin{array}{c}Z^{\{0\}}-Z^{\{1\}}\\ Z^{\{1\}}-Z^{\{2\}}\\ Z^{\{2\}}-Z^{\{3\}}\\ 
\vdots\\ Z^{\{m-1\}}-Z^{\{m\}}\end{array}\right].
\label{hcs2}
\end{equation}
Here the $m\times (m+1)$ differencing matrix is
\begin{equation}
\Delta=
\left[\begin{array}{rrrrrrrrr} 
-1&1&0&0& \cdots &0& 0&0&0\\
0& -1&1& 0& \cdots & 0&  0& 0&0\\ 
\vdots&\vdots&\vdots&\vdots&\ddots&\vdots&\vdots&\vdots&\vdots\\
 0& 0& 0& 0&\cdots & 0& -1&1& 0\\
 0& 0& 0& 0&\cdots & 0& 0& -1& 1
\end{array}\right].
\label{hcs4}
\end{equation}
Substituting (\ref{bb18}) into (\ref{hcs2}) we find
\begin{equation}
|R]=-4\pi \Delta \dot M|B]-4\pi \Delta \ddot N|B^{\{\rm in\}}\rangle.
\label{hcs6}
\end{equation}
According to (\ref{hcs6}), the net heating rate $|R]$ is determined by the Planck intensities $|B]$ of the  clouds, given by (\ref{bb14}) and by the downward and upward Planck intensities of the incoming radiation $|B^{\{\rm in\}}\rangle$, given by (\ref{hi20}).

An interesting special case of (\ref{hcs6}) is radiative equilibrium, when the net heating and cooling rates of all clouds vanish and $|R]=\breve 0$. Then we can use (\ref{hcs6}) to find the Planck intensities $|B]$ of clouds that are in radiative equilibrium with the Planck intensities $|B^{\{\rm in\}}\}$ of incoming radiation,
\begin{eqnarray}
|B]&=&-(\Delta \dot M)^{-1}\Delta \ddot N|B^{\{\rm in\}}\rangle\nonumber\\
&=&-(\Delta \dot M) \backslash \Delta \ddot N|B^{\{\rm in\}}\rangle.
\label{hcs8}
\end{eqnarray}
  Since the matrix  $\Delta\dot M$ is a relatively small  array of  $m\times m$ numbers, in contrast  to the $2nm\times 2nm$ array
 $(\hat 1-X)$, the advantage of using a backslash numerical computation instead of a matrix inverse computation in (\ref{hcs8}) is much more modest than for (\ref{fs42}) or (\ref{fs44}).
\begin{figure}[t]
%\postscriptscale{stackap5}{1.1}
\includegraphics[height=100mm,width=1\columnwidth]{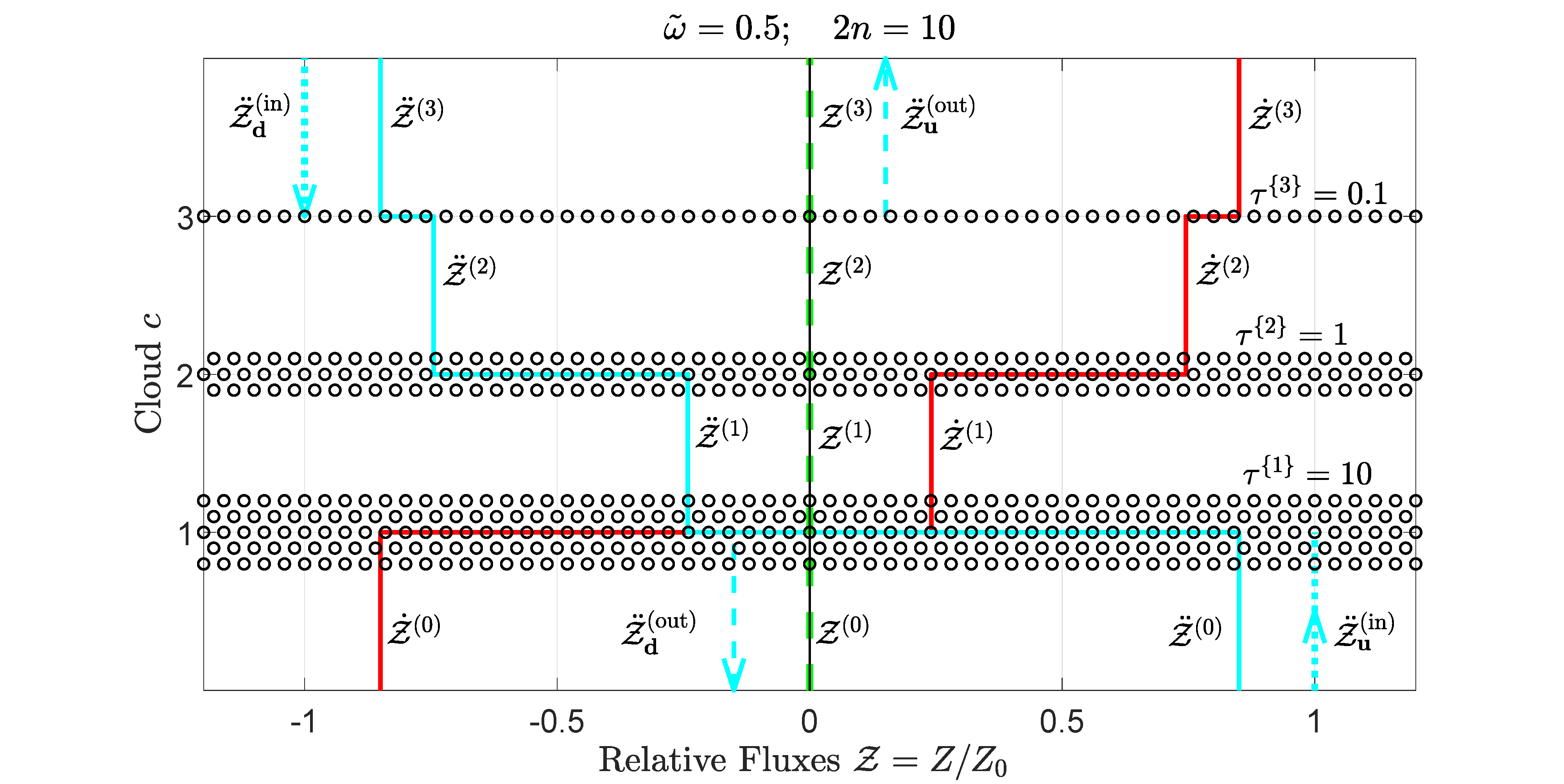}% Here is how to import EPS art
\caption {Three isothermal, Rayleigh scattering clouds. All have the same single-scattering albedo, $\tilde\omega = 0.5$. The optical depths are $\tau^{\{1\}}=10$, $\tau^{\{2\}}=1$ and $\tau^{\{3\}}=0.1$.  The clouds and the incoming radiation have equal temperatures and are in thermal equilibrium. The radiation is modeled with $2n=10$ stream pairs.
The continuous red line shows the relative gap fluxes $\dot{\mathcal{Z}}^{(g)}=\dot Z^{(g)}/Z_0$ of thermal radiation emitted by cloud particulates and gas molecules. The reference flux $Z_0$ was given by (\ref{nex2}). The continuous cyan lines show the relative gap fluxes, $\ddot{\mathcal{Z}}^{(g)}$,  due to scattering of incoming radiation, and the dashed green line is the algebraic sum, $\mathcal{Z}^{(g)}=\dot {\mathcal{Z}}^{(g)}+\ddot {\mathcal{Z}}^{(g)}$, of the gap fluxes  from both sources. See the text for a detailed discussion.}
\label{stackap5}
\end{figure}  
\subsection{3-cloud examples\label{ex}}
\begin{figure}[t]
%\postscriptscale{stackap5u}{1.1}
\includegraphics[height=100mm,width=1\columnwidth]{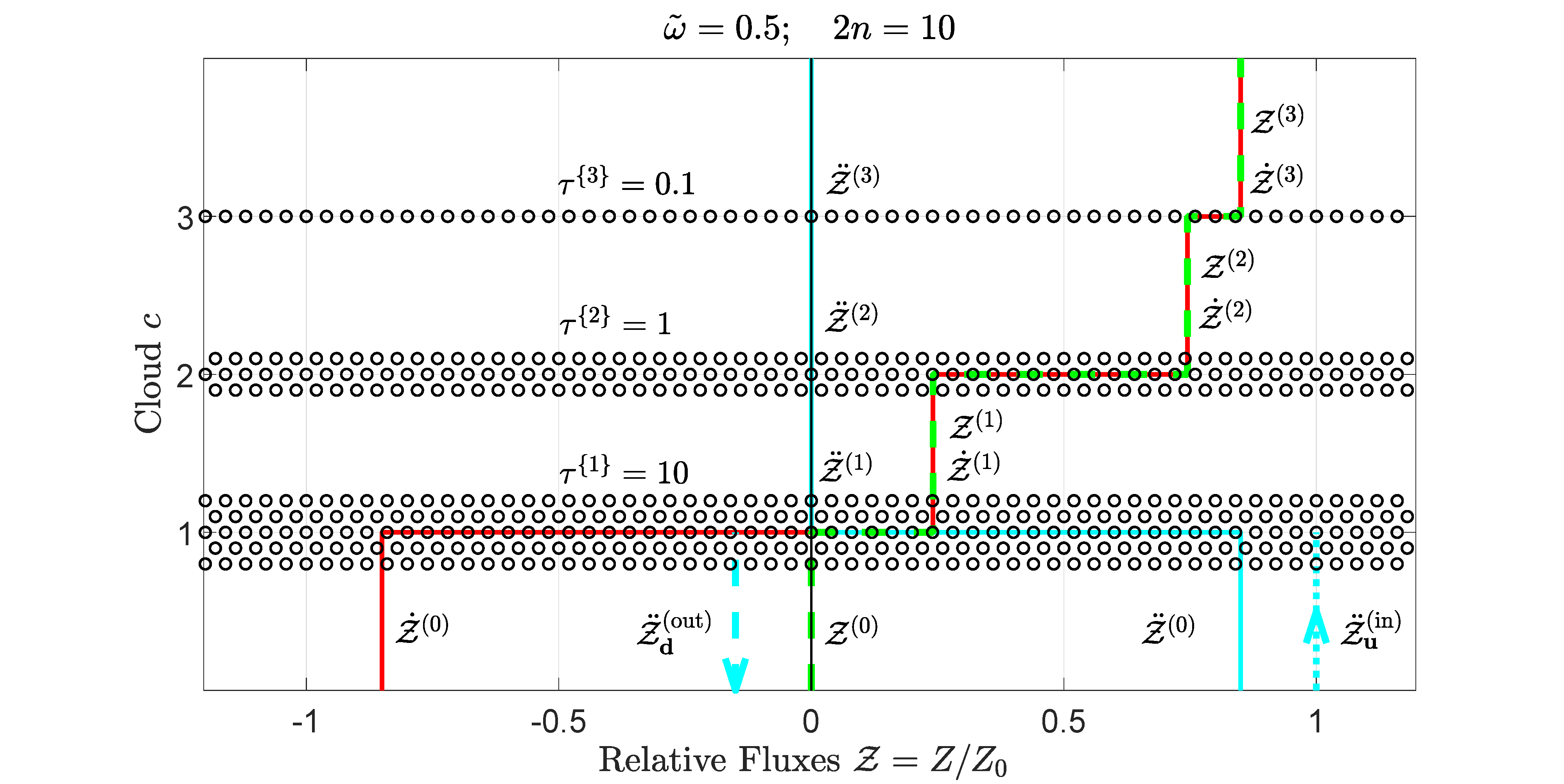}% Here is how to import EPS art
\caption {The same three clouds as for Fig. \ref{stackap5} but with no downward incoming radiation.  A negligible fraction of the upward incoming radiation, incident onto the bottom of the stack and shown as the dotted cyan line, penetrates the bottom cloud, with its large optical depth $\tau^{\{1\}}=10$. Some 15\% is reflected and 85\% is absorbed and converted to heat. Because there is no downward incoming radiation to heat them, all of the clouds are radiatively cooled.  See the text for a more detailed discussion of the figure.}
\label{stackap5u}
\end{figure}

\begin{figure}[t]
%\postscriptscale{stackap5re}{1.1}
\includegraphics[height=100mm,width=1\columnwidth]{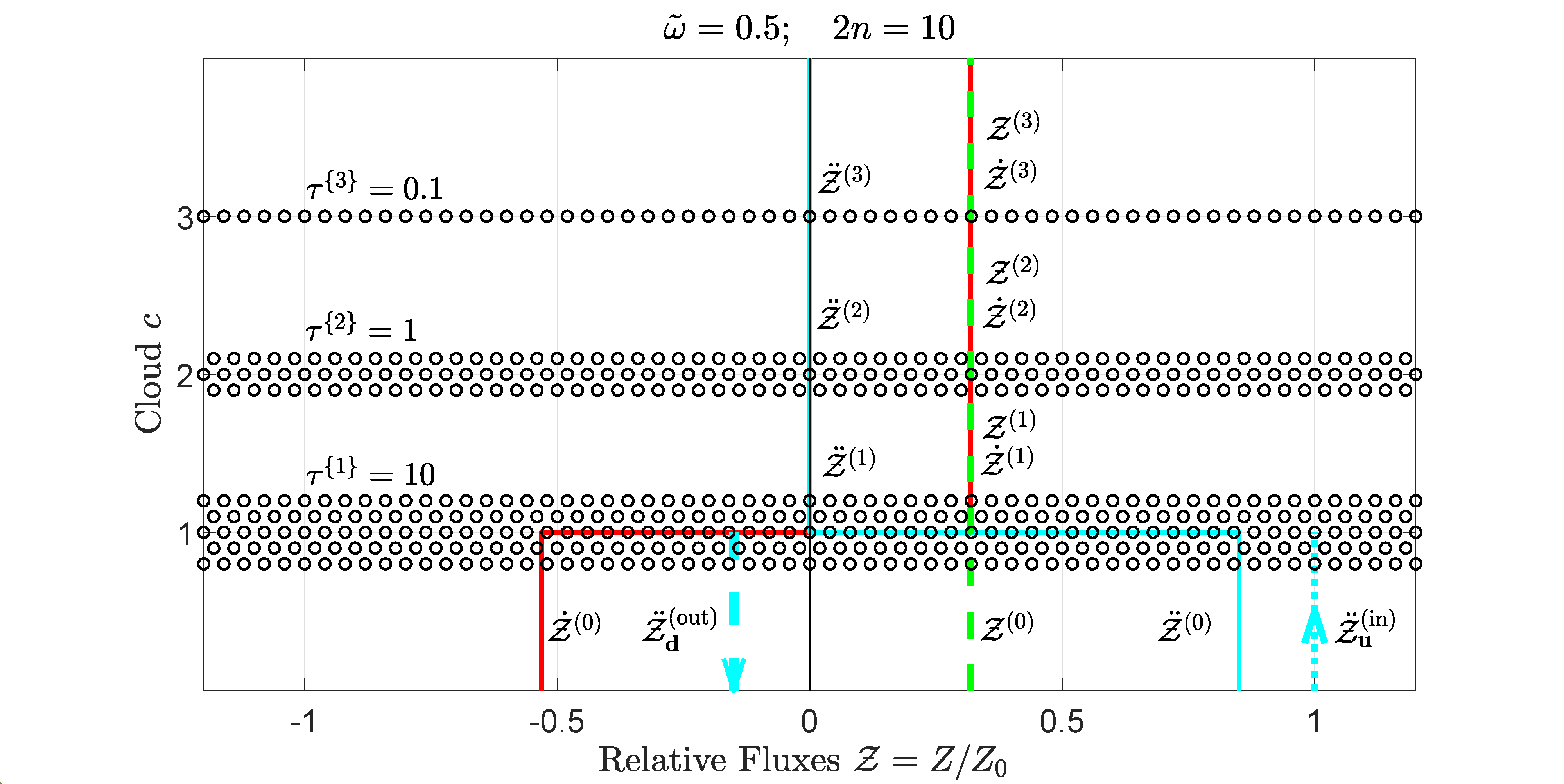}% Here is how to import EPS art
\caption {The fluxes of cloud stack of Fig. \ref{stackap5u} after the clouds have cooled to radiative equilibrium.
The clouds neither heat nor cool, but they are not in thermal equilibrium like those of Fig. \ref{stackap5}, since they have different temperatures, corresponding to the Planck intensities of (\ref{ex24}).  See the text for a more detailed discussion of the figure.}
\label{stackap5re}
\end{figure}

In this section we discuss numerical solutions of (\ref{bb18}) for an $m$-cloud stack that are analogous to those of Section {\bf \ref{nex}} for  a single cloud.   We cite fluxes  in units of the reference flux $Z_0$ and Planck intensities in units of the reference Planck intensity $B_0$ of (\ref{nex2}).
In Figs. \ref{stackap5}, \ref{stackap5u} and \ref{stackap5re} we show 
stacks of $m=3$ isothermal, homogeneous clouds, with optical depths
\begin{equation}
\left[\begin{array}{c} \tau^{\{1\}}\\ \tau^{\{2\}}\\ \tau^{\{3\}}\end{array}\right]=\left[\begin{array}{c} 10\\ 1\\ 0.1\end{array}\right].
\label{ex2}
\end{equation}
As for the examples of Section {\bf \ref{nex}}, the single-scattering albedo is $\tilde \omega = 0.5$, and there is Rayleigh scattering with the phase matrix $\hat p$ of (\ref{et26}). 
Modeling the radiation with $2n = 10$ streams as in Section {\bf\ref{nex}}, we find that the upward
Planck emissivities $\varepsilon^{\{c\}}_{\bf u}$ of  (\ref{ter8}) and the
upward Planck albedos $\omega^{\{c\}}_{\bf u}$ of (\ref{pa10}) of the three clouds, if each were isolated, would be 

\begin{equation}
\left[\begin{array}{c} \varepsilon^{\{1\}}_{\bf u}\\ \varepsilon^{\{2\}}_{\bf u}\\ \varepsilon^{\{3\}}_{\bf u}\end{array}\right]=\left[\begin{array}{c} 0.8499\\ 0.5568 \\ 0.0920 \end{array}\right]\quad\hbox{and}\quad
\left[\begin{array}{c} \omega^{\{1\}}_{\bf u}\\ \omega^{\{2\}}_{\bf u}\\ \omega^{\{3\}}_{\bf u}\end{array}\right]=\left[\begin{array}{c} 0.1501\\ 0.4432 \\ 0.9080 \end{array}\right].
\label{ex4}
\end{equation}
The bottom cloud with optical thickness $\tau^{\{1\}}=10$, has the same, relatively small Planck albedo, 
$\omega^{\{1\}}_{\bf u}=0.1501$, as an infinitely thick cloud with $\tau = \infty$. The Planck albedo of the middle cloud with optical thickness $\tau^{\{2\}}=1$ is larger,  $\omega^{\{2\}}_{\bf u}=0.4432$, the same as (\ref{nex12}).  More incident photons are transmitted through the middle cloud or reflected, rather than being absorbed. The Planck albedo of the top cloud with the smallest optical thickness $\tau^{\{3\}}=0.1$ is largest,  $\omega^{\{3\}}_{\bf u}=0.9080$. Most incident photons are transmitted through the optically thin top cloud and few are reflected. Only a small fraction of photons are absorbed.

If each of the three isothermal clouds were alone in cold space, and each had the reference Planck intensity $B_0$ of (\ref{nex2}),  their cooling rates $ \dot C^{\{c\}}$ of (\ref{h8}), due to unhindered thermal emission to space, would be
\begin{equation}
\left[\begin{array}{c}\dot C^{\{1\}}\\ \dot C^{\{2\}}\\\dot C^{\{3\}}\end{array}\right]=
2\left[\begin{array}{c} \varepsilon^{\{1\}}_{\bf u}\\ \varepsilon^{\{2\}}_{\bf u}\\ \varepsilon^{\{3\}}_{\bf u}\end{array}\right]Z_0=\left[\begin{array}{c}
1.6998\\    1.1136 \\    0.1840 \end{array}\right]Z_0
\label{ex7a}
\end{equation}
Noting that the reference flux $Z_0$ of (\ref{nex2}) is the same as the blackbody flux $\dot Z^{(bb)}$ of (\ref{bc12}) out of the top or bottom of a black cloud with Planck intensity $B_0$,
\begin{equation}
Z_0 =\dot Z^{(bb)},
\label{ex7b}
\end{equation}
we see that the nearly optically thick bottom cloud, with $\tau^{\{1\}}=10$ and  $\dot C^{\{1\}}=1.6998\, Z_0$ comes closest to the maximum cooling rate $\dot C^{\{\rm max\}}=2\, Z_0$, with blackbody fluxes out of the top and bottom of the cloud. Although it is optically thick, the bottom cloud  emits less than a black cloud because we have assumed a non-zero single-scattering albedo $\tilde \omega = 1/2$. As one can see from Fig. \ref{emis2}, this limits the emissivity of an optically thick, Rayleigh-scattering cloud to $\varepsilon_{\bf u} = 0.8499$. The middle cloud, with $\dot C^{\{2\}}=1.1136\, Z_0$, emits less cooling radiation because of its moderate optical depth, $\tau^{\{2\}}=1$. The top cloud, with $\dot C^{\{3\}}=0.1840\, Z_0$, emits the least cooling radiation because of its small optical depth, $\tau^{\{3\}}=0.1$.

We now suppose that the three clouds are stacked as shown in Figs. \ref{stackap5}, \ref{stackap5u} and \ref{stackap5re}.
 Using the optical thicknesses, single-scattering albedos and phase functions mentioned above to calculate scattering matrices $\mathcal{S}^{\{c\}}$ as outlined  in Section {\bf \ref{sm}} and
using these with formulas of Sections {\bf \ref{sa}} and {\bf \ref{fs}} to calculate 
 other necessary factors, we find that the matrix $\dot M$ of (\ref{bb18}) becomes
\begin{equation}
4\pi \dot M=\left[\begin{array}{rrr}-0.8499&-0.0000&-0.0000\\ 0.7481&-0.4871&-0.0199\\
0.2610&0.5609&-0.0772\\0.2411&0.5037&0.1051\end{array}\right]\frac{Z_0}{B_0}.
\label{ex6}
\end{equation}
Similarly, the matrix $\ddot N$ of (\ref{bb18}) becomes
\begin{equation}
4\pi \ddot N=\left[\begin{array}{rr}-0.0000&0.8499\\ -0.2411&0.0000\\
-0.7448&0.0000\\-0.8499&0.0000\end{array}\right]\frac{Z_0}{B_0}.
\label{ex8}
\end{equation}
The differencing matrix (\ref{hcs4}) becomes
\begin{equation}
\Delta=
\left[\begin{array}{rrrrr} 
-1&1&0&&0\\
0& -1&1&& 0\\
0&0& -1&&1\\  
\end{array}\right].
\label{ex10}
\end{equation}
For the example of Fig.  \ref{stackap5}
the clouds and the incoming radiation  all have the same temperature, corresponding to the Planck intensity $B_0$, so  (\ref{bb14}), which describes the Planck intensities of the clouds, becomes 
\begin{equation}
|B]=
\left[\begin{array}{c} B^{\{1\}}\\ B^{\{2\}}\\ B^{\{3\}}\end{array}\right]=\left[\begin{array}{c} 1\\ 1 \\ 1 \end{array}\right]B_0,\
\label{ex12a}
\end{equation}
and  (\ref{hi20}), which describes the Planck intensities of the half-isotropic incoming radiation (\ref{hi2}) becomes 
\begin{equation}
|B^{\{\rm in\}}\rangle=\left[\begin{array}{c} B^{\{\rm in\}}_{\bf d}\\ B^{\{\rm in \}}_{\bf u}\end{array}\right]=
 \left[\begin{array}{c} 1\\ 1\end{array}\right]B_0
\label{ex12b}
\end{equation}
The dotted cyan lines on the bottom and top of  Fig. \ref{stackap5}  indicate the upward and downward half-isotropic  incoming fluxes  $\ddot {\mathcal{Z}}^{(\rm in)}_{\bf u}=Z^{(\rm in)}_{\bf u}/Z_0$ and $\ddot {\mathcal{Z}}^{(\rm in)}_{\bf d}=Z^{(\rm in)}_{\bf d}/Z_0$  that follow from (\ref{hi6a}) and (\ref{hi6b}).
The dashed cyan lines are the reflected and transmitted incoming fluxes  $\ddot {\mathcal{Z}}^{(\rm out)}_{\bf u}=Z^{(\rm out)}_{\bf u}/Z_0$ and $\ddot {\mathcal{Z}}^{(\rm out)}_{\bf d}=Z^{(\rm out)}_{\bf d}/Z_0$  that follow from (\ref{hi10a}) and (\ref{hi10b}).
The continuous cyan line is the net of upward and downward fluxes $\ddot{\mathcal{Z}}^{(g)}=
(\ddot Z^{(g)}_{\bf d}+\ddot Z ^{(g)}_{\bf u})/Z_0$.   

The continuous red line of  Fig. \ref{stackap5} shows the  gap fluxes $\dot{\mathcal{Z}}^{(g)}= \dot Z^{(g)}_1/Z_0$  generated by thermal emission of cloud particulates and gas molecules.
For internal gaps with $g=1,2,3,\ldots,m-1$,  the scalar flux $\dot Z^{(g)}$ is given by (\ref{bb2}).
Below the stack,  $\dot Z^{(0)}$ is given by the sum on $c$ in (\ref{bb8}). Above the stack,  $Z^{(m)}$ is given by the sum on $c$ in (\ref{bb12}). 

The total gap  fluxes $\mathcal{Z}^{(g)}=(Z^{(g)}_{\bf u}+Z^{(g)}_{\bf d})/Z_0$ in the cloud stack are shown as 
the dashed green line.  $\mathcal{Z}^{(g)}=\dot{\mathcal{Z}}^{(g)}+\ddot{\mathcal{Z}}^{(g)}$ is  the algebraic sum of the continuous red and cyan lines.  For the situation illustrated by Fig.  \ref{stackap5}, with the clouds and the incoming radiation at the same temperature, the cloud stack is in both thermal and radiative equilibrium. The total flux, indicated by the dashed green line, is zero above and below the stack. It is also zero for all of the internal gaps. So 

\begin{equation}
\left[\begin{array}{c} Z^{(0)}\\ Z^{(1)}\\ Z^{(2)}\\ Z^{(3)}\end{array}\right]=\left[\begin{array}{c} \dot Z^{(0)}+ \ddot Z^{(0)}\\
 \dot Z^{(1)}+ \ddot Z^{(1)}\\ \dot Z^{(2)}+ \ddot Z^{(2)}\\ \dot Z^{(3)}+ \ddot Z^{(3)}\\ \end{array}\right]
 =\left[\begin{array}{c} 0\\ 0\\0\\ 0\end{array}\right]Z_0.
\label{ex14}
\end{equation}
In view of (\ref{ex14}) and (\ref{hcs2}) we see that there is neither radiative heating nor cooling of the clouds,
\begin{equation}
\left[\begin{array}{c} R^{\{1\}}\\ R^{\{2\}}\\ R^{\{3\}}\end{array}\right]=
\left[\begin{array}{c}Z^{(0)}-Z^{(1)}\\ Z^{(1)}-Z^{(2)}\\ Z^{(2)}-Z^{(3)}\end{array}\right]
=\left[\begin{array}{c} 0\\ 0\\ 0\end{array}\right]Z_0.
\label{ex16}
\end{equation}

Because the clouds have the same single-scattering albedo, $\tilde \omega = 0.5$, and the same Rayleigh scattering-phase matrix $\hat p$ of (\ref{et26}),  the equivalent cloud is like a single, homogeneous Rayleigh-scattering cloud with an optical thickness of $\tau = \tau^{\{1\}}+\tau^{\{2\}}+\tau^{\{3\}}=11.1$.  The scattering matrix of the cloud stack is
\begin{eqnarray}
\mathcal{S}&=&W^{[3]}=\mathcal{S}^{\{3\}}\rstar\mathcal{S}^{\{2\}}\rstar\mathcal{S}^{\{1\}}
=\left[\begin{array}{ll}\mathcal{S}_{\bf dd}&\mathcal{S}_{\bf du}\\\mathcal{S}_{\bf ud}&\mathcal{S}_{\bf uu}\end{array}\right]\nonumber\\
&\approx&\left[\begin{array}{ll}\breve 0&\mathcal{S}_{\bf du}\\\mathcal{S}_{\bf ud}&\breve 0\end{array}\right]
\label{ex17}
\end{eqnarray}
Because of its large optical depth, $\tau=11.1$,
the scattering matrix of the cloud stack hardly differs from that of an infinitely thick cloud with $\tau = \infty$. Optically thick clouds have no upward or downward transmission through them, and therefore have the block matrix elements $\mathcal{S}_{\bf dd}=\mathcal{S}_{\bf dd}=\breve 0$.
There is about 15\% reflection from the top and bottom of the stack as described by the non-zero block elements $\mathcal{S}_{\bf du}$ and $\mathcal{S}_{\bf ud}$.

Fig. \ref{stackap5u} shows the same stack of clouds as for Fig. \ref{stackap5}, but with no downward  incoming  radiation. The Planck intensity (\ref{ex12b}) changes to
\begin{equation}
|B^{\{\rm in\}}\rangle=\left[\begin{array}{c} B^{\{\rm in\}}_{\bf d}\\ B^{\{\rm in \}}_{\bf u}\end{array}\right]=
 \left[\begin{array}{c} 0\\ 1\end{array}\right]B_0,
\label{ex18}
\end{equation}
but the Planck intensity $|B]$ of (\ref{ex12a}) remains the same.
Solving (\ref{bb18}) with the Planck intensities (\ref{ex18}) and (\ref{ex12a}) gives the scalar fluxes
\begin{equation}
\left[\begin{array}{c}Z^{(0)}\\ Z^{(1)}\\ Z^{(2)}\\ Z^{(3)}\end{array}\right]=\left[\begin{array}{c}1.54\times 10^{-5}\\ 0.2411\\ 0.7448\\ 0.8499\end{array}\right]Z_0.
\label{ex20}
\end{equation}
Since the bottom cloud is so optically thick, $\tau^{\{1\}}=10$, and has the same Planck intensity as the flux coming up from below, $B^{\{1\}}=B^{\{\rm in\}}_{\bf u}=B_0$,
the scalar flux below the stack almost vanishes.
Upward incoming radiation  onto the bottom of the stack is almost balanced by thermally emitted and reflected downward radiation. 

The continuous red line of Fig. \ref{stackap5u} shows the gap fluxes $\dot{\mathcal{Z}}^{(g)}$ due to thermal emission of cloud particulates and gas molecules.  Above the bottom cloud, in the gaps $g>1$,  the thermal flux $\dot {\mathcal{Z}}^{(g)}$ is nearly equal to the total flux $ \mathcal{Z}^{(g)}$, as one can see from the near coincidence of the continuous red line and the dashed green line. The nearly optically thick bottom cloud attenuates the incoming flux $\ddot{\mathcal{Z}}^{(in)}_{\bf u}$ from below, shown as the dotted cyan line, to negligibly small values in the gaps $g>1$ above the bottom cloud.

There is radiative cooling of all three clouds of Fig.  \ref{stackap5u}. The numerical values,
\begin{equation}
\left[\begin{array}{c} R^{\{1\}}\\ R^{\{2\}}\\ R^{\{3\}}\end{array}\right]=
\left[\begin{array}{c}Z^{(0)}-Z^{(1)}\\ Z^{(1)}-Z^{(2)}\\ Z^{(2)}-Z^{(3)}\end{array}\right]
=-\left[\begin{array}{c} 0.2411\\ 0.5037\\ 0.1051\end{array}\right]Z_0,
\label{ex22}
\end{equation}
are the lengths of the horizontal segments of the dashed green lines of  Fig. \ref{stackap5u}. 
It is interesting to compare the cooling rates $-R^{\{c\}}$ of (\ref{ex22}) of the individual clouds in the stack of Fig. \ref{stackap5u} with the cooling rates $\dot C^{\{c\}}$ of (\ref{ex7a}) for the  clouds if they had the same temperatures but were isolated and could radiate freely to empty space.
The cooling rate $-R^{\{1\}}=0.2411\,Z_0$ of the bottom cloud in the stack of Fig. \ref{stackap5u} is only 14\% of the isolated-cloud cooling rate $\dot C^{\{1\}}=1.6998 \,Z_0$ of (\ref{ex7a}). The reason is the heating of the bottom cloud of Fig. \ref{stackap5u} by incoming radiation from below, and the heating from above by thermal radiation and reflected radiation of the two higher clouds. Because of heating from above and below, the cooling rate $-R^{\{2\}}=0.5037\,Z_0$ of the middle  cloud in the stack of Fig. \ref{stackap5u} is 45\% of   the isolated-cloud cooling rate $\dot C^{\{2\}}=1.1136 \,Z_0$ of (\ref{ex7a}). Because of heating from below, the cooling rate $-R^{\{3\}}=0.1051\,Z_0$ of the top  cloud in the stack of Fig. \ref{stackap5u} is  57\% of   the isolated-cloud cooling rate $\dot C^{\{3\}}=0.1840 \,Z_0$ of (\ref{ex7a}).

Without some non-radiative source of energy, for example the vertical convection that often characterizes the daytime atmosphere of the Earth, or the ultraviolet solar heating of the stratosphere, the temperatures of the clouds of Fig. \ref{stackap5u} would decrease until the radiative cooling from emission is exactly balanced by heating from absorption.   Fig. \ref{stackap5re} shows what happens if the clouds of  Fig. \ref{stackap5u} cool to radiative equilibrium.  Then  (\ref{hcs8}) implies that the Planck intensities of the clouds will be
\begin{equation}
|B]=
\left[\begin{array}{c} B^{\{1\}}\\ B^{\{2\}}\\ B^{\{3\}}\end{array}\right]=\left[\begin{array}{c} 0.6251\\ 0.2994\\ 0.1623\end{array}\right]B_0.
\label{ex24}
\end{equation}
From inspection of (\ref{ex24}) we see that the top cloud cools the most, with the Planck intensity dropping to $B^{\{3\}}= 0.1623\, B_0$. Very little incoming radiation gets through the bottom two clouds to keep  the top cloud warm.  The bottom cloud, which is heated directly by the incoming radiation, cools the least, and maintains a relatively large Planck intensity $B^{\{1\}}= 0.6251\, B_0$. 

\begin{figure}[t]
%\postscriptscale{ten1}{1}
\includegraphics[height=100mm,width=1\columnwidth]{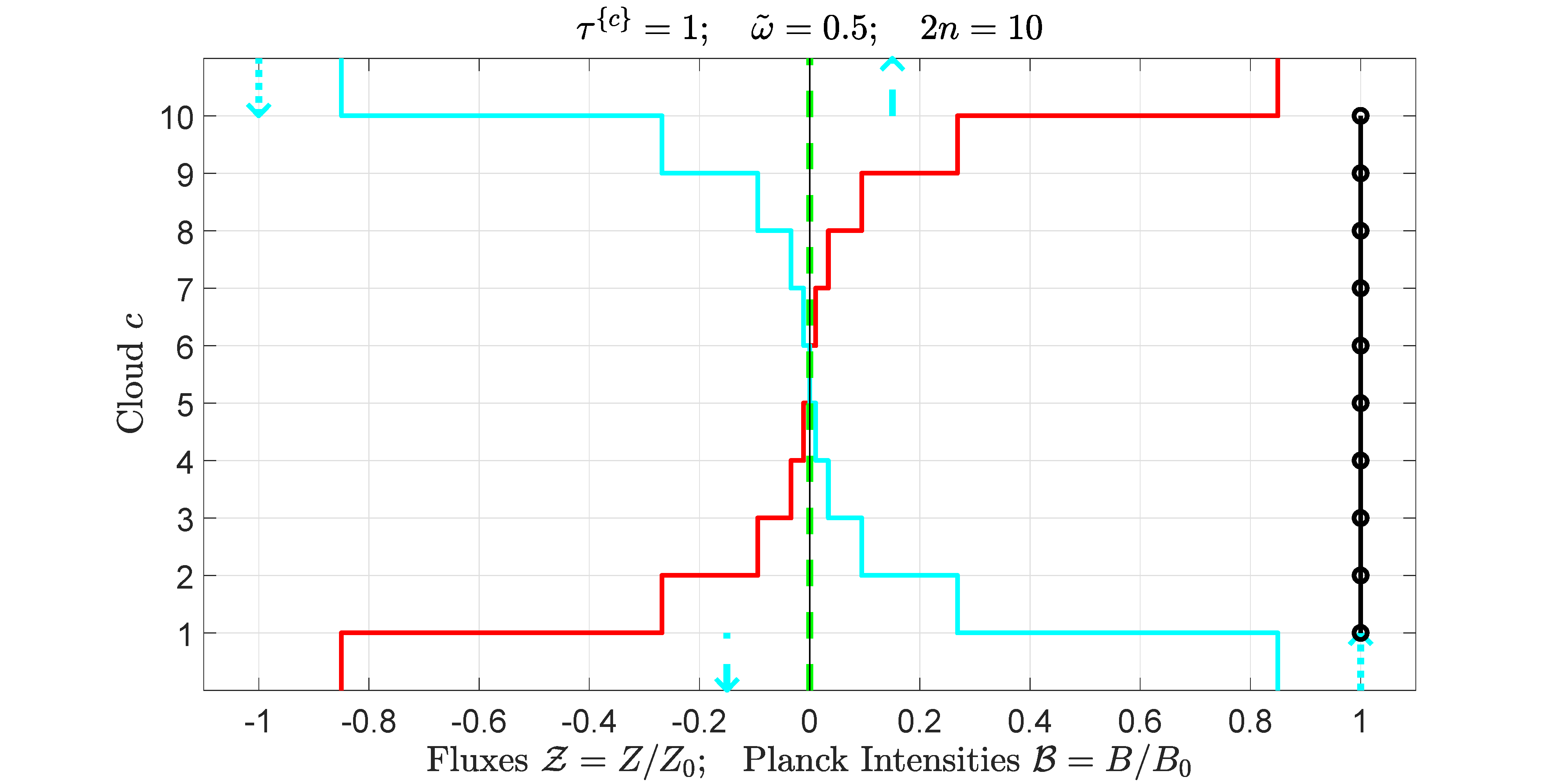}% Here is how to import EPS art
\caption{ A  stack of ten isothermal clouds in thermal equilibrium with incoming radiation. Each cloud has the same optical depth, $\tau^{\{c\}}=1$. The color coding is the same as for Fig. \ref{stackap5}; the continuous red lines are relative flux produced by thermal emission of cloud particulates and gas molecules, the continuous cyan lines are the relative flux due to incoming radiation, and the dashed green line is the net flux. The small black circles give the relative Planck intensities, $\mathcal{B}^{\{c\}}$, of the clouds. 
For this example where all the clouds are assumed to have the same temperature as the incoming radiation, $\mathcal{B}^{\{c\}}=1$.  See the text for a more detailed discussion of the figure.}
\label{ten1}
\end{figure}
Substituting (\ref{ex24}) into (\ref{bb18}) we find
\begin{equation}
\left[\begin{array}{c}Z^{(0)}\\ Z^{(1)}\\ Z^{(2)}\\ Z^{(3)}\end{array}\right]=\left[\begin{array}{c}0.3186\\ 0.3186\\ 0.3186\\ 0.3186\end{array}\right]Z_0.
\label{ex26}
\end{equation}
The gap fluxes $Z^{(\rm g)}$ are identical for radiative equilibrium, and
the net absorption $R^{\{c\}}$ of each cloud vanishes. 
\begin{equation}
\left[\begin{array}{c} R^{\{1\}}\\ R^{\{2\}}\\ R^{\{3\}}\end{array}\right]=
\left[\begin{array}{c}Z^{(0)}-Z^{(1)}\\ Z^{(1)}-Z^{(2)}\\ Z^{(2)}-Z^{(3)}\end{array}\right]
=\left[\begin{array}{c} 0\\ 0\\ 0\end{array}\right]Z_0.
\label{ex28}
\end{equation}
\subsection{10-cloud examples\label{ten}}
\begin{figure}[t]
%\postscriptscale{ten2}{1}
\includegraphics[height=90mm,width=1\columnwidth]{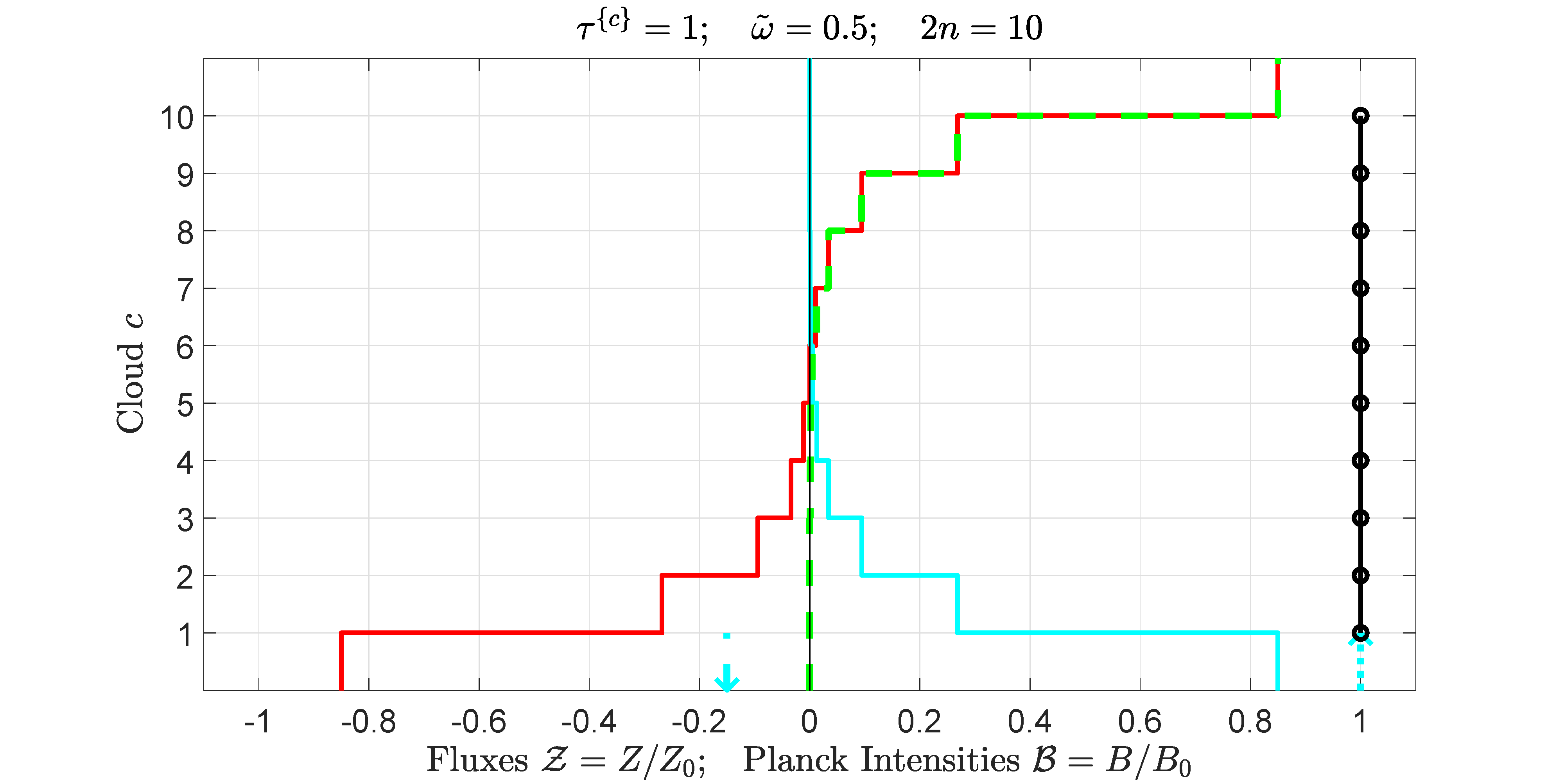}% Here is how to import EPS art
\caption{ Like Fig. \ref{ten1} but with the downward incoming radiation removed. Above the fifth cloud, little remains of the incoming radiation incident on the cloud bottom, and denoted by the continuous cyan line.  The radiation in the higher gaps is almost all thermally emitted by cloud particulates and gas molecules, denoted by the continuous red line. The top few clouds have large radiative cooling rates which are equal to the horizontal discontinuity of the net flux, denoted by  the dashed green line.  The bottom few clouds are nearly in thermal equilibrium with the incoming radiation incident from above and below, so the net flux in the lowest gaps, and below the cloud, is nearly zero. The cooling rates of the lower clouds, the horizontal discontinuities of the dashed green line, are also very small compared to the cooling rate of the top cloud. See the text for a more detailed discussion of the figure.}
\label{ten2}
\end{figure}
\begin{figure}[t]
%\postscriptscale{ten3}{1}
\includegraphics[height=100mm,width=1\columnwidth]{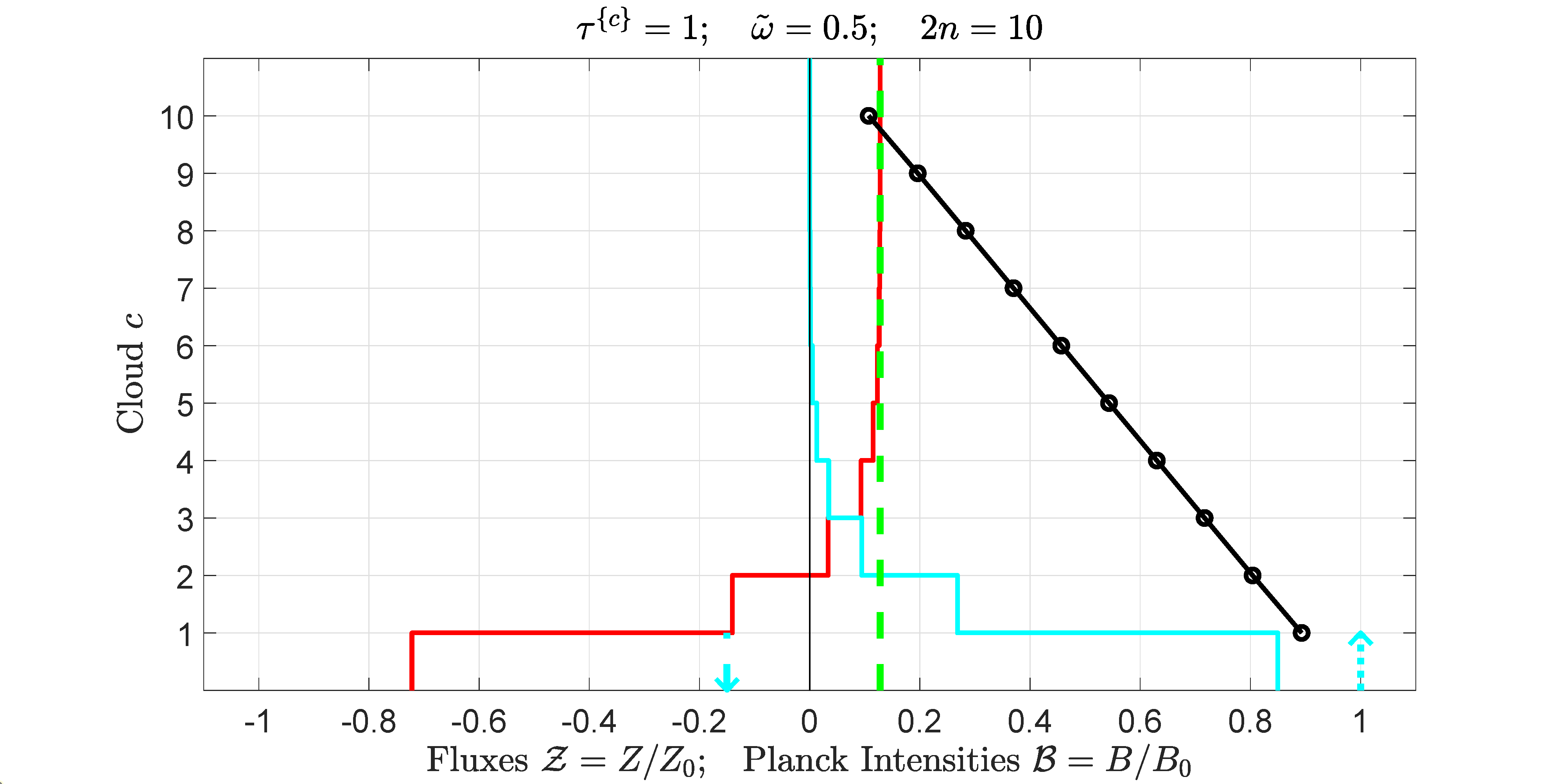}% Here is how to import EPS art
\caption{ Like Fig. \ref{ten2} but the ten clouds have cooled to radiative equilibrium. The relative  Planck intensities $\mathcal{B}^{\{c\}}$ decrease nearly linearly from the bottom cloud to the top cloud.  Just below the bottom cloud, the net flux is the difference between upward incoming radiation, indicated by the dotted cyan line, reflected incoming radiation, indicated by the dashed cyan line, and downward thermal radiation, indicated by the red continuous line. The radiation to space from the top cloud is almost all thermally generated since so little incoming radiation (the continuous cyan line) is transmitted from below through the optically thick stack, with $\tau_m =10$.  See the text for a more detailed discussion of the figure. \label{ten3}}
\end{figure}
Some 
final examples of radiation transfer in cloud stacks are shown in Fig. \ref{ten1} to Fig. \ref{ten3}. To minimize clutter, the small circles used to indicate cloud layers in earlier figures have been omitted from the figures.  The stacks are composed  of $m=10$ Rayleigh-scattering clouds, with single-scattering albedos $\tilde\omega = 0.5$ as in the previous examples, but with each cloud in the stack having the same unit optical depth, $\tau^{\{c\}} =1$.  

Fig. \ref{ten1} shows full thermal equilibrium. Each cloud has a Planck intensity $B^{\{c\}}$ equal to the reference intensity $B_0$ of (\ref{nex2}),
 $B^{\{c\}}=B_0$. The upward and downward parts of the the half-isotropic incoming radiation are also both equal to the reference intensity, $B^{\{\rm in\}}_{\bf u} =B^{\{\rm in\}}_{\bf d}=B_0$.  Most of the incoming radiation, shown as the continuous cyan lines, is attenuated by the middle of the stack, between clouds $c=5$ and $c=6$. In accordance with thermal equilibrium, the total flux, shown as the dashed green line, is zero in all the gaps and above or below the stack. The flux out of the top or bottom of the stack is about 85\% from thermal emission, the continuous red lines, and about 15\% from reflection, the dashed cyan lines.  The small black circles are the relative Planck intensities $\mathcal{B}^{\{c\}}=B^{\{c\}}/B_0$ of the 10 clouds.

Fig. \ref{ten2} shows what happens to the situation of Fig. \ref{ten1} if the incoming radiation onto the top of the stack is removed.  Below the bottom of the cloud and in the lower gaps there is still almost complete mutual cancellation of negative flux due to thermal emission, shown as the continuous red line, and positive flux due to the transmission of upward incoming radiation, and  shown as the continous cyan line. For gaps above the middle of the stack, the cyan incoming radiation from the bottom has been attenuated to negligible values, but the flux from thermal emission of cloud particulates and gas molecules gets larger, as there is less and less cancellation of upward and downward thermal flux in the gaps.  The flux emerging from the top of the stack is about  85\% (the Planck emissivity $\varepsilon_{\bf u}$) of the flux from a blackbody of the same temperature as the cloud stack.  

The horizontal discontinuities of the green dashed  line,  representing the total flux, give the net cloud cooling rates $-R^{\{c\}}$. The cooling maximizes for the top cloud, $c=10$, which thermally radiates to cold space above, with no downward back radiation to compensate for the losses. The cooling rate of the  next highest cloud, $c=9$, is much smaller than that of the top cloud, because of heating by back radiation from the top cloud. The cooling rates of lower clouds are progressively smaller and almost negligible for $c\le 5$.

Fig. \ref{ten3} shows what happens when the clouds of Fig. \ref{ten2} are allowed to cool to radiative equilibrium, where the thermal emissive cooling rate of each cloud is exactly counterbalanced by absorption of radiation coming in from above and below. The relative Planck  intensities, $\mathcal{B}^{\{c\}}=B^{\{c\}}/B_0$, decrease very nearly linearly from the bottom to the top of the stack. 
\section{A Double Cloud\label{dc}} 
In this final section, we show that the fundamental equation (\ref{fs4}) for the unknown  intensities $|U]$ can be written in closed form for $m=2$.  That expression permits a recursive determination of the equivalent scattering matrix $\mathcal{S}^{\{\rm ev\}}$ and thermal source vector $|\dot J^{\{\rm ev\}}\}$ for a stack of $m>2$ clouds,  for which the scattering matrices $\mathcal{S}^{\{c\}}$ and thermal source vectors  $|\dot J^{\{c\}}\}$ for each cloud $c$ of the stack are known.
\subsection{The matrix $(\hat 1-X)^{-1}$ for $m=2$ \label{sx}}
The inverse matrix, $(\hat 1-X)^{-1}$, needed to evaluate (\ref{fs8}) and (\ref{fs22}), can be written as the geometric series
\begin{equation}
(\hat 1-X)^{-1}=\sum_{k=0}^{\infty} X^k =\hat 1+X+X^2+X^3+\cdots.
\label{svm16}
\end{equation}
For $m=2$ we will evaluate $(\hat 1-X)^{-1}$ by summing the series (\ref{svm16}) in closed form.
$X$ becomes the $4\times 4$ block matrix (\ref{sa16}).
We write the zeroth power of $X$ as  an  identity matrix
\begin{equation}
X^0=\hat 1 =\left[\begin{array}{llll} 
\mathcal{M}_{\bf d}&\breve 0 &\breve 0&\breve 0\\ 
\breve 0&\mathcal{M}_{\bf u }&\breve 0&\breve0\\
\breve 0&\breve 0&\mathcal{M}_{\bf d}& \breve 0\\ 
\breve 0&\breve 0&\breve 0&\mathcal{M}_{\bf u}\\
\end{array}\right].
\label{gm30}
\end{equation}
Squaring (\ref{sa16}) we find
\begin{equation}
X^2 =
\left[\begin{array}{llll} 
\mathcal{S}^{\{2\}}_{\bf d u}\mathcal{S}^{\{1\}}_{\bf u d}&\breve 0&\breve 0 &\breve 0\\ 
\breve 0&\mathcal{S}^{\{1\}}_{\bf u d}\mathcal{S}^{\{2\}}_{\bf d u}&\breve0&\breve 0\\
\breve 0& \mathcal{S}^{\{1\}}_{\bf dd}\mathcal{S}^{\{2\}}_{\bf d u}&\breve 0&\breve 0\\ 
\mathcal{S}^{\{2\}}_{\bf u u}\mathcal{S}^{\{1\}}_{\bf u d}&\breve 0&\breve 0&\breve 0\\
\end{array}\right].
\label{gm34}
\end{equation}
Cubing  (\ref{sa16}) we find
\begin{equation}
X^3 =
\left[\begin{array}{llll} 
\breve 0&\mathcal{S}^{\{2\}}_{\bf d u}\mathcal{S}^{\{1\}}_{\bf u d}\mathcal{S}^{\{2\}}_{\bf d u}&\breve 0&\breve 0 \\ 
\mathcal{S}^{\{1\}}_{\bf u d}\mathcal{S}^{\{2\}}_{\bf d u}\mathcal{S}^{\{1\}}_{\bf u d}&\breve0&\breve 0&\breve 0\\
\mathcal{S}^{\{1\}}_{\bf dd}\mathcal{S}^{\{2\}}_{\bf d u}\mathcal{S}^{\{1\}}_{\bf u d}&\breve 0&\breve 0 &\breve 0\\ 
\breve 0&\mathcal{S}^{\{2\}}_{\bf u u}\mathcal{S}^{\{1\}}_{\bf u d}\mathcal{S}^{\{2\}}_{\bf d u}&\breve0&\breve 0\\
\end{array}\right].
\label{gm36}
\end{equation}
The fourth power of (\ref{sa16}) is 
\begin{equation}
X^4 =
\left[\begin{array}{llll} 
\mathcal{S}^{\{2\}}_{\bf d u}\mathcal{S}^{\{1\}}_{\bf u d}\mathcal{S}^{\{2\}}_{\bf d u}\mathcal{S}^{\{1\}}_{\bf u d}&\breve0&\breve 0&\breve 0\\
\breve 0&\mathcal{S}^{\{1\}}_{\bf u d}\mathcal{S}^{\{2\}}_{\bf d u}\mathcal{S}^{\{1\}}_{\bf u d}\mathcal{S}^{\{2\}}_{\bf d u}&\breve 0&\breve 0 \\ 
\breve 0&\mathcal{S}^{\{1\}}_{\bf d d}\mathcal{S}^{\{2\}}_{\bf d u}\mathcal{S}^{\{1\}}_{\bf u d}\mathcal{S}^{\{2\}}_{\bf d u}&\breve 0&\breve 0 \\ 
\mathcal{S}^{\{2\}}_{\bf u u}\mathcal{S}^{\{1\}}_{\bf u d}\mathcal{S}^{\{2\}}_{\bf d u}\mathcal{S}^{\{1\}}_{\bf u d}&\breve0&\breve 0&\breve 0\\
\end{array}\right].
\label{gm38}
\end{equation}
The fifth power of (\ref{sa16}) is 
\begin{equation}
X^5 =
\left[\begin{array}{llll} 
\breve 0&\mathcal{S}^{\{2\}}_{\bf d u}\mathcal{S}^{\{1\}}_{\bf u d}\mathcal{S}^{\{2\}}_{\bf d u}\mathcal{S}^{\{1\}}_{\bf u d}\mathcal{S}^{\{2\}}_{\bf d u}&\breve 0&\breve 0 \\ 
\mathcal{S}^{\{1\}}_{\bf u d}\mathcal{S}^{\{2\}}_{\bf d u}\mathcal{S}^{\{1\}}_{\bf u d}\mathcal{S}^{\{2\}}_{\bf d u}\mathcal{S}^{\{1\}}_{\bf u d}&\breve0&\breve 0&\breve 0\\
\mathcal{S}^{\{1\}}_{\bf d d}\mathcal{S}^{\{2\}}_{\bf d u}\mathcal{S}^{\{1\}}_{\bf u d}\mathcal{S}^{\{2\}}_{\bf d u}\mathcal{S}^{\{1\}}_{\bf u d}&\breve0&\breve 0&\breve 0\\
\breve 0&\mathcal{S}^{\{2\}}_{\bf u u}\mathcal{S}^{\{1\}}_{\bf u d}\mathcal{S}^{\{2\}}_{\bf d u}\mathcal{S}^{\{1\}}_{\bf u d}\mathcal{S}^{\{2\}}_{\bf d u}&\breve 0&\breve 0 \\ 
\end{array}\right].
\label{gm40}
\end{equation}
The sixth power of (\ref{sa16}) is 
\begin{equation}
X^6 =
\left[\begin{array}{llll}
\mathcal{S}^{\{2\}}_{\bf d u}\mathcal{S}^{\{1\}}_{\bf u d}\mathcal{S}^{\{2\}}_{\bf d u}\mathcal{S}^{\{1\}}_{\bf u d}\mathcal{S}^{\{2\}}_{\bf d u}\mathcal{S}^{\{1\}}_{\bf u d}&\breve0&\breve 0&\breve 0\\ 
\breve 0&\mathcal{S}^{\{1\}}_{\bf u d}\mathcal{S}^{\{2\}}_{\bf d u}\mathcal{S}^{\{1\}}_{\bf u d}\mathcal{S}^{\{2\}}_{\bf d u}\mathcal{S}^{\{1\}}_{\bf u d}\mathcal{S}^{\{2\}}_{\bf d u}&\breve 0&\breve 0 \\ 
\breve 0&\mathcal{S}^{\{1\}}_{\bf d d}\mathcal{S}^{\{2\}}_{\bf d u}\mathcal{S}^{\{1\}}_{\bf u d}\mathcal{S}^{\{2\}}_{\bf d u}\mathcal{S}^{\{1\}}_{\bf u d}\mathcal{S}^{\{2\}}_{\bf d u}&\breve 0&\breve 0 \\ 
\mathcal{S}^{\{2\}}_{\bf u u}\mathcal{S}^{\{1\}}_{\bf u d}\mathcal{S}^{\{2\}}_{\bf d u}\mathcal{S}^{\{1\}}_{\bf u d}\mathcal{S}^{\{2\}}_{\bf d u}\mathcal{S}^{\{1\}}_{\bf u d}&\breve0&
\breve 0&\breve 0\\ \end{array}\right].
\label{gm42}
\end{equation}
From inspection of (\ref{sa16}) and (\ref{gm30}) -- (\ref{gm42}) we see that the sum (\ref{svm16}) becomes
\begin{equation}
\sum_{k=0}^{\infty}X^k=(\hat 1-X)^{-1} =
\left[\begin{array}{rrrr}
\mathcal{Q}^{\{21\}}_{\bf d}&\mathcal{S}^{\{2\}}_{\bf d u}\mathcal{Q}^{\{12\}}_{\bf u}&\breve 0&\breve 0 \\\mathcal{S}^{\{1\}}_{\bf u d}
\mathcal{Q}^{\{21\}}_{\bf d}&\mathcal{Q}^{\{12\}}_{\bf u} &\breve 0&\breve 0 \\ 
\mathcal{S}^{\{1\}}_{\bf d d}\mathcal{Q}^{\{21\}}_{\bf d}&\mathcal{S}^{\{1\}}_{\bf d d}\mathcal{S}^{\{2\}}_{\bf d u}\mathcal{Q}^{\{12\}}_{\bf u}&\mathcal{M}_{\bf d}&\breve 0 \\ 
\mathcal{S}^{\{2\}}_{\bf u u}\mathcal{S}^{\{1\}}_{\bf u d}\mathcal{Q}^{\{21\}}_{\bf d}& \mathcal{S}^{\{2\}}_{\bf uu}\mathcal{Q}^{\{12\}}_{\bf u}  &\breve 0&\mathcal{M}_{\bf u}\\ 
\end{array}\right].
\label{gm44}
\end{equation}
The reverberation matrices that occur in (\ref{gm44}),
\begin{equation}
 \mathcal{Q}_{\bf d}^{\{21\}} =\sum_{q=0}^{\infty}\left(\mathcal{S}^{\{2\}}_{\bf d u} \mathcal{S}^{\{1\}}_{\bf ud}\right)^q
 =\left (\mathcal{M}_{\bf d}-\mathcal{S}^{\{2\}}_{\bf d u} \mathcal{S}^{\{1\}}_{\bf ud}\right)^{-1},
\label{gm46}
\end{equation}
and
\begin{equation}
 \mathcal{Q}_{\bf u}^{\{12\}} =\sum_{q=0}^{\infty}\left(\mathcal{S}^{\{1\}}_{\bf  u d} \mathcal{S}^{\{2\}}_{\bf du}\right)^q
 =\left (\mathcal{M}_{\bf u}-\mathcal{S}^{\{1\}}_{\bf  u d} \mathcal{S}^{\{2\}}_{\bf du}\right)^{-1}.
\label{gm48}
\end{equation}
account for multiple reflections of radiation in the gap between the two clouds. 
The matrices $\mathcal{Q}_{\bf d}^{\{21\}}$ and  $\mathcal{Q}_{\bf u}^{\{12\}}   $ are similar  to the quality factor or  ``$Q$'' of a resonant electromagnetic or mechanical system\,\cite{Q}. The elements of the reverberation matrices can become much larger than $1$ for clouds that are optically thick enough and lossless enough to permit large numbers of up and down reflections of radiation in the gap between a lower and upper cloud before the radiation is absorbed or transmitted. The rightmost terms of (\ref{gm46}) and (\ref{gm48}) denote pseudoinverse matrices, like those discussed in Section {\bf 2.1} of reference\,\cite{WH3}.

Reverberation matrices  satisfy {\it pushthrough identities}\,\cite{Pushthrough}
\begin{equation}
\mathcal{Q}_{\bf d}^{\{21\}}\mathcal{S}^{\{2\}}_{\bf d u}=\mathcal{S}^{\{2\}}_{\bf d u}\mathcal{Q}_{\bf u}^{\{12\}},
\label{gm50}
\end{equation}
and
\begin{equation}
\mathcal{S}^{\{1\}}_{\bf u d}\mathcal{Q}_{\bf d}^{\{21\}}=\mathcal{Q}_{\bf u}^{\{12\}}\mathcal{S}^{\{1\}}_{\bf u d}.
\label{gm52}
\end{equation}
The indices of the scattering matrices  $\mathcal{S}^{\{2\}}_{\bf d u}$ or $\mathcal{S}^{\{1\}}_{\bf u d}$ do not change as they are ``pushed through'' the reverberation matrices  $\mathcal{Q}_{\bf u}^{\{12\}}$ or $\mathcal{Q}_{\bf d}^{\{21\}}$.   The indices of  $\mathcal{Q}_{\bf u}^{\{12\}}$ or $\mathcal{Q}_{\bf d}^{\{21\}}$  toggle between their two possible values when the matrices $\mathcal{S}^{\{2\}}_{\bf d u}$ or $\mathcal{S}^{\{1\}}_{\bf u d}$  are pushed through them. From (\ref{gm46}) and (\ref{gm48}) we see that
\begin{equation}
\left[\begin{array}{c}\mathcal{Q}_{\bf d}^{\{21\}}\\ \mathcal{Q}_{\bf u}^{\{12\}}\end{array}\right]\to
\left[\begin{array}{c}\mathcal{M}_{\bf d}\\ \mathcal{M}_{\bf u}\end{array}\right]   \quad\hbox{if}\quad
\mathcal{S}^{\{2\}}_{\bf d u}\to \breve 0\quad\hbox{ and/or}\quad  \mathcal{S}^{\{1\}}_{\bf ud}\to \breve 0,
\label{gm54}
\end{equation}
If the gap-facing side of one or both clouds is non-reflective, there is no enhancement of the gap intensity by reverberation.
\subsection{Discrete Green's matrices  \label{scg}}
We use (\ref{gm44}) and (\ref{sa24}) to write the block matrix $G$ of (\ref{fs8}) for the discrete Green's matrices as
\begin{eqnarray}
G&=&(\hat 1-X)^{-1}P \nonumber\\
&=&\left[\begin{array}{rrrr}
\mathcal{Q}^{\{21\}}_{\bf d}&\mathcal{S}^{\{2\}}_{\bf d u}\mathcal{Q}^{\{12\}}_{\bf u}&\breve 0&\breve 0 \\\mathcal{S}^{\{1\}}_{\bf u d}
\mathcal{Q}^{\{21\}}_{\bf d}&\mathcal{Q}^{\{12\}}_{\bf u} &\breve 0&\breve 0 \\ 
\mathcal{S}^{\{1\}}_{\bf d d}\mathcal{Q}^{\{21\}}_{\bf d}&\mathcal{S}^{\{1\}}_{\bf d d}\mathcal{S}^{\{2\}}_{\bf d u}\mathcal{Q}^{\{12\}}_{\bf u}&\mathcal{M}_{\bf d}&\breve 0 \\ 
\mathcal{S}^{\{2\}}_{\bf u u}\mathcal{S}^{\{1\}}_{\bf u d}\mathcal{Q}^{\{21\}}_{\bf d}& \mathcal{S}^{\{2\}}_{\bf uu}\mathcal{Q}^{\{12\}}_{\bf u}  &\breve 0&\mathcal{M}_{\bf u}\\ 
\end{array}\right]\left[\begin{array}{cccc} 
\breve 0&\breve 0&\hat 1&\breve 0\\ 
\breve 0&\hat 1&\breve 0&\breve 0\\ 
\hat 1&\breve 0&\breve 0&\breve 0\\
\breve 0&\breve 0&\breve 0&\hat 1\\
\end{array}\right]\nonumber\\
&=&
\left[\begin{array}{rrrr}\breve 0 &\mathcal{S}^{\{2\}}_{\bf d u}\mathcal{Q}^{\{12\}}_{\bf u}&\mathcal{Q}^{\{21\}}_{\bf d}&\breve 0\\
\breve 0&\mathcal{Q}^{\{12\}}_{\bf u}&\mathcal{S}^{\{1\}}_{\bf u d}\mathcal{Q}^{\{21\}}_{\bf d}&\breve 0\\
\mathcal{M}_{\bf d}&\mathcal{S}^{\{1\}}_{\bf d d}\mathcal{S}^{\{2\}}_{\bf d u}\mathcal{Q}^{\{12\}}_{\bf u}&\mathcal{S}^{\{1\}}_{\bf d d}\mathcal{Q}^{\{21\}}_{\bf d}&\breve 0 \\
\breve 0&\mathcal{S}^{\{2\}}_{\bf uu}\mathcal{Q}^{\{12\}}_{\bf u} &\mathcal{S}^{\{2\}}_{\bf u u}\mathcal{S}^{\{1\}}_{\bf u d}\mathcal{Q}^{\{2 1\}}_{\bf d}&\mathcal{M}_{\bf u}\\
\end{array}\right]\nonumber\\
&=&\left[\begin{array}{ll}G^{[1 1\}}&G^{[1 2\}}\\G^{[2 1\}}&G^{[2 2\}}\end{array}\right].
\label{scg2}
\end{eqnarray}
The  reverberation matrices $\mathcal{Q}_{\bf d}^{\{21\}}$ and  $\mathcal{Q}_{\bf u}^{\{12\}}$  of (\ref{scg2})  were given by (\ref{gm46}) and (\ref{gm48}).
From inspection of (\ref{scg2}) we see that the elements of the Green's coefficient matrix are

\begin{equation}
G^{[11\}}=
\left[\begin{array}{rrrr}\breve 0 &\mathcal{S}^{\{2\}}_{\bf d u}\mathcal{Q}^{\{12\}}_{\bf u}\\
\breve 0&\mathcal{Q}^{\{12\}}_{\bf u}\\
\end{array}\right]\quad G^{[12\}}=
\left[\begin{array}{rrrr}\mathcal{Q}^{\{21\}}_{\bf d}&\breve 0\\ \mathcal{S}^{\{1\}}_{\bf u d}\mathcal{Q}^{\{21\}}_{\bf d}&\breve 0\\
\end{array}\right],
\label{scg4}
\end{equation}
and
\begin{equation}
G^{[2 1\}}=
\left[\begin{array}{rrrr}
\mathcal{M}_{\bf d}&\mathcal{S}^{\{1\}}_{\bf d d}\mathcal{S}^{\{2\}}_{\bf d u}\mathcal{Q}^{\{12\}}_{\bf u} \\
\breve 0&\mathcal{S}^{\{2\}}_{\bf uu}\mathcal{Q}^{\{12\}}_{\bf u} 
\end{array}\right]\quad 
G^{[2 2\}}=\left[\begin{array}{rrrr}
\mathcal{S}^{\{1\}}_{\bf d d}\mathcal{Q}^{\{21\}}_{\bf d}&\breve 0 \\
\mathcal{S}^{\{2\}}_{\bf u u}\mathcal{S}^{\{1\}}_{\bf u d}\mathcal{Q}^{\{2 1\}}_{\bf d}&\mathcal{M}_{\bf u}\\
\end{array}\right].
\label{scg6}
\end{equation}

From (\ref{fs12}) and (\ref{scg6}) we see that the thermal source vector of the equivalent cloud is
\begin{equation}
|\dot J\}=|\dot U^{[2]}\}= G^{[21\}}|\dot J^{\{1\}}\}+G^{[2 2\}}|\dot J^{\{2\}}\}
\label{scg8}
\end{equation}
or
\begin{equation}
\left[\begin{array}{c}|\dot U^{[2]}_{\bf d}\}\\ |\dot U^{[2]}_{\bf u}\}\end{array}\right]=
\left[\begin{array}{c}|\dot J_{\bf d}^{\{1\}}\}+\mathcal{S}^{\{1\}}_{\bf d d}\mathcal{S}^{\{2\}}_{\bf d u}\mathcal{Q}^{\{12\}}_{\bf u}|\dot J_{\bf u}^{\{1\}}\}
+\mathcal{S}^{\{1\}}_{\bf d d}\mathcal{Q}^{\{21\}}_{\bf d}|\dot J_{\bf d}^{\{2\}}\}
\\ 
\mathcal{S}^{\{2\}}_{\bf uu}\mathcal{Q}^{\{12\}}_{\bf u}|\dot J_{\bf u}^{\{1\}}\}+\mathcal{S}^{\{2\}}_{\bf u u}\mathcal{S}^{\{1\}}_{\bf u d}\mathcal{Q}^{\{2 1\}}_{\bf d}|\dot J_{\bf d}^{\{2\}}\}+|\dot J_{\bf u}^{\{2\}}\}
\end{array}\right].
\label{scg10}
\end{equation}
The thermal source vector $|\dot J\}=|\dot U^{[2]}\}=|\dot I^{(\rm out)}\}$ of the equivalent cloud, given by  (\ref{scg8}),  is a linear combination of  the thermal source vectors, $|\dot J^{\{1\}}\}$ and  $|\dot J^{\{2\}}\}$ of the two component clouds.
\subsection{Scattering coefficients\label{sc}}
We can use (\ref{gm44}) and (\ref{sa30}) to write the block matrix $W$ of (\ref{fs22}) for the scattering coefficients as
\begin{eqnarray}
W&=&(\hat 1-X)^{-1}Y =\left[\begin{array}{l}W^{[1]}\\ W^{[2]}\end{array}\right]\nonumber\\
&=&\left[\begin{array}{rrrr}
\mathcal{Q}^{\{21\}}_{\bf d}&\mathcal{S}^{\{2\}}_{\bf d u}\mathcal{Q}^{\{12\}}_{\bf u}&\breve 0&\breve 0 \\\mathcal{S}^{\{1\}}_{\bf u d}
\mathcal{Q}^{\{21\}}_{\bf d}&\mathcal{Q}^{\{12\}}_{\bf u} &\breve 0&\breve 0 \\ 
\mathcal{S}^{\{1\}}_{\bf d d}\mathcal{Q}^{\{21\}}_{\bf d}&\mathcal{S}^{\{1\}}_{\bf d d}\mathcal{S}^{\{2\}}_{\bf d u}\mathcal{Q}^{\{12\}}_{\bf u}&\mathcal{M}_{\bf d}&\breve 0 \\ 
\mathcal{S}^{\{2\}}_{\bf u u}\mathcal{S}^{\{1\}}_{\bf u d}\mathcal{Q}^{\{21\}}_{\bf d}& \mathcal{S}^{\{2\}}_{\bf uu}\mathcal{Q}^{\{12\}}_{\bf u}  &\breve 0&\mathcal{M}_{\bf u}\\ 
\end{array}\right]
\left[\begin{array}{ll}\mathcal{S}^{\{2\}}_{\bf d d}&\breve 0\\
\breve 0&\mathcal{S}^{\{1\}}_{\bf u u}\\
\breve 0&\mathcal{S}^{\{1\}}_{\bf d u} \\
\mathcal{S}^{\{2\}}_{\bf ud}&\breve 0 \\
\end{array}\right],\nonumber\\
&=&
\left[\begin{array}{rrr}\mathcal{Q}^{\{21\}}_{\bf d}\mathcal{S}^{\{2\}}_{\bf d d}&&
\mathcal{S}^{\{2\}}_{\bf d u}\mathcal{Q}^{\{12\}}_{\bf u}\mathcal{S}^{\{1\}}_{\bf u u} \\  
\mathcal{S}^{\{1\}}_{\bf u d}\mathcal{Q}^{\{21\}}_{\bf d}\mathcal{S}^{\{2\}}_{\bf d d}
 && \mathcal{Q}^{\{12\}}_{\bf u} \mathcal{S}^{\{1\}}_{\bf u u} \\
\mathcal{S}^{\{1\}}_{\bf d d}\mathcal{Q}^{\{21\}}_{\bf d}\mathcal{S}^{\{2\}}_{\bf d d}&& \mathcal{S}^{\{1\}}_{\bf d u}+\mathcal{S}^{\{1\}}_{\bf d d}\mathcal{S}^{\{2\}}_{\bf d u}\mathcal{Q}^{\{12\}}_{\bf u} \mathcal{S}^{\{1\}}_{\bf u u} \\
\mathcal{S}^{\{2\}}_{\bf u d}+
\mathcal{S}^{\{2\}}_{\bf u u}\mathcal{S}^{\{1\}}_{\bf u d}\mathcal{Q}^{\{21\}}_{\bf d}\mathcal{S}^{\{2\}}_{\bf d d}&& \mathcal{S}^{\{2\}}_{\bf uu}\mathcal{Q}^{\{12\}}_{\bf u} \mathcal{S}^{\{1\}}_{\bf u u}\\
\end{array}\right].
\label{sc4}
\end{eqnarray}
We see that the scattering coefficients for the  internal gap between the first and second cloud, the top half of the $4\times 2$ block matrix on the last line of (\ref{sc4}), is

\begin{eqnarray}
W^{[1]}&=&
\left[\begin{array}{rrr}\mathcal{Q}^{\{21\}}_{\bf d}\mathcal{S}^{\{2\}}_{\bf d d}&&
\mathcal{S}^{\{2\}}_{\bf d u}\mathcal{Q}^{\{12\}}_{\bf u}\mathcal{S}^{\{1\}}_{\bf u u} \\  
\mathcal{S}^{\{1\}}_{\bf u d}\mathcal{Q}^{\{21\}}_{\bf d}\mathcal{S}^{\{2\}}_{\bf d d}
 && \mathcal{Q}^{\{12\}}_{\bf u} \mathcal{S}^{\{1\}}_{\bf u u} \\
\end{array}\right].
\label{sc6}
\end{eqnarray}
Eqs. (\ref{sc4}) and (\ref{fs30}) give the scattering coefficient of the equivalent cloud as the star product of the scattering matrices $\mathcal{S}^{\{2\}}$ and $\mathcal{S}^{\{1\}}$ of the individual clouds
\begin{equation}
W^{[2]}=\mathcal{S}^{\{\rm ev\}}=\mathcal{S}^{\{2\}}\rstar\mathcal{S}^{\{1\}}.
\label{sc7}
\end{equation}
\subsection{Star products}
We see from inspection of (\ref{sc7}) and  (\ref{sc4}) that the {\it star product} of the two scattering matrices $\mathcal{S}^{\{2\}}$ and $\mathcal{S}^{\{1\}}$ of individual clouds is given by the explicit formula
\begin{eqnarray}
\mathcal{S}^{\{2\}}\rstar\mathcal{S}^{\{1\}}
&=&\left[\begin{array}{rrr}
\mathcal{S}^{\{1\}}_{\bf d d}\mathcal{Q}^{\{21\}}_{\bf d}\mathcal{S}^{\{2\}}_{\bf d d}&& \mathcal{S}^{\{1\}}_{\bf d u}+\mathcal{S}^{\{1\}}_{\bf d d}\mathcal{S}^{\{2\}}_{\bf d u}\mathcal{Q}^{\{12\}}_{\bf u} \mathcal{S}^{\{1\}}_{\bf u u} \\
\mathcal{S}^{\{2\}}_{\bf u d}+
\mathcal{S}^{\{2\}}_{\bf u u}\mathcal{S}^{\{1\}}_{\bf u d}\mathcal{Q}^{\{21\}}_{\bf d}\mathcal{S}^{\{2\}}_{\bf d d}&& \mathcal{S}^{\{2\}}_{\bf uu}\mathcal{Q}^{\{12\}}_{\bf u} \mathcal{S}^{\{1\}}_{\bf u u}\\
\end{array}\right].
\label{sc8}
\end{eqnarray}
The  function  (\ref{sc8}) of two $2\times 2$ block matrices, $\mathcal{S}^{\{2\}}$ and $\mathcal{S}^{\{1\}}$ is often called a {\it Redheffer star product} after R. Redheffer\cite{Redheffer}, who first introduced it.  For three arbitrary matrices, $A$, $B$ and $C$, the Redheffer star product obeys the associative law of multiplication,
\begin{equation}
A \rstar B \rstar C =(A \rstar B)\rstar C = A\rstar(B\rstar C),
\label{sc10}
\end{equation}
but not the distributive or commutative laws.  
To our knowledge, the star product rule (\ref{sc8}) for calculating the equivalent scattering matrix for two clouds in series was first given Eq.  (235) of reference\,\cite{WH1}. There it was used to prove Eq. (254) of reference\,\cite{WH1}, which we rewrite as
\begin{equation}
\lvec\mu_i|\mathcal{S}|\mu_{i'})\ge 0.
\label{sc12}
\end{equation}
The elements of the scattering matrix in $\mu$-space are nonnegative.
\section{Summary}
We have shown how to use $2n$-stream multiple scattering theory to analyze 
radiation transfer through a stack of $m$ clouds.
This is a long paper which uses unfamiliar but powerful notation. To help those who are interested, we summarize the contents here.  

After a brief review of prior work in Section {\bf \ref{hist}}, we begin 
 Section~{\bf \ref{in}} with a discussion of the {\it equation of transfer}, the integro-differential equation (\ref{in10}) that describes the  rate of change of the monochromatic intensity $I(\mu,\tau)$ with vertical optical depth $\tau$, defined by (\ref{int0}). We assume axial symmetry, so there is no dependence of quantities on the azimuthal angle $\phi$, but there can be arbitrary dependence on the zenith angle $\theta$, or the corresponding direction cosine $\mu = \cos\theta$. The fundamental parameters of the equation of transfer (\ref{in10}) are  the single-scattering albedo $\tilde \omega$,  the scattering phase function $p(\mu,\mu')$,   and the Planck intensity $B=B(\tau)$ of (\ref{et10}) which describes the local intensity of thermally emitted radiation.

In Section {\bf \ref{int}} we review the $2n$-stream method for solving the equation of radiative transfer.
As illustrated in Fig. \ref{streams}, we represent axially symmetric radiation with $2n$ streams, directed along the colatitude angles, $\theta_i = \cos^{-1}\mu_i$. In accordance with (\ref{int2}), the Gauss-Legendre  direction cosines, $\mu_i=\cos\theta_i$, are the zeros of the Legendre polynomial $P_{2n}(\mu)$. These are illustrated in Fig. \ref{streams2}, where we also show the stream weights $w_i$, which can be evaluated with (\ref{int12}).  To simplify notation, we think of the weighted intensity $w_iI(\mu_i) =\lvec\mu_i|I\}$ of the $i$th stream as an element of the $2n\times 1$ intensity vector  $|I\}$ of (\ref{int14}).  

It is convenient to describe the intensity vector and related radiation-transfer vectors and matrices  with the aid of stream basis functions,   $|\mu_i)$ and $\lvec\mu_i|$ of (\ref{int16}) and (\ref{int18}).  As shown in (\ref{int20}) -- (\ref{int24}), these are $2n$-dimensional, orthonormal right and left eigenvectors of the direction cosine matrix $\hat\mu$ of (\ref{int26}). The Gauss-Legendre direction cosines $\mu_i$ are the eigenvalues of $\hat\mu$.  The upward and downward parts,  $\hat\mu_{\bf u}$ and $\hat\mu_{\bf d}$ of $\hat\mu$, are defined by (\ref{int54}) and (\ref{int56}).

In Section {\bf \ref{mm}} we review the multipole basis $\lvec l|$ and $|l)$ for describing the directional properties of radiation transfer in terms of  the intensity multipole moments $I_l=\lvec l|I\}$ of (\ref{mm4}).
The overlap coefficients $\lvec l|\mu_i)=P_l(\mu_i)$ and  $\lvec \mu_i|l)=w_i(2l+1)P_l(\mu_i)$, needed to transform vector amplitudes from the stream basis to the multipole basis, and vice versa, are given by (\ref{mm6}) and (\ref{mm8}). We also review the $2n$-stream approximations $E_{q}^{\{n\}}(\tau)$ of (\ref{mm22}) to the exponential integral functions $E_q(\tau)$ of (\ref{mm24}). The exponential integral functions conveniently account for the contribution of slant  paths to vertical radiation transfer of heat in purely absorptive atmospheres.

The abstract vector form of the equation of transfer for the $2n$-stream model.  is given in Section {\bf\ref{et}} by (\ref{et2}). This can be written as a set of $2n$ coupled linear equations. 
These are much easier to solve than the integro-differential equation (\ref{in10}) for the full continuum of direction cosines $\mu$.  The equation of transfer (\ref{et2}) is  parameterized by the $2n\times 2n$ exponentiation rate matrix, $\hat \kappa$,  of (\ref{et12}), the product of the direction secant matrix $\hat\varsigma$ of (\ref{int28}) and the multipole damping  matrix $\hat\eta$ of (\ref{et16}).  For the $2n$-stream model, the scattering phase function $p(\mu,\mu')$ of (\ref{in14}) is replaced by the scattering phase matrix, $\hat p=2\sum_{l=0}^{2n-1}p_l|l)\lvec l|$, of (\ref{et22}) which is diagonal in multipole space and parameterized by $2n$ multipole coefficients $p_l=p_0,p_1,p_2,\ldots,p_{2n-1}$. Phase matrices for isotropic scattering, Rayleigh scattering, forward and backward scattering  are given by (\ref{et24})  --  (\ref{et30}).

In Section {\bf\ref{flx}} we discuss the vector flux $|Z\}=4\pi\hat\mu|I\}$ of (\ref{flx2}), which describes the contributions of each of the $2n$ streams to vertical radiative heat flow. The scalar flux, $Z=\lvec 0|Z\}=Z_{\bf d}+Z_{\bf u}$, is the sum of of a negative downward part $Z_{\bf d}=\lvec 0|\mathcal{M}_{\bf d}|Z
\}$ of (\ref{flx8}) and a positive upward part $Z_{\bf u}=\lvec 0|\mathcal{M}_{\bf u}|Z
\}$ of (\ref{flx10}). 

In Section {\bf\ref{oir}} we discuss how the intensity vector $|I^{(\rm in)}\}$ of incoming radiation, and the intensity vector $|I^{(\rm out)}\}$ of outgoing radiation are related to the intensity vector $|I^{(0)}\}$ of radiation below the  cloud stack and the intensity vector $|I^{(m}\}$ for radiation above. The relationships are given by (\ref{oir2}), (\ref{oir6}), (\ref{oir10}) and  (\ref{oir12}). The corresponding relations for vector and scalar fluxes, $|Z\}$ and $Z$ are reviewed in the final paragraphs of the section.

In Section {\bf\ref{exin}} we discuss the partitioning of cloud intensities, $|I\}=|\dot I\}+|\ddot I\}$, into a part  $|\dot I\}$, produced by thermal emission of cloud particulates and gas molecules, and a part $|\ddot I\}$, produced by external radiation incident onto the top and bottom of the cloud stack. We denote thermally generated parts with a single dot, and externally generated parts by a double dot. In  Eq.  (\ref{exin10}), $|\ddot I^{(\rm out)} \}=\mathcal{S}|\ddot I^{(\rm in)} \}$,  the scattering matrix $S$ for a  cloud is defined as the  matrix coefficient of proportionality  between incoming intensity vector $|\ddot I^{(\rm in)} \}$ from external sources and the scattered outgoing intensity vector $|\ddot I^{(\rm out)} \}$.  Similarly, in Eq. (\ref{exin12}), $|\ddot Z^{(\rm out)}\}=\Omega|\ddot Z^{(\rm in)}\}$, the albedo matrix 
$\Omega$ is defined as the matrix coefficient of proportionality between the  incoming  flux vector $|\ddot Z^{(\rm in)}\}$ from external sources and the scattered, outgoing flux   vector $|\ddot Z^{(\rm out)}\}$. The albedo matrix  is a similarity transformation (\ref{exin14}) of the scattering matrix,  $\Omega =(\hat\mu_{\bf u}-\hat \mu_{\bf d})\mathcal{S}(\hat\mu_{\bf u}-\hat \mu_{\bf d})^{-1}$.

Eq.  (\ref{exin22}) shows that for a non isothermal cloud, the source vector, $|\dot J\}=|\dot I^{(\rm out)}\}=\int_0^{\tau}d\tau' G(\tau')|0)B(\tau')$, is the superposition  contributions from infinitesimal isotropic sources, the products  of the  infinitesimal optical depths of the source $d\tau'$, the monopole basis vector $|0)$ of (\ref{mm10}), and the Planck intensities $B(\tau')$ at optical depth $\tau'$ of the source above the bottom of the cloud. Before reaching the top or bottom surface of the cloud, of total optical depth $\tau$, the infinitesimal contributions are multiplied by  the Green's matrix $G(\tau')$.  For isothermal clouds, with constant Planck intensity $B$, the source vector simplifies to (\ref{exin20}), $|\dot J\}=\mathcal{E}|0)B$, where the emissivity matrix $\mathcal{E}$ is related to the scattering matrix by Kirchhoff's radiation law of (\ref{exin24}), $\mathcal{E}=\hat 1 -\mathcal{S}$.

In Section {\bf\ref{sm}} we review how to calculate the scattering matrix $\mathcal{S}$ for a single homogeneous cloud, that is,  a cloud with an exponentiation-rate matrix $\hat \kappa$ that is independent of optical depth $\tau'$ above the bottom. In (\ref{sm2}) we define the  inverse of the exponentiation-rate matrix $\hat \kappa$ as the penetration-length matrix  $\hat \lambda=\hat\kappa^{-1}=\sum_{i=1}^{2n}\lambda_i|\lambda_i)\lvec \lambda _i|$.  The eigenvalues of $\hat\lambda$ are $\lambda_i$ and the right and left eigenvectors are $|\lambda_i)$ and $\lvec\lambda_i|$, the $\lambda$-space bases. The eigenvectors  are used to construct   the overlap matrix $\mathcal{C}$ of (\ref{sm18}) between $\mu$-space bases $\lvec \mu_{i'}|$  and $\lambda$-space bases $|\lambda_{i})$. We use $\mathcal{C}$  together with the eigenvalues $\lambda_i=1/\kappa_i$ to construct incoming and outgoing matrices $\mathcal{I}$ and $\mathcal{O}$ of (\ref{sm24}) and (\ref{sm26}). In turn, these are used in (\ref{sm0})
 to write the scattering matrix as $\mathcal{S}=\mathcal{O}\mathcal{I}^{-1}$. 

In Section {\bf \ref{gf}} we show that the continuous Green's matrix
$G(\tau')$ for homogeneous clouds is a linear combination (\ref{gf2}) of the matrices $\mathcal{R}(\tau')$ of (
\ref{gf4}) and  $\mathcal{Q}(\tau')$ of (\ref{gf6}), which are 
constructed from the overlap matrix $\mathcal{C}$ of (\ref{sm18}) in much the same way as   $\mathcal{I}$ and $\mathcal{O}$.  

In Section {\bf\ref{bc}} we discuss the simple properties of black clouds, with vanishing scattering matrices, $\mathcal{S}=\breve 0$. The emissivity matrices of black clouds are identity matrices, $\mathcal{E}=\hat 1$. For black clouds, the upward and downward Planck emissivities are both unity,  $\varepsilon_{\bf u}=\varepsilon_{\bf d}=1$.   
In Section {\bf\ref{ter}} we discuss the  upward and downward Planck emissivity, $\varepsilon_{\bf u}=\lvec 0|\hat\mu_{\bf u}\mathcal{E}|0)/\lvec 0|\hat\mu_{\bf u}|0)$ of  (\ref{ter8}) and $\varepsilon_{\bf d}=\lvec 0|\hat\mu_{\bf d}\mathcal{E}|0)/\lvec 0|\hat\mu_{\bf d}|0)$ of  (\ref{ter12}), 
for non-black clouds.   Fig. \ref{emis2} shows how the Planck emissivities $\varepsilon_{\bf u}=\varepsilon_{\bf d}$ of an optically thick, homogeneous cloud depend on the single-scattering albedo $\tilde\omega$ and on the scattering phase matrices of  $\hat p$ of (\ref{et24}) -- (\ref{et30}).

In Section {\bf\ref{qir}} we discuss {half isotropic} incoming radiation (\ref{hi2}), $|\ddot I^{(\rm in)}\}=\mathcal{M}_{\bf d}|0)B^{\{\rm in\}}_{\bf d}+\mathcal{M}_{\bf u}|0)B^{\{\rm in\}}_{\bf u}$, which is parameterized by only two numbers, the Planck intensities $B^{\{\rm in\}}_{\bf d}$ and $B^{\{\rm in\}}_{\bf u}$ of downwelling and upwelling half-isotropic intensity incident on the top and bottom of the cloud stack. 

In Section {\bf\ref{hc}} we show that external radiation incident on   the top and bottom of an isolated cloud $c$ causes the radiative heating, $\ddot H^{\{c\}}
=\ddot Z^{(\rm in)}- \ddot Z^{(\rm out)}=\lvec 0|\mathcal{A}^{\{c\}}|\ddot Z^{(\rm in)}\}$,  of (\ref{h4}).  According to (\ref{h6}), the absorptivity matrix,  $\mathcal{A}^{\{c\}}=\hat 1-\Omega^{\{c\}}$, is the complement of the albedo matrix, $\Omega^{\{c\}}$. The absorptivity matrix is a similarity transformation of the emissivity matrix, $\mathcal{A}^{\{c\}}=(\hat\mu_{\bf u}-\hat \mu_{\bf d})\mathcal{E}^{\{c\}}(\hat\mu_{\bf u}-\hat \mu_{\bf d})^{-1}$. 
Thermal emission by particulates and gas molecules of the isolated cloud causes the radiative cooling $\dot C^{\{c\}}=\dot Z^{(1)}-\dot Z^{(0)}$ of (\ref{h8}). 
According to (\ref{h2}), the net heating rate of a isolated cloud  is $R^{\{c\}} =\ddot H^{\{c\}} -\dot C^{\{c\}}$,  the excess of the heating rate $\ddot H^{\{c\}}$ over the cooling rate $\dot C^{\{c\}}$.

In Section {\bf \ref{re}} we show that for a fixed intensity $|\ddot I^{(\rm in)}\}$ of incoming radiation, a cloud can heat or cool until the net heating rate vanishes, $R^{\{c\}}=0$, a state of {\it radiative equilibrium}. For an isolated homogeneous and isothermal cloud, heated by half isotropic incoming radiation of (\ref{hi2}), we showed in (\ref{re8}) that  the Planck intensity of a cloud in radiative equilibrium is $B^{\{1\}}=(B^{\{\rm in\}}_{\bf u}+B^{\{\rm in\}}_{\bf d})/2$, the average Planck intensity of the upward and downward incoming radiation.

 Fig. \ref{one1} of Section {\bf \ref{nex}}  shows a single cloud in complete thermal equilibrium with incoming radiation of the same temperature.  In Fig. \ref{one2}  the downward incoming radiation of Fig. \ref{one1} has been removed. This diminishes the heating rate $\ddot H^{\{c\}}$ by a factor of 2 but does not affect the cooling rate $\dot C^{\{c\}}$.  So the cooling rate $\dot C^{\{c\}}$ becomes twice as large as the heating rate $\ddot H^{\{c\}}$, and the net heating $R^{\{c\}}=\ddot H^{\{c\}}-\dot C^{\{c\}}$ of  (\ref{h2}) is negative. There is net cooling
of the cloud and unless non radiative heating sources are present, the temperature and Planck intensity of the clould will decrease.   Fig. \ref{one3} shows what happens if the cloud of Fig. \ref{one2} is allowed to cool to radiative equilibrium.  Then the Planck intensity $B^{\{1\}}$ of the cloud is halved, in accordance with (\ref{re8}). The cooling rate $\dot C^{\{c\}}$ becomes equal to the one-sided heating rate  $\ddot H^{\{c\}}$ and the net heating rate vanishes, $R^{\{c\}}=\ddot H^{\{c\}}-\dot C^{\{c\}}=0$.

In Section {\bf\ref{cs}} we discuss  stacks of $m>1$ clouds.  The generalization of thermal source vectors $|\dot J\}$ for a single cloud to $|\dot J^{\{c\}}\}$ for cloud $c$ in the stack  is given by (\ref{rrs6}) for an arbitrary temperature distribution inside each cloud, or by (\ref{rrs8}) if each cloud is isothermal.  The  gap intensities $|I^{(g)}\}$ are defined by (\ref{rrs9a}). The  intensity vectors $|I^{(0)}\}$ and $|I^{(m)}\}$ for the gaps below and above the stack  are special cases given by  (\ref{oir10}) and (\ref{oir12}).
The basic equations of radiative transfer, (\ref{ic2}) -- (\ref{ic18}),  for a stack of $m>1$ clouds are illustrated by Fig. \ref{seriesclouds6}.

In Section {\bf\ref{sa}} we write the basic equations  (\ref{ic2}) -- (\ref{ic18})  as the formally simpler stack equation, $| U]=X| U]+ P|\dot J]+Y|\ddot I^{\{\rm in\}}\}$ of  (\ref{sa2}).
 The {\it unknown intensity} stack vector $|U]$ of (\ref{sa4})  has $m$ block elements, the $2n\times 1$ vectors  $|U^{[x]}\}$, indexed by $x=1,2,3,\ldots,m$. For $x=1,2,3,\ldots, m-1$, the unknown intensity element $|U^{[x]}\}$ describes intensity in the gap between the cloud $x$ and the cloud $x+1$ and is equal to the gap intensity element of (\ref{rrs9a}), $|U^{[x]}\}=|I^{(x)}\}$.  For the special case of $x=m$, the element  $|U^{[x]}\}$  is  the same as the outgoing intensity vector (\ref{oir6}) of $|I^{(\rm out)}\}$ or the equivalent thermal source vector $|\dot J^{\{\rm ev\}}\}$ of   the stack, $|U^{[m]}\}=|I^{(\rm out)}\}=|\dot J^{\{\rm ev\}}\}$.

The generation matrix $X$ that occurs in  (\ref{sa2}) is constructed from the scattering matrices $S^{\{c\}}$ of the individual clouds, as shown by (\ref{sa12}). 
The permutation matrix $P$ that occurs in  (\ref{sa2})   determines how the upward and downward parts of the stack vector $|\dot J]$ of thermal sources, given by  (\ref{sa20}),  contribute to the unknown intensities $|U^{[x]}\}$ in accordance with (\ref{sa18}).   As shown in (\ref{sa20}) $|\dot J]$ is simply a concatenation of the thermal source vectors  $|\dot J^{\{c\}}\}$ of the individual clouds. The structure of $P$ is shown by (\ref{sa22}).  In accordance with (\ref{sa26}), the insertion matrix $Y$ of (\ref{sa2})  describes how much of the incoming radiation $|\ddot I^{(\rm in)}\}$ contributes to the intensity in the gaps $g=1$ above the bottom cloud and $g=m-1$ below the top cloud, and also to the intensities scattered from the bottom and top of the stack. As shown by (\ref{sa28}), $Y$ is constructed from the scattering matrices 
$\mathcal{S}^{\{1\}}$ and $\mathcal{S}^{\{m\}}$ and is independent of the scattering matrices $\mathcal{S}^{\{c\}}$ of any interior clouds with $ 1<c< m$.

As described in  Section {\bf{\ref{te}}}, for thermal equilibrium, with the same Planck intensity $B$ for all the clouds and for the external radiation, the stack vector $|U]$ for ``unknown'' intensities is actually known exactly from basic thermodynamics and given by,  $|U]=|0]B$. The monopole stack vector $|0]$  is defined by (\ref{te4}) as a concatenation of $m$ copies of the $2n\times 1$ monopole stack vectors  $|0)$ of (\ref{mm10}). The thermal equilibrium limit of the stack equation provides a simple identity, $(\hat 1-X- P[\mathcal{E}])|0]=Y|0)$ of (\ref{te12}) that can be used as a numerical consistency check of the matrices $X$, $P$, $Y$ and  the stack emission matrix $[\mathcal{E}]$ of (\ref{te8}).

In Section {\bf \ref{fs}} we discuss formal solutions for the unknown intensity $|U]$ of the stack equation (\ref{sa2}). According to (\ref{fs4}),  the solution is $|U]=G|\dot J]+W|\ddot I^{(\rm in)}\}$.  The first term, $G|\dot J]$, which represents thermally generated radiation, is the product of the Green's matrix  $G$ of (\ref{fs8}), a concatenation of the discrete Green's matrices of (\ref{fs8}), and the thermal source stack vector $|\dot J]$ of (\ref{sa20}). The second term, $W|\ddot I^{(\rm in)}\}$, is the product of the  scattering-coefficient  matrix $W$ of (\ref{fs22}) and the incoming intensity $|I^{\{\rm in\}}\}$.

The Green's coefficients $G^{[x, c\}}$ of (\ref{fs6}) quantify how much the thermal source vector $|\dot J^{\{c\}}\}$ contributes to  $|\dot U^{[x]}\}$.  In accordance with (\ref{fs12}), the equivalent thermal source vector of the entire cloud stack is $|\dot J^{\{\rm ev\}}\} =\sum_{c=1}^{m}G^{[m, c\}}|\dot J^{\{c\}}\}$, a weighted average of the thermal source vectors of the individual clouds. 

The scattering coefficient $W^{[x]}$ of (\ref{fs20}) can be thought of as a generalized scattering matrix that   quantifies how much the incoming radiation  $|\ddot I^{(\rm in)}\}$ contributes to  $|\ddot U^{[x]}\}$, the  intensity in the internal gap $x$, where $1\le x\le m-1$.  According to   (\ref{sa8}) and (\ref{exin10}), for $x=m$, 
the unknown intensity vector is simply the outgoing, scattered intensity, $|\ddot U^{[m]}\}=W^{[m]}|\ddot I^{\{\rm in\}}\}   =   |\ddot I^{(\rm out)}\}=\mathcal{S}^{\{\rm ev\}}|\ddot I^{\{\rm in\}}\} $. Therefore
the scattering coefficient  $W^{[m]}$ is equal to the equivalent scattering matrix $\mathcal{S}^{\{\rm ev\}}$ of the entire cloud stack, $W^{[m]}=\mathcal{S}^{\{\rm ev\}}=\mathcal{S}^{\{m\}}\rstar\mathcal{S}^{\{m-1\}}\rstar\cdots\rstar\mathcal{S}^{\{2\}}\rstar
\mathcal{S}^{\{1\}}$, in accordance with (\ref{fs32}).  $\mathcal{S}^{\{\rm ev\}}$ is the Redheffer star product of the scattering matrices of the individual clouds, as we discuss in the later  Section {\bf \ref{dc}}. The Kirchhoff identity
$\sum_{c=1}^m G^{[m c\}}\mathcal{E}^{\{c\}}+
\mathcal{S}^{\{m\}}\rstar\mathcal{S}^{\{m-1\}}\rstar\cdots\rstar\mathcal{S}^{\{2\}}\rstar
\mathcal{S}^{\{1\}}=\hat 1$ of (\ref{fs40})  provides a useful consistency check for the cloud scattering matrices 
$\mathcal{S}^{\{c\}}$, emissivity matrices,  $\mathcal{E}^{\{c\}}=\hat 1-\mathcal{S}^{\{c\}}$, and Green's matrices $ G^{[m c\}}$ of the cloud stack.

In Section {\bf\ref{gi}} we review how the the elements of the unknown-intensity stack vector $|U]$ can be used to construct  the  stack vector $|I)$ for gap intensities $|I^{(g)}\}$ of (\ref{gi2}). $|I)$ has $m+1$ block elements, one more  than  the $m$ elements of the  stack vector $|U]$ of (\ref{sa4}).

In  Section {\bf \ref{bb}}  we outline one of the simplest ways to characterize a stack of clouds. We assume each cloud $c$ is isothermal, with the Planck intensity $B^{\{c\}}$.  We concatenate these to form the  $m\times 1$ scalar stack vector $|B]$ of (\ref{bb14}).  We assume the  incoming intensity $|\ddot I^{(\rm in)}\}$ is the half isotropic radiation of (\ref{hi2}) with the downward and upward Planck intensities,  $B^{\{\rm in\}}_{\bf d}$ and  $B^{\{\rm in\}}_{\bf u}$, which we concatenate as the $2\times 1$ stack vector $|B^{\{\rm in\}}\rangle$ of (\ref{hi20}). Then the stack vector $|Z)$ of scalar fluxes  is given by the formally simple equation
$|Z)=4\pi\dot M|B]+4\pi\ddot N|B^{\{\rm in\}}\rangle$ of (\ref{bb18}).  The relatively small coupling matrices $\dot M$ 
and $\ddot N$ are given by (\ref{bb5a}), (\ref{bb8a}) and (\ref{bb12a}).

In Section {\bf\ref{hcs}} we show that for the simple model of Section  {\bf \ref{bb}} the net  heating rates $R^{\{c\}}$ of the clouds $c$, concatenated as the $m\times 1$ stack vector $|R]$ of (\ref{hcs2}), are given by the formally simple expresssions 
of (\ref{hc2}) and (\ref{hcs6}), $|R]= -\Delta|Z)=-4\pi \Delta \dot M|B]-4\pi \Delta \ddot N|B^{\{\rm in\}}\rangle$. The differencing matrix $\Delta$ is given by (\ref{hcs4}). In radiative equilibrium, when all heating rates vanish and $|R]=\breve 0$, the Planck brightnesses of the isothermal clouds are given by (\ref{hcs8}) as $|B]=-(\Delta \dot M)^{-1}\Delta \ddot N|B^{\{\rm in\}}\rangle$. 

Numerical examples of radiation transport by 3-cloud stacks are discussed in Section {\bf\ref{ex}}.  Fig. \ref{stackap5} shows a 3-cloud stack in complete thermal equilibrium with incoming  radiation. In Fig. \ref{stackap5u}  the downward incoming radiation of Fig. \ref{stackap5} has been removed. This diminishes the heating rates for all the clouds, but the intrinsic cooling rates $\dot C^{\{c\}}$ remain the same. So all three clouds have a net cooling, $R^{\{c\}}<0$.
Fig. \ref{stackap5re} shows what happens if the clouds of Fig. \ref{stackap5u} are allowed to cool to radiative equilibrium,  when $R^{\{c\}}=0$. The  Planck intensities $B^{\{c\}}$ of (\ref{ex24}) for radiative equilibrium are calculated with (\ref{hcs8}). 
In Section {\bf \ref{ten}} we show
examples of radiation transfer in a 10-cloud stack.  Figs. \ref{ten1}  -- \ref{ten3}, are similar to the corresponding figures  Fig. \ref{stackap5} --  Fig. \ref{stackap5re} of a 3-cloud stack.  

In Section {\bf \ref{dc}} we give closed-form expressions for the properties of the equivalent cloud made up of $m=2$ single clouds. We show explicitly in (\ref{sc7}) that the scattering matrix $\mathcal{S}^{\{\rm ev\}}$ for the double cloud is the Redheffer star product  $\mathcal{S}^{\{\rm ev\}}=\mathcal{S}^{\{2\}}\rstar\mathcal{S}^{\{1\}}$  of the scattering matrix  $\mathcal{S}^{\{1\}}$   of the lower cloud and   $\mathcal{S}^{\{2\}}$  of the upper cloud. The elements of star-product matrices include factors of the reverberation matrices 
$\mathcal{Q}^{\{21\}}_{\bf d}$ of (\ref{gm46}) and $\mathcal{Q}^{\{12\}}_{\bf u}$ of (\ref{gm48}), which quantify the buildup of intensity due to multiple upward and downward scattering in the gap between the clouds.

The numerical methods described in this paper are computationally efficient and readily model  cloud stacks with $m\gg 10$. Each cloud of the stack can have its own, temperature $T^{\{c\}}$, single-scattering albedo
 $\tilde\omega^{\{c\}}$, scattering phase matrix $\hat p^{\{c\}}$ and optical depth $\tau^{\{c\}}$. For thermal radiation, some ``clouds'' can have transmission, absorption and emission but negligible scattering, like layers of clear air containing the main greenhouse gases, water vapor and carbon dioxide.  Clouds which also include water droplets, ice crystallites,  or other particulates can have transmission, absorption, scattering and emission. 

Inspection of satellite images of the Earth shows that  half or more of Earth's atmosphere can be approximated as stacks of clouds.  The $2n$-stream methods outlined here can be used to model radiation transfer, including the heating or cooling of the various clouds. Regions of Earth's atmosphere which are well approximated as stacks of clouds are equivalent to a single inhomogeneous cloud, with its bottom at the Earth's surface and its top at the mesopause, above which changes in the radiation flux are neglible. The scattering matrix $\mathcal{S}^{\{\rm ev\}}$ of the equivalent cloud tells how much of the radiation generated at the surface is transmitted to space without absorption or reflection back to the surface. The thermal source vector $|\dot J^{\{\rm ev\}}\}$ of the equivalent cloud tells how much of the thermal radiation generated inside the clouds is emitted to space above or to the surface below.
\section*{Acknowledgements} We are grateful to Professor Todd Palmer for a careful review of an early draft of this paper, and for pointing out related literature.
 The Canadian Natural Science and Engineering Research  Council provided financial support for one of us.

\end{document}